\documentclass[reqno]{amsart}
\usepackage{times}
\usepackage{amsmath}
\usepackage{amsfonts}
\usepackage{amssymb}
\usepackage{latexsym}
\usepackage{graphicx}
\usepackage{bm}
\usepackage{verbatim}
\usepackage{cite}

\author[Phillies]{George D. J. Phillies}
\address{Department of Physics\\
Worcester Polytechnic Institute\\
Worcester, MA 01609}
\title[Hydrodynamic Scaling Model]{The Hydrodynamic Scaling Model \\ for the Dynamics\\of Non-Dilute Polymer Solutions: \\ A Comprehensive Review}
\email{phillies@4liberty.net}
\urladdr{http://users.wpi.edu/~phillies}
\date{\today}

\begin{document}

\begin{abstract}
This article presents a comprehensive review of the Hydrodynamic Scaling Model for the dynamics of polymers in dilute and nondilute solutions. The Hydrodynamic Scaling Model differs from some other treatments of non-dilute polymer solutions in that it takes polymer dynamics up to high concentrations to be dominated by solvent-mediated hydrodynamic interactions, with chain crossing constraints presumed to create at most secondary corrections. Many other models take the contrary stand, namely that chain crossing constraints dominate the dynamics of nondilute polymer solutions, while hydrodynamic interactions only create secondary corrections. This article begins with a historical review.  We then consider single-chain behavior, in particular the Kirkwood-Riseman model; contradictions between the Kirkwood-Riseman and more familiar Rouse-Zimm models are emphasized. An extended Kirkwood-Riseman model that gives interchain hydrodynamic interactions is developed and applied to generate pseudovirial series for the self-diffusion coefficient and the low-shear viscosity.  To extrapolate to large concentrations, rationales based on self-similarity and on the Altenberger-Dahler Positive-Function Renormalization Group are developed and applied to the pseudovirial series for $D_{s}$ and $\eta$. Based on the renormalization group method, a two-parameter temporal scaling \emph{ansatz} is invoked as a path to determining the frequency dependences of the storage and loss moduli.  A short description is given for each of the individual papers that developed the Hydrodynamic Scaling Model.   Phenomenological evidence supporting aspects of the model is noted. Finally, directions for future development of the Hydrodynamic Scaling Model are presented.

\end{abstract}

\maketitle

\tableofcontents

\section{Introduction}

\subsection{The Hydrodynamic Scaling Model}

The nature of the dynamics of polymers in non-dilute solutions remains a substantial challenge for chemical physics.  At one time, it appeared likely that the tube model for polymer melts could be applied to polymer solutions\cite{doi1986a}. Moderately more recent reviews\cite{lodge1990a,skolnick1990a,phillies1986a} instead concluded that reptation/tube/\-scaling models are not applicable to polymer solutions, at least for solutions of polymers in commonly studied concentration and molecular weight ranges.  A recent monograph-length examination of a wide range of polymer solution properties\cite{phillies2011a} recently came to a similar conclusion. The nature of polymer motion in polymer melts and covalently crosslinked gels remains a separate issue not within the remit of this review.

Fortunately, there is an alternative to reptation/tube/scaling models, namely the Hydrodynamic Scaling Model.  The hydrodynamic scaling and reptation/scaling models differ in that reptation models of polymer solutions treat the chain crossing constraint as the dominant interaction and hydrodynamic interactions as providing secondary corrections, while the Hydrodynamic Scaling Model takes hydrodynamic forces between chains to be the dominant interaction and chain crossing constraints as perhaps providing secondary corrections.

This article presents a systematic review of the Hydrodynamic Scaling Model for the dynamics of non-dilute polymer solutions. The model has previous been presented in an extended series of papers\cite{phillies1986a,phillies2011a,phillies1985y,phillies1987a,phillies1987b,phillies1988a,phillies1988b,phillies1988c,phillies1989a,phillies1990a,phillies1991a,yu1991a,nelson1991a,phillies1992a,phillies1993a,phillies1993b,phillies1995a,phillies1995b,ngai1996a,phillies1997a,phillies1997b,phillies1998a,phillies1999a,phillies1999b, phillies2002a,phillies2002b,phillies2002c,phillies2002d,merriam2004a}.  The objective here is to present the results of these papers in a coherent way, showing what has been calculated thus far and what remains to be accomplished.

The Hydrodynamic Scaling Model arises from the Kirkwood-Riseman model\cite{kirkwood1948a} for the dynamics of a single neutral polymer molecule in a simple solvent. Hydrodynamic scaling transcends the earlier work of Kirkwood and Riseman by including hydrodynamic interactions between different polymer molecules. Five major components of the model are readily identified:

\textbf{First,} the Hydrodynamic Scaling Model presumes that the dominant interactions between neutral polymers in solution are the solvent-mediated hydrodynamic forces. Chain-crossing constraints are taken to provide at most secondary corrections. Because hydrodynamic forces are strong, nearby segments of different polymer molecules move in unison with each other, so the effects of chain crossing constraints are greatly reduced. When two chains are close to each other, each chain drags the other along, rather than each chain acting as a stationary obstacle to block the other chain's movements.

\textbf{Second,} following the Kirkwood-Riseman\cite{kirkwood1948a} model, each polymer chain is treated as a line of frictional centers ("beads") separated by a series of frictionless links ("springs"). The hydrodynamic interactions between beads on different chains are taken to be described by the Oseen tensor\cite{zimm1956a} and its modern short-range extensions\cite{kynch1958a}.

\textbf{Third,} the above assumptions are used to obtain a pseudovirial expansion for the concentration dependence of each transport coefficient, as a power series in concentration.

\textbf{Fourth,} to extend the model to elevated concentrations, recourse is had to self-similarity \cite{phillies1987a} or to renormalization group methods\cite{phillies1998a}. The renormalization group method of choice is the Altenberger-Dahler Positive Function Renormalization Group\cite{altenberger1997a,altenberger1997b,altenberger2001a,altenberger2001b,altenberger2002a}. Altenberger and Dahler developed this group from Shirkov's general treatment of renormalization analysis, based on functional self-similarity\cite{shirkov1984a,bogoliubov1983a,shirkov1988a}.  While renormalization group methods are indirect, they allow one to extrapolate lower-order pseudovirial expansions to elevated concentrations.

\textbf{Fifth,} the quantitative success of the Hydrodynamic Scaling Model is in part based on polymer statics.  In particular, it has been predicted theoretically\cite{daoud1975a} and demonstrated experimentally\cite{daoud1975a,king1985a} that in solution polymer coils contract as the polymer concentration is increased. This fairly modest degree of chain contraction has a substantial effect on the predicted concentration dependences of the polymer transport coefficients.

The Hydrodynamic Scaling Model was first used to treat the self-diffusion coefficient $D_{s}$ of polymers in solution, predicting the functional form for the dependence of $D_{s}$ on polymer concentration $c$ and polymer molecular weight $M$.  Physical interpretations and predictions of numerical values for the functional form's parameters have been provided\cite{phillies1987a,phillies1988c,phillies1989a,phillies1990a}. The model has been extended to consider the effect of polymer concentration on the mobility of individual beads of a polymer chain and on the mobility of small probe molecules in the surrounding solution\cite{phillies1993b}. An extended calculation predicted the low-shear viscosity of non-dilute polymer solutions\cite{phillies2002b}.  Consideration of the inferred fixed-point-structure of the renormalization group led to an \emph{ansatz}\cite{phillies1999a} that qualitatively determines the frequency dependences of the storage and loss moduli. The validity of the  Hydrodynamic Scaling Model  is shown by a huge mass of experimental data, as found in our companion volume \emph{Phenomenology of Polymer Solution Dynamics}\cite{phillies2011a,phillies2011b}. In the following, discussion of experiments will be limited to results that test particular aspects of the Hydrodynamic Scaling Model.

\subsection{ Reptation/Scaling and Hydrodynamic Scaling Models Compared}

This Section compares the reptation/scaling and hydrodynamic scaling models.  The major emphasis is on points where the two models are entirely different. Failure to recognize the great disparities between the two models occasionally leads to confusion in the literature. Readers should recognize that there are large numbers of modestly different reptation/scaling treatments and several different hydrodynamic treatments.

The core physical difference between the reptation/scaling and hydrodynamic scaling treatments is that the models do not agree as to which forces dominate polymer solution dynamics.  Many models\cite{doi1986a} assume that at elevated concentrations chain crossing (topological) constraints (``entanglements'') between polymer chains are the dominant physical interactions. In these models, hydrodynamic interactions between chains serve primarily to dress the bare monomer drag coefficients. Hydrodynamic Scaling Models assert to the contrary that hydrodynamic forces are dominant. In these models, excluded-volume and chain-crossing constraints are taken to provide only secondary corrections to the hydrodynamic interactions. The Hydrodynamic Scaling Model is not unique in assuming the dominance of hydrodynamic interactions. Oono's renormalization group treatment of mutual diffusion shares with the Hydrodynamic Scaling Model the assumption that hydrodynamic forces are dominant\cite{oono1986a}.

Corresponding to the assumptions as to the nature of the dominant forces, there are assumptions as to the concentration ranges in which the models are valid. Reptation/scaling models require that the concentration is large enough that neighboring polymer coils overlap each other and form \emph{entanglements}, circumstances where chain crossing constraints are particularly significant.  As a result, there is a lowest concentration $c^{*}$, the overlap concentration, below which tube model/reptation models are inappropriate.  Tube models describe small concentrations $c < c^{*}$ as constituting the \emph{dilute} regime,, while concentrations $c > c^{*}$ include the overlapping \emph{semidilute}, \emph{entangled}, and \emph{concentrated} regimes. There are no entanglements in the hydrodynamic scaling model, whose validity extends up from extreme dilution toward the melt.  However, within the hydrodynamic scaling model,  there is expected to be a transition concentration regime above which typical gaps between polymer chains are similar in size to individual polymer molecules; at larger concentrations, it appears inappropriate to describe solvent dynamics in terms of continuum fluid mechanics.

Entanglement-based models were originally applied to described the diffusion of a single polymer molecule, the \emph{probe} chain, through a chemically cross-linked gel, the polymer \emph{matrix}.  In a cross-linked gel, the chains of the matrix can not move over large distances\cite{degennes1971a}. Probe chains must thread their way through the matrix, like a very long snake threading its way through a grove of bamboo.. To transfer the entanglement model from probe chains in a crosslinked gel to probe chains in a polymer solution, it was hypothesized that the motions of probe chains in crosslinked gels and in polymer solutions can be given the same description.  Unlike a gel, in solution the matrix chains are free to move. Entanglement-based models assume that on the time scales of interest, these being the time scales on which the probe chains move, the matrix chains are effectively stationary. The entangled matrix chains of a polymer solution are said to form a \emph{transient lattice} or \emph{pseudogel} that constrains probe chain motions in the same way that a true crosslinked gel constrains probe chain motions, namely the probe chain can only move parallel to its own chain contour. There is no transient lattice or pseudogel in the context of the Hydrodynamic Scaling Model. Chains are free to translate and to rotate around their center of masses.

It is implicitly assumed in the tube/reptation models that when the probe chain encounters a matrix chain, the probe chain does not drag the matrix chain along; instead, the probe chain is brought to a stop by the matrix chain.  No rationale for this implicit assumption is provided\cite{philliesdaoud1978}. It is thus assumed in reptation-scaling models that a long polymer chain in a nondilute polymer solution can only move through solution in the ways that the chain can move through a true cross-linked gel, namely over large distances the probe chain only moves parallel to its own length. Hydrodynamic Scaling models make an opposite assumption, namely that when polymer chains encounter each other they tend to move in parallel directions, so they do not block each others' motions.

Many entanglement-based models incorporate a second, independent assumption, the \emph{scaling} assumption, which proposes that polymer transport coefficients such as the self-diffusion coefficient are assumed to depend on solution properties via scaling laws, e.~g.,
\begin{equation}
      D_{s}(c,M) = D_{cm} c^{\nu} M^{\gamma},
    \label{eq:scalinglaw}
\end{equation}
where here $\nu$ and $\gamma$ are scaling exponents.  The business of entanglement models and experimental studies is then to obtain the exponents $\nu$ and $\gamma$.  Presumably a complete model would also compute the scaling prefactor $D_{cm}$ and supply the ranges of $c$ and $M$ for which the model should be accurate, but much early work treats $D_{cm}$ as an undetermined constant.

The Hydrodynamic Scaling Model instead usually makes predictions in terms of stretched exponentials
\begin{equation}
      D_{s}(c,M) = D_{o} \exp(- \alpha c^{\nu} M^{\gamma}).
    \label{eq:seeq}
\end{equation}
This function form arises theoretically from the Altenberger-Dahler Positive Function Renormalization Group to extrapolate $D_{s}(c,M)$ to larger concentrations, as treated in Section 6.

\subsection{Historical Aside\label{ss:hist}}

The Hydrodynamic Scaling Model arose from a series of entirely empirical observations. Experimental studies of the diffusion of microscopic polystyrene latex spheres (as probes) through solutions of non-neutralized polyacrylic acid, poly-ethyl-ene oxide, and bovine serum albumin (as matrices) found\cite{lin1984a,lin1984b,ullmann1983a,ullmann1983b,ullmann1985a,phillies1985z} that the concentration dependence of the probe's diffusion coefficient $D_{p}$ could be described to good accuracy by stretched exponentials in polymer concentration, \emph{viz.},
\begin{equation}
      D_{p}(c) = D_{o} \exp(- \alpha c ^{\nu})
      \label{eq:stretchedexp}
\end{equation}
Here $c$ is the polymer concentration, $D_{o}$ is the probe diffusion coefficient in the limit of low concentration, and in the original work $\alpha$ and $\nu$ were fitting parameters.  Comparison of these experimental results\cite{phillies1985y} revealed that $\nu$ was consistently in the range 0.5-1.0, while over two orders of magnitude in polymer molecular weight $M$ one had
\begin{equation}
      \alpha \sim M^{\gamma}
      \label{eq:alphaandM}
\end{equation}
for $\gamma = 0.9\pm 0.1$. Measurements with different probe sizes found that $\alpha$ is approximately independent of probe sphere radius $R$.

Furthermore, in most of these systems $D_{p}$ did not track the solution viscosity via $D_{p} \sim \eta^{-1}$. In this non-Stokes-Einsteinian behavior, probes diffused faster than expected from their known sizes and the solution viscosity. Obvious artifacts, including polymer adsorption by the spheres and polymer-driven sphere aggregation, would cause the spheres to diffuse slower than expected, indicating that this non-Stokes-Einsteinian behavior was not simply an artifact. Non-Stokes-Einsteinian behavior, which was noticed well before eq.\ \ref{eq:stretchedexp} and the dependences of $\alpha$ and $\nu$ on $M$ and $R$ were identified, was the driving motivation for the early\cite{lin1984a,lin1984b,ullmann1983a,ullmann1983b,ullmann1985a,phillies1985z} experimental work.

Eq.\ \ref{eq:stretchedexp} was then compared\cite{phillies1986a} with published studies of the polymer self-diffusion coefficient $D_{s}$, finding that $D_{s}(c)$ uniformly follows a similar equation
\begin{equation}
     D_{s}(c) = D_{o} \exp(- \alpha c ^{\nu}).
   \label{eq:stretchedexpDs}
\end{equation}
This equation was therefore identified\cite{phillies1986a} as the \emph{universal scaling equation} for polymer self-diffusion. The functional form of eq.\ \ref{eq:stretchedexp} has since been tested\cite{phillies1988a,phillies1995b,phillies1997a} against literature reports of the polymer solution viscosity $\eta$, sedimentation coefficient $s$, rotational diffusion coefficient $D_{r}$, and dielectric relaxation time $\tau_{r}$. In each case these transport coefficients have stretched-exponential concentration dependences with various prefactors and exponents $\alpha$ and $\nu$.

Several features of eq.\ \ref{eq:stretchedexp} as revealed in refs.\  \cite{phillies1986a} and \cite{phillies1985y}  were not in accord with expectations from entanglement-based models of polymer solution dynamics.  In particular: (i) The concentration dependence was found to be a stretched exponential in $c$, not the expected power law in $c$; (ii) The concentration dependence was described over all concentrations studied by a single set of parameters $(\alpha, \nu)$, with no indication of a transition in dynamic behavior between a "dilute" regime (in which hydrodynamics was expected to dominate) and a ``semidilute'' regime (in which polymer coils overlapped and entanglements were proposed to dominate); (iii) For probe diffusion (spheres diffusing through random-coil polymers), in the semidilute regime $D_{p}(c)$ was found to be dependent, not independent,  of polymer molecular weight; (iv) In the semidilute regime, $\alpha$ was found to be nearly independent of probe radius, while is had been expected to have a strong dependence on probe radius, and (v) $D_{p}$ of large probes was expected to be determined by the macroscopic solution viscosity. Furthermore, (vi) In the dilute solution regime, $D_{s}(c) $ is often proposed in the context of reptation/scaling models to be nearly independent of $c$.  None of expectations (i)-(vi) were met in the systems studied.

How might this set of discrepancies between the universal scaling equation \ref{eq:stretchedexp} and expectations based on entanglement models be resolved?  First, one could always propose that the agreement between the universal scaling equation and the particular data sets with which it had been compared was a curiosity, an empirical coincidence having no real importance. In that case, the equation would be an accident having no relationship to fundamental theoretical considerations. Second, one could propose that the agreement arose because the universal scaling equation is remarkably flexible. This second proposal encounters the information-theoretic obstacle that the equation has three free parameters (and the measurable zero-concentration limiting constant $D_{o}$), so it therefore can cover neither more nor less of the possibly solution space than can any other reasonable three-parameter equation.

Finally, the criticism was advanced that equation \ref{eq:stretchedexp} is purely empirical and has no physical content. This final criticism led to the clear recommendation\cite{ware} that proponents of eq.\ \ref{eq:stretchedexp} needed to find an \emph{ab initio} theoretical derivation of eq.\ \ref{eq:stretchedexp}, preferably a derivation that reveals the physical interpretations of $\alpha$ and $\nu$. The remainder of this article reviews the research program that generated the requested derivation. We present the papers that supplied that derivation, ending antiquated suggestions that the universal scaling equation and its parameters are purely empirical and have no physical interpretation.

\subsection{Precis of the Work\label{ss:precis}}

This section presents an outline of the remainder of this article.

Section \ref{s:singlechain} considers the Kirkwood-Riseman and Rouse-Zimm models for the dynamics of a single polymer chain. The two sets of models start at the same point, describing a polymer chain as a set of hydrodynamic beads connected by non-frictional links. However, their descriptions of how polymer chains move in solution are radically contradictory. In the Kirkwood-Riseman model, polymer coils in solution translate and rotate; internal modes are neglected.  Rouse-Zimm chains translate and have internal modes, but as shown below cannot rotate.

In Section 2, we first discuss the less-studied Kirkwood-Riseman model, because the Kirkwood-Riseman model provides the foundation for calculating hydrodynamic interactions between polymer chains. The drag coefficient of a Kirkwood-Riseman polymer is calculated.  We then present the Rouse and Zimm models. Hiding in their seductively simple mathematical derivations is a curiosity: Rouse-Zimm chains do not rotate, and thus cannot perform the core motion of Kirkwood-Riseman model chains.

Section 3 presents our extended Kirkwood-Riseman model. The extension calculates chain-chain hydrodynamic interactions.  It thus provides the physical basis for the hydrodynamic scaling model.    Section 3.1 presents the modern bead-bead hydrodynamic interaction tensors including short range and three-bead interactions. Section 3.2 shows how to move from bead-bead to chain-chain hydrodynamic interactions in the context of the Kirkwood-Riseman model.

Section 4 uses the extended Kirkwood-Riseman model to calculate, through $\mathcal{O}(c^{2})$, the concentration dependence of the polymer self-diffusion coefficient.
Section 5 uses the model to calculate the concentration dependence of the viscosity. Section 5.1 calculates the flow field $\bm{u}^{(1)}$ created by the scattering of a shear field $\bm{u}^{(0)}$ by a polymer chain, and the additional flow field $\bm{u}^{(2)}$ created by the scattering of flow field $\bm{u}^{(1)}$ by a second polymer. Section 5.2 calculates the power dissipated by various polymer chains exposed to flow fields $\bm{u}^{(0)}$, $\bm{u}^{(1)}$, and $\bm{u}^{(2)}$. Section 5.3 calculates the total shear field that would be determined experimentally as a result of those flow fields, leading to a determination in Section 5.4 of the intrinsic viscosity  and the Huggins coefficient for the extended Kirkwood-Riseman model. Sections 3.1 and 4.1 consider some of the ways in which short-range hydrodynamic interactions modify polymer dynamics.

Section 6 considers paths for extending the hydrodynamic calculation of pseudovirial coefficients, as seen in Sections 4 and 5, to determine polymer dynamics at elevated concentrations.  Section 6.1 considers self-similarity rationales.  Section 6.2 develops the mathematical basis for the alternative approach, the Altenberger-Dahler Positive Function Renormalization Group. Section 7 then uses the Positive-Function Renormalization Group to extend the calculations of sections 4 and 5 to large concentrations. The universal scaling equation for polymer self-diffusion is obtained.

Section 8 presents an \emph{ansatz} for computing the frequency dependences of the bulk and shear moduli.  The \emph{ansatz}, \emph{Two-Parameter Temporal Scaling}, arises from the inferred fixed point structure of the Positive-Function Renormalization Group calculation of the shear viscosity.

Section 9 offers single-paragraph summaries, in publication order, of the theoretical and phenomenological papers that describe the Hydrodynamic Scaling Model.  Section 10 summarizes experimental results testing various aspects of the Hydrodynamic Scaling Model. The tests confirm the validity of the model. Section 11 discusses the results here and considers consider where the Hydrodynamic Scaling Model has gaps and omissions, thereby identifying a few directions for future research.

\section{Single-Chain Behavior\label{s:singlechain}}

This section discusses models for single-chain polymer motion.   There are two major classes of models, namely models based on the Kirkwood-Riseman\cite{kirkwood1948a} treatment, and models based on the treatments of Rouse\cite{rouse1953a} and Zimm\cite{zimm1956a}. Qualitatively, the two classes of model supply radically different descriptions for chain motion in dilute solution. The Hydrodynamic Scaling Model is based on extensions of the Kirkwood-Riseman model, while in contrast many tube/reptation models reference the original Rouse treatment.   A major emphasis of this Section is therefore to alert readers familiar with Rouse and Zimm models as to the very different way in which Kirkwood and Riseman described the movements of an individual polymer coil.

In all of these models, a polymer chain is treated as a series of \emph{beads}, pairs of beads being connected by \emph{links}. The polymer interacts hydrodynamically with the solvent via the beads, each of which acts as a small sphere or point that applies a frictional force on the solvent. The links are hydrodynamically inert.  They serve to control the distances between the beads. In the Rouse and Zimm models, the beads are abstractions representing the hydrodynamic friction of a subsection of the polymer, while the links are treated as subsections of the polymer chain, each subsection being barely long enough to have a gaussian distribution of lengths. In the original Kirkwood-Riseman model, the beads were taken to be monomer units, while the links were the covalent bonds connecting one monomer to the next. In some modern applications of the Kirkwood-Riseman model, the beads and links are interpreted in the Rouse and Zimm sense.

In the Rouse and Zimm models, each subsection acts as a Hookian spring. Each subsection generates an attractive force on the two beads to which it is attached.  The force has magnitude $k \ell$, where $k$ is an effective spring constant and $\ell$ is the distance between the two beads; the force acts along the line of centers connecting the beads. In these models the unstretched (rest) length of each subsection is zero.

In the original Kirkwood-Riseman model, the links are covalent bonds having rigid lengths and bond angles, but perhaps a potential energy for torsion. Within the model, the effect of the links is to determine the statistico-mechanical distribution functions for the distances between pairs of beads along the polymer chain.  Because the beads of the original Kirkwood-Riseman model are monomers, the number of beads in a Kirkwood-Riseman chain can be very large, much larger than the number of beads in a Rouse or Zimm model for the same polymer. For beads that are well separated along the chain, in the Kirkwood-Riseman model the distribution function for the bead-bead distance is assumed to be a Gaussian.

These models for polymer dynamics make contradictory assumptions as to how polymer chains move in solution. In the Kirkwood-Riseman model, the interesting motions of the beads are described as \emph{whole body motion}.  In whole body motion, the polymer beads may experience equal linear displacements, and they may rotate around the polymer center of mass, but the displacements and rotations are such that the chain motion does not alter the relative positions of the polymer beads. The phrase \emph{whole body motion} does not mean that the polymer coil is  mechanically rigid. A full description of the motions of $N$ polymer beads requires $3N$ coordinates. The whole body motion description extracts from these $3N$ coordinates a set of six collective coordinates, describing whole-body translations and rotations, with the remaining motions being described as the \emph{internal modes}.

Kirkwood and Riseman are entirely specific that the polymer coil in their model has internal motions, so that the relative positions of beads fluctuate with respect to each other. However, in the Kirkwood-Riseman model the whole-body motions assumed to dominate polymer solution dynamics. Internal motions are taken to provide corrections to the dominant chain motions, the whole body displacements. The internal motions are coarse-grained out, so bead velocities are approximated with the components created by polymer translational and angular velocities. Kirkwood and Riseman did not compute the magnitude of the internal mode corrections.

In contrast to the Kirkwood-Riseman model, the Rouse and Zimm models assume that the beads move relative to each other.  The relative motions of the beads are driven by attractive forces between adjoining beads as created by the links.  These relative motions are described by the Rouse-Zimm polymer internal modes, and are taken to dominate polymer solution dynamics. Rouse and Zimm model polymer coils do perform whole-body translation, but translation does not to contribute to the polymer solution's viscosity. It is a curiosity not generally remarked upon (see Subsection \ref{ss:rousewrong}, below) that Rouse-model polymers can not rotate.

In discussing these models, we emphasize two major issues.  First, the Kirkwood-Risemann and Rouse-Zimm models invoke entirely contradictory descriptions of the important aspects of polymer dynamics.  Second, the solutions to the Rouse and Zimm models are inconsistent with basic laws of mechanics and cannot possibly be correct for a real polymer.

\subsection{Kirkwood-Riseman Model\label{ss:krm}}

We first consider the Kirkwood-Riseman model\cite{kirkwood1948a}, whose \emph{ansatz} provides the basis of the Hydrodynamic Scaling Model.  The Kirkwood-Riseman model is much less discussed than are the Rouse and Zimm models and their extensions, in part because it is more demanding mathematically and in part because Kirkwood and Riseman use a less familiar notation.  This presentation of the Kirkwood-Riseman model has therefore been reset in a more modern form.

The Kirkwood-Riseman model describes a chain of $N$ beads connected by links having length $b_{0}$. The links are covalent bonds, with adjoining links separated by a rigid angle $\theta$. Successive three-bead planes are related by a torsion angle $\phi$. In the original model, the potential energy was taken to be independent of the angle $\phi$.  The effective bond length, the contribution of each link to the distance between distant beads, is
\begin{equation}
   b = \left(\frac{1 + \langle \cos(\phi)\rangle}{1 - \langle \cos(\phi)\rangle}  \right)  \left(\frac{ 1 - \cos(\theta)}{1 + \cos(\theta)} \right) b_{0}.
   \label{eq:bb0}
\end{equation}

For beads $\ell$ and $s$ that are well-separated, Kirkwood and Riseman supply several average values, notably
\begin{align}
    \langle \mid R_{\ell s}\mid^{2} \rangle  &= \mid \ell - s\mid b^{2} \\
    \langle \mid R_{0 \ell}\mid^{2} \rangle &= b^{2} \left(\frac{12 \ell^{2} + N^{2} - 2 N + 1}{12(N-1)}\right)\\
    \langle \bm{R}_{0\ell} \cdot  \bm{R}_{0s} \rangle &= \frac{b^{2}}{N-1}\left(\frac{\ell^{2} + s^{2}}{2} - \frac{N-1}{2}\mid\ell - s\mid + \frac{(N-1)^{2}}{2}  \right)\\
    \left\langle \frac{1}{R_{\ell s} }\right\rangle &= \frac{6}{\sqrt{\pi} b \mid \ell - s \mid^{1/2}}.
     \label{eq:distances}
\end{align}
Here beads $\ell$ and $s$ have locations $\bm{r}_{\ell}$ and $\bm{r}_{s}$, $\bm{R}_{\ell s} = \bm{r}_{s} -\bm{r}_{\ell}$ is the vector from bead $\ell$ to bead $s$, $R_{\ell s} = |\bm{R}_{\ell s}|$, and $\bm{r}_{0}$ is the location of the center of mass of the polymer, so that $\bm{R}_{0 \ell}$ is the vector from the center of mass to bead $\ell$.  The final equation assumes that $R_{\ell s}$ has a normal distribution.

The Kirkwood-Riseman model assumes that polymer beads have a long range hydrodynamic interaction described by the Oseen tensor
\begin{equation}
     \bm{T}_{ij}(\bm{r}_{ij})  = \frac{1}{8 \pi \eta_{0} r_{ij}}(\bm{I} + \bm{\hat{r}}_{ij}\bm{\hat{r}}_{ij}),
    \label{eq:Oseen}
\end{equation}
which gives the fluid flow created at a point $\bm{r}_{j}$ by a force $\bm{F}_{i}$ applied to the solution at point $\bm{r}_{i}$.  The vector from point $i$ to point $j$ is $\bm{r}_{ij}$, with magnitude $r_{ij} = | \bm{r}_{ij}|$ and corresponding unit vector $\bm{\hat{r}}_{ij}= \bm{r}_{ij}/r_{ij}$. Here $\eta_{0}$ is the solvent viscosity. In eq.\ \ref{eq:Oseen} and its associated notation, there is no assumption that there is a polymer bead at point $\bm{r}_{j}$.  The theoretical model treats the force as a point source, and assumes that the presence of the polymer has no effect on the solvent's viscosity, an assumption that is known experimentally to be incorrect\cite{solvent}. The fluid flow induced at $\bm{r}_{j}$ by $\bm{F}_{i}$ is
\begin{equation}
            \bm{v'}(\bm{r}_{j})  = \bm{T}_{ij}(\bm{r}_{ij}) \cdot \bm{F}_{i}({\bf r}_{i}).
    \label{eq:oseenflow}
\end{equation}

Within the model, the forces $\bm{F}_{i}$ arise because the beads are moving with respect to the fluid.  If a bead is stationary with respect to the local fluid flow, it exerts no force on the fluid.  The force exerted on the fluid by a bead $\ell$ is determined by the velocity $\bm{u}_{\ell}$ of the bead, the velocity $\bm{v}(\bm{r}_{\ell})$ that the fluid would have had, at the point $\bm{r}_{\ell}$, if the bead were not present, and the drag coefficient $\xi$ of the bead, namely
\begin{equation}
          \bm{F}_{i} = \xi(\bm{u}_{\ell} - \bm{v}(\bm{r}_{\ell}))
\label{eq:oseenforce}
\end{equation}
Because the beads are treated as points, a single bead is assumed to exert no torque on the surrounding fluid.

We now come to the modelled dynamics of the polymer.  The beads are taken to lie along a Gaussian chain, meaning that on the average their concentration declines with the distance, from the center of mass, as a Gaussian in that distance.  The velocities of the individual beads are taken to be determined entirely by the time-dependent chain center-of-mass velocity $\bm{V}(t)$  and chain rotational velocity $\bm{\Omega}(t)$ as
\begin{equation}
           \bm{u}_{\ell}(t) = \bm{V}(t) + \bm{\Omega}(t) \times \bm{R}_{0\ell}
\label{eq:beadvelocity1}
\end{equation}
$\bm{u}_{\ell}$, as given by equation \ref{eq:oseenforce}, is the velocity that the bead $\ell$ would have, if it were part of a rigid body that had translational velocity $\bm{V}$ and rotational velocity $\bm{\Omega}$.  We therefore describe the chain motions as whole-body translation and whole-body rotation. As noted above, Kirkwood and Riseman recognized that polymer molecules also have internal coordinates whose fluctuations contribute to the bead velocities, but those fluctuations were as an approximation neglected.

What forces act on a polymer chain?  The model assumption is that, in the absence of external forces, over long times the polymer's translational and rotational accelerations must both average to zero. Under these conditions the long-time averages of the sum of the forces and of the sum of the torques must both vanish. The zero-force and zero-torque conditions determine the response of the polymer to an external force or to an external torque.

As an example of the effect of hydrodynamic interactions, we consider the drag coefficient (and hence the diffusion coefficient) of a polymer chain.  The analysis of Zwanzig\cite{zwanzig1969a} is followed.  Note that Kirkwood and Riseman took $\bm{F}_{j}$ to be the force on the solvent, while Zwanzig takes $\bm{F}_{j}$ to be the force on the bead, so the papers have sign differences. We have a polymer chain whose beads have arbitrary velocities $\bm{u}_{\ell}$, while the fluid at $\bm{r}_{\ell}$ has an unperturbed velocity $\bm{v}_{\ell}^{0}$. The hydrodynamic interactions perturb  the fluid flow at $\bm{r}_{\ell}$, so the actual fluid velocity at $\bm{r}_{\ell}$ is
\begin{equation}
      \bm{v}_{\ell} =  \bm{v}_{\ell}^{0} + \sum_{k \neq \ell =1}^{N} \bm{T}_{\ell k} \cdot \bm{F}_{k}.
      \label{eq:perturbedvelocity}
\end{equation}

However, the hydrodynamic force that a bead $k$ exerts on the solvent is
\begin{equation}
      \bm{F}_{k} = f (\bm{u}_{k} - \bm{v}_{k}),
      \label{eq:solventforce}
\end{equation}
$f$ being the drag coefficient of a single bead. Combining the above two equations,
\begin{equation}
      \bm{v}_{\ell} =  \bm{v}_{\ell}^{0} - f \sum_{k \neq \ell =1}^{N} \bm{T}_{\ell k} \cdot f (\bm{v}_{k} - \bm{u}_{k}) .
      \label{eq:perturbedvelocity2}
\end{equation}
Subtracting  $\bm{u}_{\ell}$ from each side of the equation,
\begin{equation}
      \bm{v}_{\ell} -\bm{u}_{\ell} =  \bm{v}_{\ell}^{0}-\bm{u}_{\ell} - f\sum_{k \neq \ell =1}^{N} \bm{T}_{\ell
k} \cdot f (\bm{v}_{k} - \bm{u}_{k}) .
      \label{eq:perturbedvelocity3}
\end{equation}
which allows us to write
\begin{equation}
      \bm{v}_{\ell}^{0} -\bm{u}_{\ell} =   f \sum_{k  =1}^{N} \bm{\mu}_{\ell k} \cdot (\bm{v}_{k} - \bm{u}_{k}) .
      \label{eq:perturbedvelocity4}
\end{equation}
The new matrix $\bm{\mu}$ is
\begin{equation}
           \bm{\mu}_{\ell k} = \frac{\bm{I} \delta_{\ell k}}{f} + \bm{T}_{\ell k}
     \label{eq:muzwanzigdef}
\end{equation}
where the rule $\bm{T}_{kk} = 0$ has been applied and $\bm{I}$ is the $3 \times 3$ identity matrix.

Matrix inversion gives the $\bm{v}_{k} - \bm{u}_{k}$ in terms of the $\bm{v}_{\ell}^{0} -\bm{u}_{\ell}$ and the inverse of $\bm{\mu}$, namely
\begin{equation}
         \bm{v}_{k} - \bm{u}_{k} = f^{-1} \sum_{\ell=1}^{N} (\bm{\mu}^{-1})_{k\ell} \cdot (\bm{v}_{\ell}^{0} -\bm{u}_{\ell})
     \label{eq:actualvelocities}
\end{equation}
so the force on a bead $k$ due to its hydrodynamic interactions with the solvent becomes
\begin{equation}
      - \bm{F}_{k} \equiv f(\bm{v}_{k} - \bm{u}_{k}) = - \sum_{\ell=1}^{N} (\bm{\mu}^{-1})_{k\ell} \cdot (\bm{v}_{\ell}^{0} -\bm{u}_{\ell}).
      \label{eq:Fkvalue}
\end{equation}
The minus sign appears because $\bm{F}_{k}$ is the force of the bead on the solvent, not vice versa.

The drag coefficient $f_{c}$ of the polymer chain is obtained by choosing all bead velocities to be equal to $\bm{u}_{0}$ and the unperturbed fluid velocity to be zero, and calculating the total of the drag forces on all beads of the chain, leading to
\begin{equation}
      - \sum_{k=1}^{N} \bm{F}_{k} \equiv  f_{c} \bm{u}_{0} = \sum_{k=1}^{N}  \sum_{\ell=1}^{N} (\bm{\mu}^{-1})_{k\ell} \cdot \bm{u}_{0}.
      \label{eq:dragcoefficient}
\end{equation}
Bead-bead hydrodynamic interactions as described by the Oseen tensor thus perturb the drag coefficient of the whole chain.

\subsection{Rouse and Zimm models\label{eq:rzm}}

This Section presents the Rouse and Zimm models.  The model's odd features appear in the next Section. The original Rouse and Zimm models were quite elaborate in their derivations, but as is so often the case with derivations the passage of time has led to substantial simplifications in the presentation of the calculation.  I follow here the elegantly clear treatment of Doi and Edwards\cite{doi1986a41}. The polymer is approximated as a line of $N$ beads whose positions are $(\bm{R}_{1}, \bm{R}_{2}, \ldots \bm{R}_{N})$, respectively. The beads are free to move with respect to each other, and do not have any excluded-volume interactions. Each bead interacts with the solvent; each bead has a hydrodynamic drag coefficient $f$. (Some authors use $N+1$ beads labelled $\{0,1,\ldots,N\}$.)

In the Rouse model, the distribution $P(r)$ of distances $r$  between a pair of neighboring beads is taken to be a Gaussian $P(r) \sim \exp(- a r^{2})$.  Corresponding to this distribution, there is a matching potential of average force $W(r) = - k_{B}T \ln(P(r))$, with $k_{B}$ being Boltzmann's constant and $T$ being the absolute temperature. $W(r)$ is therefore a quadratic in $r$, namely
\begin{equation}
      W(r) = + \frac{1}{2} k r^{2},
      \label{eq:rousepoaf}
\end{equation}
$k$ being a constant. Within the model, the 'spring constant' $k$ and the mean-square bead separation $b^{2}$ are related by
\begin{equation}
      k = 3 k_{B}T/b^{2}.
      \label{eq:rousekdefinition}
\end{equation}
Corresponding to the potential of average force, each linked pair of beads $(i, i+1)$ is subject to an attractive Hooke's-Law force having magnitude $|k (\bm{R}_{i+1}-\bm{R}_{i})|$.

In the Rouse model, all bead motions are massively overdamped, so that inertia is neglected. The beads move with the terminal velocities determined by the spring and hydrodynamic drag forces. For beads $i, i \in (2, N-1)$, the equations of motion of the beads are therefore
\begin{equation}
     f \frac{d \bm{R}_{i}(t)}{dt} = -k(2 \bm{R}_{i} - \bm{R}_{i-1}  - \bm{R}_{i+1}) + \bm{F}_{i} (t)
     \label{eq:rousemostbeads}
\end{equation}
 while the end beads satisfy equations
\begin{equation}
     f \frac{d \bm{R}_{N}(t)}{dt} = -k(\bm{R}_{N} - \bm{R}_{N-1}) + \bm{F}_{N} (t)
     \label{eq:rouseendbeads}
\end{equation}
and correspondingly for bead $1$. The above two equations are an elaborate way to write that the total force on each bead vanishes; the mechanical and hydrodynamic drag forces  must sum to zero if inertia is negligible.  $\bm{F}_{i}(t)$ is the random force on bead $i$, physically arising from interactions with the solvent. In the Rouse model, the random forces on different beads are not correlated with each other.  In the Zimm model, the random forces $\bm{F}_{i}$ are cross-correlated, and the drag coefficients $f$ are replaced with a hydrodynamic interaction tensor.

Eqs.\ \ref{eq:rousemostbeads} and \ref{eq:rouseendbeads} represent a set of $N$ vector equations and therefore $3N$ scalar equations. Within this model, equations corresponding to different cartesian axes are uncoupled. Noting that the $x$-component direction cosine for the vector between beads $i$ and $i+1$ is  $(x_{i+1}-x_{i})/(\mid (\bm{R}_{i+1}-\bm{R}_{i})\mid )$, and similarly for the $y$-component and the $z$-component, the $N-2$ vector equations \ref{eq:rousemostbeads} may be replaced by three sets of $N-2$ scalar equations, viz., $N-2$ equations
\begin{equation}
     f \frac{d x_{i}(t)}{dt} = -k(2 x_{i} - x_{i-1}  - x_{i+1}) + F_{ix} (t)
     \label{eq:rousemostbeadsx}
\end{equation}
for the $x$ coordinates, and matching sets of $N-2$ equations for the $y$ and $z$ coordinates.  Here $x_{i}$ and $F_{ix}$ are the $x$ coordinate of particle $i$ and the $x$ component of the thermal force on particle $i$.

For bead 1, the corresponding equation is
\begin{equation}
     f \frac{d x_{1}(t)}{dt} = k(x_{2} - x_{1}) + F_{1x} (t)
     \label{eq:rouseendbeadsx}
\end{equation}
and correspondingly for bead $N$. The Rouse model thus yields a set of $3N$ coupled first-order linear differential equations.  However, as noted by Rouse, the equations for the $x$, $y$, and $z$ coordinates are entirely uncoupled, and are the same except for coordinate label, so one only needs to solve a set of $N$ coupled equations, and that once, to have the complete solution.  It should be stressed, however, that the Rouse model \emph{is not a one-dimensional model. It is a three-dimensional model.} We largely discuss the solutions for the $x$ coordinate, but the solutions for the $y$ and $z$ coordinate are the same except for the coordinate label.

The $x$-coordinate solutions are sometimes written in a generalized vector form, the generalized vector being $\bm{X}(t) = (x_{1}(t), x_{2}(t),\ldots x_{N}(t))$, namely
\begin{equation}
        \bm{H} \cdot \frac{d \bm{X}(t)}{dt} = - \bm{A} \cdot \bm{X}(t),
      \label{eq:rousexvectorform}
\end{equation}
in which in the Rouse model the hydrodynamic interaction matrix $\bm{H}$ is $H_{ij} = f \delta_{ij}$, $\delta_{ij}$ being the Kronecker delta. In the Zimm model $\bm{H}$ has a more complex form reflecting bead-bead hydrodynamic interactions.  The interaction matrix $\bm{A}$ is
\begin{equation}
      \bm{A} = \begin{Bmatrix} 1 & -1 & 0 &0 & \ldots & 0 & 0 & 0 \\  -1 & 2 & -1 &0 & \ldots & 0 & 0 & 0 \\ 0 &  - 1 & 2 & - 1  & \ldots & 0 & 0 & 0  \\  & & & & & & &  \\  0 & 0 & 0 &0 & \ldots & -1 & 2 & -1  \\  0 & 0 & 0 &0 & \ldots & 0 & -1 & 1 \end{Bmatrix}.
      \label{eq:rouseinteractionmatrix}
\end{equation}

Equation \ref{eq:rouseendbeadsx} is a set of $N$ coupled linear first-order differential equations. These equations were solved by Rouse. Its solutions are a set of $N$ eigenmodes $\bm{Q}_{i}$, each mode having a corresponding eigenvalue $q_{i}$ and (with one exception) a corresponding relaxation time $\tau_{i}$.    For each coordinate, there is a single mode $\bm{Q}_{0}$ having eigenvalue $0$ (and, hence, no value for $\tau_{0}$)) and $x_{i}(t) = x(t) \ \forall \ t$. Corresponding to each Cartesian coordinate there is also a series of $N-1$ modes $\bm{Q}_{n}$ with relaxation times
\begin{equation}
    \tau_{n}  = \frac{f}{8 k \sin^{2}(n \pi/ 2N},
    \label{eq:rouseeigenvalues}
\end{equation}
for $n \in (1, 2, \ldots, N-1)$.

The normal mode amplitudes $C_{i}$ for the $x$-coordinate modes can be calculated from the coordinates $x_{i}$ of the $N$ beads as
\begin{equation}
      C_{i} =  \frac{1}{N} \sum_{n=1}^{N} x_{n} \cos\left(\frac{i \pi (n-1/2)}{N}    \right).
      \label{eq:rouseamplitudes}
\end{equation}
Corresponding equations give the amplitudes of the $y$ and $z$ normal mode amplitudes in terms of the $y_{i}$ and $z_{i}$, respectively.

Correspondingly, the displacements $x_{i}$ of the individual atoms are determined by the amplitudes of the normal modes as
\begin{equation}
     x_{i} = C_{0} + 2 \sum^{N-1}_{n=1} C_{n} \cos\left(\frac{n \pi (i-1/2)}{N}   \right).
     \label{eq:rousedisplacements}
\end{equation}
Equations identical to eq.\ \ref{eq:rousedisplacements}, except for the coordinate label and the values of the normal mode amplitudes, describe the $y_{i}$ and the $z_{i}$.

By setting $C_{i} =1$ and $C_{j} = 0$ for $j \neq i$, eq.\ \ref{eq:rousedisplacements} can be used to determine the representations $\{x_{1i}, x_{2i}, \ldots, x_{Ni}\}$  in particle position space of the Rouse model eigenvectors. A representative eigenvector is then
\begin{equation}
    \bm{Q}_{ix}= \{x_{1i}, x_{2i}, \ldots, x_{Ni}\},
    \label{eq:eigenmode}
\end{equation}
    where $x_{1i}$ is the displacement in the $x$ direction of atom $1$ in mode $i$.  In the mode $Q_{ix}$, the atoms all have $y$ and $z$ displacements, but all $y$ and $z$ displacements are zero, and similarly for the modes  $Q_{iy}$ and $Q_{iz}$.  There are a total of $3N$ modes, with $Q_{0x}$, $Q_{0y}$, and $Q_{0z}$ being the molecular uniform translations and the other $3N-3$ modes describing molecular motions in which the atoms move with respect to each other.  Any set $(Q_{ix}, Q_{iy}, Q_{iz})$ of eigenvectors with the same $i$ have the same eigenvalue $q_{i}$, so all of their linear combinations are also eigenvectors having eigenvalue $q_{i}$. As the closing and central point, the Rouse model for an $N$ bead polymer chain has three modes with eigenvalue zero and $3N-3$ modes with non-zero eigenvalues in which the beads move with respect to each other.

\subsection{Unacceptable Properties of the Rouse Model\label{ss:rousewrong}}

This Section demonstrates a quaint physical property of Rouse chains: Rouse chains can not rotate. As a result, the Kirkwood-Riseman and Rouse-Zimm models are totally contradictory in their physical description of how polymer molecules contribute to viscosity.  In the Kirkwood-Riseman model, polymer chains in a shear field translate and rotate, dissipation arising from their rotation motion.  In the Rouse and Zimm models, dissipation is due entirely to the internal modes; the chains cannot rotate at all.  An explanation for this irrotational oddity is suggested.

An interesting mathematical comparison, showing that the Rouse model has gone astray, is provided by the Wilson-Decius-Cross treatment of molecular vibrations. Wilson, Decius, and Cross\cite{wilson1941a} describe an $N$-atom molecule as having $3N$ equilibrium atomic coordinates $r_{1}$, $r_{2},\ldots,r_{3N}$, which may be written as the $3N$-dimensional vector $\bm{r} = \{\bm{r}_{1}, \bm{r}_{2},\ldots, \bm{r}_{3N}\}$. When the molecule vibrates, its atoms have displacements $\bm{R}_{1}(t), \bm{R}_{2}(t),\ldots,\bm{R}_{3N}(t)$ from their equilibrium locations, which may be written as the time-dependent displacement vector  $\bm{R}(t)= \{\bm{R}_{1}(t), \bm{R}_{2}(t),\ldots,
\bm{R}_{3N}(t)\}$. While the molecule is vibrating, the coordinates of the $N$ atoms are therefore $\bm{r} +\bm{R}(t)$.

The equilibrium position of the atoms is a potential energy minimum, so the first derivatives of the molecular potential energy $U$ with respect to the $R_{i}$ must be zero. The second derivatives of $U$ with respect to the displacements are the matrix $\bm{V}$, whose components are
\begin{equation}
     V_{ij} = \frac{\partial^{2} U}{\partial R_{i} \partial R_{j}}.
     \label{eq:wilsonforcematrix}
\end{equation}
In the Wilson model, the potential energy is expanded to be quadratic in molecular displacements, with terms in chemical bond stretches, bond bendings, bond torsions, and atomic out-of-plane motions, leading to $3N$ molecular equations of motion.
\begin{equation}
    m_{i} \frac{\partial^{2} R_{i}(t)}{\partial^{2} t} = - \sum_{j=1}^{3N} V_{ij}R_{j}(t)
    \label{eq:molecularvibration}
\end{equation}
Here $m_{i}$ is the mass of the atom associated with coordinate $i$. The right hand side of the equation gives the forces on atom $i$ due to displacements of all atoms from their equilibrium positions. The expansion of $U$ to quadratic terms is with rare exceptions adequate so long as the $R_{i}(t)$ are small, as is the case for thermal vibrations of conventional molecules at room temperature.

Equation \ref{eq:molecularvibration} is a set of coupled linear differential equations, very much like eq.\ \ref{eq:rousexvectorform} except that the $f_{i}$ have been replaced by the $m_{i}$. Also, the time derivatives are now second rather than first order, so the corresponding atomic motions are oscillatory, rather than having the relaxational motions of the Rouse beads.

The solutions of the Wilson molecular model are well-known.  There are three whole-body translations.  There are three whole-body rotations. The six whole-body motions  do not change the distances or angles between any of the atoms, so corresponding to the whole-body motions there are no restoring forces. The eigenvalues (oscillation frequencies) corresponding to these six modes are therefore all zero. Finally, in the Wilson model there are $3N-6$ internal modes corresponding to the molecular vibrations.

Two-thirds of a century ago, the presence of degenerate zero eigenvalues created technical difficulties with solving the corresponding eigenvector-eigenvalue problem by numerical means. Wilson\cite{wilson1941a} removed the difficulty by identifying an appropriate complete set of internal coordinates, and a matrix method for replacing the $3N$ equations of eq.\ \ref{eq:molecularvibration} with a new set of $3N-6$ equations that only described the internal vibrations and only had non-zero eigenvalues.  The molecule could still translate and vibrate, but the coordinates describing those translations and vibrations occupy a 6-dimensional subspace that is orthogonal to the subspace in which the internal vibrations occur. The corresponding solution process only involved a non-singular matrix and was thus straightforward to solve. (Furthermore, in-period, matrix inversion had to be done by hand or on very limited digital computers, so the reduction from $3N$ to $3N-6$ coordinates meant a major reduction in the demanded calculational effort.)

The mathematical forms describing the Wilson-Decius-Cross molecular vibration model and the  Rouse polymer chain model are substantially similar. Their solutions should therefore in key respects be substantially similar. In particular, rigid-body translations and rotations do not change the relative positions of the beads (or atoms), so therefore in each model there should be six modes having their eigenvalues equal to zero.  Unfortunately, this expectation is not satisfied. For an isolated triatomic molecule, there are nine modes: six whole-body-motions have zero vibrational frequencies, and three modes (two stretching, one bending) describe internal vibrations. For a three-bead polymer, there are three translational modes with an eigenvalue equal to zero, and six stretching modes with finite relaxation times.

There is no possible doubt that triatomic molecules only have three modes.  The question then is: How can the Rouse model for a three-bead system have six vibrational modes?

The first part of the answer is that Rouse-model chains cannot rotate.  You can put a Rouse chain in a sheared fluid flow, and, as shown by Rouse, the Rouse-model chain will respond.  However, the response is not rotation.  The demonstration of this surprising fact is actually entirely straightforward.  A rigid-body rotation does not change the relative position of the beads in a chain. Rotating a chain creates no internal forces, so the eigenvalue corresponding to whole-chain rotation must be zero. However, we have a complete list of the Rouse model's modes.  There are indeed three modes with eigenvalue zero; they are the three translational modes. The remaining modes all have non-zero eigenvalues; their relaxation times are non-zero finite. Rotations and translations are orthogonal, so a rotational mode cannot receive a contribution from any of the three translational modes.  Any rotational mode of a Rouse model can only be constructed from the $3N-3$ internal modes, all of which have finite relaxation times.  No combination of modes with finite relaxation times can have the infinite relaxation time (zero eigenvalue) of a rotation mode. Therefore there is no way to write a linear combination of Rouse modes that corresponds to rotation.  Rouse chains therefore cannot rotate.

How is it possible for a polymer chain to be unable to rotate?  To give credit where it is due, when I described this conundrum to a former student, Paul Whitford, he gave an immediate response: "It must be a point!"\cite{paul2014a}  Indeed, the path that shows that the Rouse chain is a point is revealed by comparing the construction of the coordinates and the force constant matrices in the Wilson-Decius-Cross and Rouse-Zimm models.  Consider two mass points connected by a spring. In both models, the location $\bm{R}_{i}$  of mass point $i$ is described as the sum of a vector $\bm{O}_{i}$  from the origin to the equilibrium position of mass $i$ , plus a displacement vector $\bm{r}_{i}$  from the equilibrium position of mass $i$ to its actual position, i. e.,
\begin{equation}
    \bm{R}_{i} =  \bm{O}_{i}  +  \bm{r}_{i}.
    \label{eq:position}
\end{equation}
In the Wilson-Decius-Cross molecular model, the directions of the spring forces are calculated using the equilibrium positions  $\bm{O}_{i}$ of the mass points, the vibrational displacements $\bm{r}_{i}$ of the mass points from their equilibrium positions being approximated as being negligible relative to the equilibrium distance between the points.  Correspondingly, the force between two mass points lies along the line of centers connecting the equilibrium positions of the two mass points.  In the Rouse-Zimm models, the directions of the spring forces between two polymer beads are entirely determined by the displacements $\bm{r}_{i}$  of the polymer beads from their equilibrium positions.  In terms of eq.\ \ref{eq:position}, in the Rouse-Zimm models  $\bm{R}_{i} = 0$  for all $i$.  Rouse-Zimm beads do not have positions that are partially fixed relative to each other, so they do not form an extended object that can rotate. Indeed, if all bead displacements $\bm{r}_{i}$  in a Rouse-Zimm chain are set to zero, all bead locations are the same. The rest configuration of a Rouse-Zimm chain is a point.

The actual number of Rouse modes in a realistic polymer is extremely large. Does it matter that the Rouse model does not capture three of them well? It might be proposed that if only a few of the Rouse model modes were incorrect, the consequences would not be substantial.  That argument might be acceptable if a few very short-lived modes were incorrect.  However, the modes that are missing are the three modes corresponding to whole body rotation. For some polymers, whole body rotation is the dominant mode for dielectric relaxation and, as shown by Kirkwood and Riseman, for viscous dissipation, so whole body rotation can not be neglected.

Readers will note that Rouse modes are sometimes useful in calculations on polymer dynamics.  How is this possible, if the underlying model is problematic?  The answer is that the Rouse modes are also a discrete spatial fourier transform of the particle positions.  The original bead positions form a complete orthogonal set of coordinates, valid for describing the positions of the beads and the chain conformation. Any new complete orthogonal set of linear combinations of bead positions is equally usable as a description of the chain conformation; the Rouse coordinates are just such a set.  To the extent that a new set of coordinates is chosen more or less well, the new coordinates may be more or less convenient for calculating chain dynamics.

\section{Extended Kirkwood-Riseman Model\label{s:extendedKRg}}

Here we consider the extension of the Kirkwood-Riseman model to treat multiple polymer chains. The calculation refers to time scales sufficiently long that polymer inertia can be neglected. The solvent is treated as a continuum fluid. Each polymer chain is treated as a line of beads that interacts with the solvent by applying to the solvent a series of point forces.  The point forces create solvent flows and hydrodynamic forces on other polymer beads, the flows and forces being described by mobility tensors $\bm{\mu}_{ij}$. Beads on each chain are linked by springs; a spring is a hydrodynamically-inert coupler that determines the distribution of bead-bead distances. We consider only \emph{ghost chains} that can pass through each other; excluded-volume interactions only serve to set minimum distances of approach between pairs of beads. Chain motions are approximated by whole-chain translation and rotation; internal modes that change the shape of a chain have not yet been included in the Hydrodynamic Scaling Model.

\subsection{Bead-Bead Hydrodynamic Interactions\label{ss:beadbead}}

The effects of hydrodynamic interactions are usefully described by
mobility tensors $\mu_{ij}$. These tensors give the hydrodynamic force on a bead (or chain) $i$ due to the force a bead (or chain) $j$ exerts on the solvent; $i = j$ is allowed.  Separate expressions are needed for the self ($i=j$) and distinct ($i \neq j)$ components of $\mu_{ij}$. For the calculations here, we begin with the $\bm{\mu}_{ij}$ that relate the force on bead $i$ to the force that bead $j$ applies to the solvent. After some work, we end with a second set of mobility tensors that give the force and torque on a chain $i$ due to a force or torque applied to the solution by a chain $j$.

The mobility tensors are specifically of interest because they determine the self-diffusion coefficient via
\begin{equation}
      D_{s} = \frac{1}{3} k_{B} T \mathrm{trace}(\bm{\mu}_{ii}).
      \label{eq:Dsdef}
\end{equation}
Here $k_{B}$ is Boltzmann's constant and $T$ is the absolute temperature. For spheres in solution, the mobility tensors can be expanded as power series in $a/r$, $a$ being a sphere radius and $r$ being the distance between the spheres, as developed by Kynch\cite{kynch1958a}, Mazur and van Saarloos\cite{mazur1982a}, this author\cite{phillies1982a}, and Ladd\cite{ladd1989a}. Part of the expansion improves the accuracy of the hydrodynamic interaction tensor for spheres that are close to each other.  Other extensions describe additional interactions between three or more spheres. The lowest-order approximation to the hydrodynamic interaction between two spheres is the Oseen tensor. The $\mu_{ij}$ can be expanded as\cite{kynch1958a,mazur1982a,phillies1982a}
\begin{equation}
\bm{\mu}_{ii}=\frac{1}{f_o}\left( (\bm{I}+\sum_{l, l \neq  i}\bm{b}_{il}+\sum_{\substack{m, m \neq i {\rm\, or\, } l \\ l \neq i}} \bm{b}_{iml}+...\right)
\label{eq:23}
\end{equation}
for the self terms and
\begin{equation}
\bm{\mu}_{ij}=\frac{1}{f_o}\left(\bm{T}_{ij} + \sum_{\substack{i, j, m \\ i, j, m \text{ distinct }} }\bm{T}_{imj}+...\right), \quad i\neq j
\label{eq:24}
\end{equation}
for the distinct terms.

The leading terms of the $\bm{b}$ and $\bm{T}$ tensors are\cite{mazur1982a}
\begin{equation}
\bm{b}_{il}  = -\frac{15}{4}\left(\frac{a}{r_{il}}\right)^{4}\bm{\hat{r}}_{il}\bm{\hat{r}}_{il}
   \label{eq:bilspheres}
\end{equation}

\begin{displaymath}
\bm{b}_{iml}  = \frac{75a^7}{16r_{im}^{2}r_{il}^{2}r_{ml}^{3}} \{ [1-3(\bm{\hat{r}}_{im}\cdot \bm{\hat{r}}_{ml})^{2}][1-3(\bm{\hat{r}}_{ml}\cdot \bm{\hat{r}}_{li})^{2}]
\end{displaymath}
\begin{equation}
+6(\bm{\hat{r}}_{im}\cdot \bm{\hat{r}}_{ml})(\bm{\hat{r}}_{ml}\cdot \bm{\hat{r}}_{li})^{2}-6(\bm{\hat{r}}_{im}\cdot \bm{\hat{r}}_{ml})(\bm{\hat{r}}_{ml}\cdot \bm{\hat{r}}_{li})(\bm{\hat{r}}_{li}\cdot \bm{\hat{r}}_{im}) \} \bm{\hat{r}}_{im} \bm{\hat{r}}_{li}
    \label{eq:bilmspheres}
\end{equation}

\begin{equation}
\bm{T}_{ij}  = \frac{3}{4}\frac{a}{r_{ij}}[ \bm{I}+\bm{\hat{r}}_{ij}\bm{\hat{r}}_{ij}]
  \label{eq:Oseen2}
\end{equation}

\begin{equation}
\bm{T}_{iml} =-\frac{15}{8}\frac{a^4}{r_{im}^{2}r_{ml}^{2}}[\bm{I}-3(\bm{\hat{r}}_{im}\cdot \bm{\hat{r}}_{ml})^{2}]\bm{\hat{r}}_{im}\bm{\hat{r}}_{ml}
    \label{eq:Oseen3}
\end{equation}
where only the lowest order term (in $\frac{a}{r}$) of each tensor is shown. See Mazur and van Saarloos\cite{mazur1982a} for the higher-order terms. Here  $\bm{I}$ is the unit tensor, $r = |\bm{r}|$, the unit vector is $\hat{\bm{r}} = \bm{r}/r$, $\eta_{o}$ is the solvent viscosity, and $ \hat{\bm{r}}\hat{\bm{r}}$ is an outer product.

$\bm{b}_{ij}$ and $\bm{T}_{ij}$ describe the hydrodynamic interactions of a pair of interacting spheres. $\bm{T}_{ij}/f_{o}$ describes the velocity induced in particle $i$ due to a force applied to particle $j$, while $\bm{b}_{ij}$ describes the retardation of a moving particle $i$ due to the scattering by particle $j$ of the wake set up by $i$. $\bm{T}_{iml}$ and $\bm{b}_{iml}$ describe interactions between trios of interacting spheres. $\bm{T}_{iml}$ describes the velocity of particle $i$ by a hydrodynamic wake set up by particle $l$, the wake being scattered by an intermediate particle $m$ before reaching $i$. $\bm{b}_{iml}$ describes the retardation of a moving particle $i$ due to the scattering, first by $m$ and then by $l$, of the wake set up by $i$.

In most of the following, the individual beads are taken to be small relative to distances between beads on different polymer chains, so only the lowest order (in $a/r$) term is used to describe the bead-bead interactions, this being the Oseen tensor of eq.\ \ref{eq:Oseen2}.

\subsection{Chain-Chain Hydrodynamic Interactions\label{ss:chainchain}}

Having considered the hydrodynamic interactions between polymer beads, we now advance to calculate the hydrodynamic interactions between pairs of polymer chains. The method of reflections is used to compute the interchain hydrodynamic interactions.   A chain whose beads move with respect to the solvent creates flows in the surrounding solvent.  These flows act on other chains.  In response to those flows, the other chains move.  Those chain motions induce additional solvent flows. The hydrodynamic equations are linear, so if a chain $A$ is subject to flows due to chains $B$ and $C$, the flow acting on chain $A$ is the sum of the flows created by $B$ acting on $A$ and by $C$ acting on $A$. Because the flow properties are linear, all hydrodynamic effects can be obtained by considering a line of chains, each chain acting on the next in the line. We say that the process is \emph{scattering}: The flow created by each chain is \emph{scattered} when it encounters the next chain in the line.  It is not assumed that each chain in a line must be different from all the other chains in a line; the line of chains may loop back on itself so that a given chain appears in the line more than once.

The chains in a line are labelled 1, 2, 3, \ldots.  The center-of-mass location of chain $j$ is the vector $\bm{a}_{j}$, the $j$ labelling which of the $N_{c}$ chains is involved. The location of a bead $i$ with respect to its chain's center-of-mass is $\bm{s}_{i}$.  Each step of the calculation here involves only beads on a single chain, so $\bm{s}_{i}$ does not need a separate label specifying the chain of which it is a part.  The vectors from the center-of-mass of each chain in the line to the next chain's center of mass are the vectors  $\bm{R_{j}}$, with $\bm{R_{j}}= \bm{a}_{j+1} - \bm{a}_{j}$. Solvent flows are denoted $\bm{u}^{(n)}(\bm{r})$; they are implicit functions of position even if no dependence on $\bm{r}$ is specified. An imposed solvent flow, such as a fluid shear field, is denoted ${\bf u}^{(0)}$; in a quiescent liquid, ${\bf u}^{(0)} = \bm{0}$. Solvent flows created by the first, second,\ldots chains in a sequence are denoted ${\bf u}^{(1)}$, ${\bf u}^{(2)}$, \ldots,
respectively.

The velocity $\bm{v}_{j}$ of a bead $j$ that is located on chain $i$ may be divided between center-of-mass motion, whole-body rotation, and internal mode motions as
\begin{equation}
     \bm{v}_{j} = \bm{V}^{(i)} + \bm{\Omega}^{(i)} \times \bm{s}_{j} + \bm{\dot{w}}_{j}.
     \label{eq:beadvelocity}
\end{equation}
Here the chain's center-of-mass velocity is $\bm{V}^{(i)}$, the chain's angular velocity around its center of mass is $\bm{\Omega}^{(i)}$, and bead motions arising from chain internal modes are denoted $\bm{\dot{w}}_{j}$.  The superscripts on $\bm{V}$ and $\bm{\Omega}$ identify the reflection that created those parts of $\bm{V}$ and $\bm{\Omega}$.

The chain center-of-mass velocity is
\begin{equation}
    \bm{V}^{(i)} = \frac{\partial \bm{a}_{i}}{\partial t}.
    \label{eq:velocity}
\end{equation}
 $\bm{V}^{(i)}$ is determined by averaging over the $N$ beads of chain $i$, namely
\begin{equation}
       \bm{V}^{(i)} =  \frac{1}{N}\sum_{j=1}^{N} \bm{v}_{j}.
     \label{eq:Vdef}
\end{equation}

The $\bm{V}^{(i)}$ and $\bm{\dot{w}}_{i}$ are independent of $\bm{\Omega}^{(i)}$, so $\bm{\Omega}^{(i)}$ can be determined from eq.\ \ref{eq:beadvelocity} as
\begin{equation}
   \frac{1}{N} \sum_{j=1}^{N} \bm{s}_{j}  \times
(\bm{\Omega}^{(i)} \times \bm{s}_{j})
        = \frac{1}{N} \sum_{j=1}^{N} \bm{s}_{j}  \times \bm{v}_{j}.
    \label{eq:omegacalc}
\end{equation}
The instantaneous-square chain radius $s^{2}$ is $ N^{-1} \sum_{j=1}^{N} s_{j}^{2}$.

The model describes the low-frequency regime. The chain linear and angular momenta fluctuate, but over time scales of interest here the fluctuations average to zero. For the same reason, contributions to fluid flow from the higher-frequency ${\bf \dot{w}}_{i}$ are not taken into account. If the fluctuations in the total linear momentum and total angular momentum of each chain average to zero, from fundamental mechanics the total force and total torque on each chain after the first must also average to zero. (The first chain in a  line may also be subject to external forces, torques, or fluid flows, and so is a special case) One obtains
\begin{equation}
   \sum_{j=1}^{N} f_{j} (\bm{v}_{j} - \bm{u}(\bm{r}_{j})) = \bm{0}
    \label{eq:zeroforce}
\end{equation}
and
\begin{equation}
 \sum_{j=1}^{N} f_{j} \bm{s}_{j} \times      (\bm{v}_{j} - \bm{u}(\bm{r}_{j})) = \bm{0}
    \label{eq:zerotorque}
\end{equation}
The four equations \ref{eq:Vdef},  \ref{eq:omegacalc}, \ref{eq:zeroforce} and \ref{eq:zerotorque} take us from the fluid velocity ${\bf u}^{(n-1)}({\bf r}_{j})$ at the beads of chain $n$ to the center-of-mass translational and rotational velocities ${\bf V}^{(n)}$ and ${\bf
\Omega}^{(n)}$ of chain $n$.  The ${\bf V}^{(n)}$ and ${\bf \Omega}^{(n)}$ depend on the relative positions of the chains.

For the calculation of the self-diffusion coefficient, the first chain in the series is presumed to have some initial velocity that corresponds to its performing translational motion.  For the calculation of the viscosity increment, the first chain in the series is in a velocity shear. As will be seen, each chain moves at the local flow velocity. Each chain rotates so as to attempt to comply at its every point with the imposed shear flow. Each chain can translate and rotate, but its local velocity cannot at every bead be the same as the velocity that the fluid would have had at the same point, if the chain were absent.

\section{Extended Kirkwood-Riseman Model: Self-Diffusion\label{s:kreds}}

We now implement the method of reflections as described above.  We begin with polymer chain 1 that has linear velocity $\bm{V}^{(1)}$ and angular velocity $\bm{\Omega}^{(1)}$ with respect to the unperturbed and hence quiescent solvent. A bead $j$ on chain 1 then has velocity $\bm{v}^{(1)}_{j} = \bm{V}^{(1)} + \bm{\Omega}^{(1)} \times \bm{s}_{j}$, plus a component corresponding to the internal modes that we are neglecting.  The flow  $\bm{u}^{(1)}$ induced at $\bm{r}$ by all $M$ beads of chain 1 is
\begin{equation}
      \bm{u}^{(1)}(\bm{r}) = \sum_{j=1}^{M} \bm{T}(\bm{r}-\bm{s}_{j}) \cdot  \bm{v}^{(1)}_{j}.
      \label{eq:flow1}
\end{equation}

In the spirit of the Kirkwood-Riseman calculation, we now average over detailed relative locations of the individual beads. Functions of the vector $\bm{s}$ from the center of mass replace functions of the bead label $j$. All sums $\sum_{j} f_{j}$ over beads are replaced with  integrals $\int d\bm{s} \, g(\bm{s})f(\bm{s})$, $\bm{s}$ being a vector from the chain center of mass to a point within the chain, $g(\bm{s})$ being the density of beads at $\bm{s}$, and $f(\bm{s})$ being the
effective drag coefficient of the beads at $\bm{s}$. The integral of $f(\bm{s})$ over the complete chain is the total drag coefficient $F_{o}$.
Correlations in the shapes of nearby chains are neglected.

A series expansion for the Oseen tensor is $\bm{T}(\bm{r}-\bm{s}) = \bm{T}(\bm{r}) - \bm{s} \cdot \bm{\nabla} + \mathcal{O}(s^{2})$, namely
\begin{equation}
      \bm{T}(\bm{r}-\bm{s}) = \frac{1}{8 \pi \eta}\left[\frac{\bm{I} + \bm{\hat{r}}\bm{\hat{r}}}{r} - \bm{\hat{r}}\frac{\bm{s}\cdot(\bm{I} -3 \bm{\hat{r}}\bm{\hat{r}} )    }{r^{2}} - \frac{\bm{s}\bm{\hat{r}}}{r^{2}}  + \frac{\bm{s}\cdot \bm{\hat{r}}}{r^{2}}\bm{I} \right] + \mathcal{O}((\frac{s}{r})^{2}).
\label{eq:oseenseries}
\end{equation}

The resulting induced flow field, to lowest order in the series expansion, is
\begin{equation}
      \bm{u}^{(1)}(\bm{r}) =  \int d\bm{s} \, g(s) \frac{f_{o}}{8 \pi \eta}\biggl[\frac{\bm{I} + \bm{\hat{r}}\bm{\hat{r}}}{r} - \bm{\hat{r}}\frac{\bm{s}\cdot(\bm{I} -3 \bm{\hat{r}}\bm{\hat{r}} )    }{r^{2}} - \frac{\bm{s}\bm{\hat{r}}}{r^{2}}  + \frac{\bm{s}\cdot \bm{\hat{r}}}{r^{2}}\bm{I} \biggr] \cdot \biggl[\bm{V}^{(1)} + \bm{\Omega}^{(1)} \times \bm{s}_{j}\biggr]
      \label{eq:u1start}
\end{equation}
In the above $\int g(s) s^{2} \, d\bm{s} = R_{g}^{2}$. Terms odd in $\bm{s}$ vanish by symmetry. $f_{o}$, the chain drag coefficient, is $6 \pi \eta R_{h}$. By direct calculation, $\int g(s) \bm{s} \cdot \bm{\hat{r}} \bm{\Omega} \times \bm{s} \, d\bm{s} = R_{g}^{2} \bm{\Omega} \times \bm{\hat{r}}/3$. Here $R_{g}$ and $R_{h}$ are the radius of gyration and the hydrodynamic radius of the chain, with additional numerical subscripts on $R_{g}$ and $R_{h}$ being used to identify which chain's radii are under consideration.

The result of these steps is
\begin{equation}
     \bm{u}^{(1)}(\bm{r}) = \frac{3}{4}\frac{R_{h1}}{r}[\frac{\bm{I} + \bm{\hat{r}}\bm{\hat{r}}}{r}] \cdot \bm{V}^{(1)} + \frac{1}{2} \frac{R_{h1}R_{g1}^{2}}{r^{2}}(\bm{\Omega}^{(1)} \times \bm{\hat{r}})
     \label{eq:u1finish}
\end{equation}
The indicated terms are the longest-range parts of the flow field created by the motions of the first chain.  By expanding $\bm{T}(\bm{r}-\bm{s})$ to higher order in $\bm{s} \cdot \bm{\nabla}$, one would obtain terms of higher order in $(R_{g}/r)^{2}$.

The calculation proceeds now by iteration.  The flow field $\bm{u}^{(1)}(\bm{r})$ exerts forces on the next chain in the series.  The zero-force and zero-torque conditions let us calculate the linear and angular velocities $\bm{V}^{(2)}$ and $\bm{\Omega}^{(2)}$ of the next chain. Under the approximation that we neglect chain internal modes, the beads of the next chain move with velocities  $\bm{v}^{(2)}_{j} = \bm{V}^{(2)} + \bm{\Omega}^{(2)} \times \bm{s}_{j}$. Those beads cannot simply move with the solvent.  As a result, the beads of chain 2 exert forces on the solvent, thereby creating a  new flow field $\bm{u}^{(2)}(\bm{r})$, where $\bm{r}$ is now measured from the center of mass of chain 2.

The force on a representative bead $i$ of chain 2, due to the flow field $\bm{u}^{(1)}(\bm{r})$  scattered by chain 1, is
\begin{equation}
         \bm{F}_{i}^{(2)} =  f_{i}(\bm{u}^{(1)}(\bm{R}_{1}+ \bm{s}_{i}) - \bm{V}^{(2)} - \bm{\Omega}^{(2)} \times \bm{s}_{i}).
\label{eq:F2istart}
\end{equation}
$f_{i}$ is the bead's drag coefficient.  The bead is at $\bm{R}_{1}+ \bm{s}_{i}$, a displacement by $\bm{s}_{i}$ from the displacement $\bm{R}_{1}$ of the center of mass of chain 2 from the center of mass of chain 1.

The zero-force and zero-torque conditions are then applied to chain 2. To do this, beads at locations $\bm{s}_{i}$ are again replaced with a bead density $g(s)$, and the flow field $\bm{u}^{(1)}(\bm{R}_{1}+ \bm{s}_{i})$ is given  a series expansion, centered on the center-of-mass of chain 2, in powers of $\bm{s} \cdot \bm{\nabla}$. The zero-force condition starts as
\begin{equation}\label{eq:F2middle}
      f_{o} \int d\bm{s} \, g(s) [\bm{u}^{(1)}(\bm{R}_{1}) + \bm{s} \cdot \bm{\nabla}\bm{u}^{(1)}(\bm{R}_{1}) - \bm{V}^{(2)} - \bm{\Omega}^{(2)} \times \bm{s}_{i}]=0
\end{equation}
while the zero-torque condition starts as
\begin{equation}\label{eq:t2start}
    f_{o} \int d\bm{s} \, g(\bm{s}) [\bm{s} \times \bm{u}^{(1)}(\bm{r}) - \bm{s} \times \bm{V}^{(2)} - \bm{s} \times (\bm{\Omega}^{(2)} \times \bm{s})]=0.
\end{equation}

After noting that everything except $\bm{s}$ itself is independent of $\bm{s}$, while terms odd in $\bm{s}$ integrate to zero, and integrating on $\bm{s}$, one finds
\begin{equation}\label{eq:v2final}
   \bm{V}^{(2)} = \bm{u}^{(1)}(\bm{R}_{1})
\end{equation}
and
\begin{equation}\label{eq:omega2middle}
  \frac{2}{3} f_{o} \bm{\Omega}^{(2)} = f_{o} \int d\bm{s} \, g(s) [\bm{s} \times (\bm{s} \cdot \bm{\nabla}_{R}) \bm{u}^{(1)}(\bm{R}_{1})],
\end{equation}
the subscript on the $\bm{\nabla}$ being the variable with respect to which the derivatives are taken. Taking the spherical averages, one finally reaches\cite{phillies1998a}
\begin{equation}\label{eq:omega2final}
  \bm{\Omega}^{(2)} = - \frac{3}{4} \frac{R_{h1}}{R_{1}^{2}}[\bm{\hat{R_{1}}} \times \bm{V}^{(1)}] - \frac{1}{4}\frac{R_{h1} R_{g1}^{2}}{R_{1}^{3}}\bm{\Omega}^{(1)} \cdot [\bm{I}- 3\bm{\hat{R_{1}}}\bm{\hat{R_{1}}}].
\end{equation}

The flow field due to scattering from chain 2 is
\begin{displaymath}
  \bm{u}^{(2)}(\bm{r}) = -\frac{9}{16} \frac{R_{h1}R_{h2} R_{g2}^{2}}{R_{1}^{2} r^{2}}[1-3(\bm{\hat{r}}\cdot\bm{\hat{R}}_{1})^{2}] ( \bm{\hat{R}}_{1} \cdot \bm{V}^{(1)})\bm{\hat{r}}+ \notag
\end{displaymath}
\begin{displaymath}
  \frac{3}{8}\frac{R_{h1}R_{h2} R_{g1}^{2}R_{g2}^{2}}{R_{1}^{3} r^{2}}
  \biggl[\bm{\hat{r}} \times \bm{{\Omega}}^{(1)} - (\bm{\hat{r}} \times \bm{\hat{R}}_{1}) \bm{\hat{R}}_{1} \cdot \bm{\Omega}^{(1)} + \bm{\hat{r}}\cdot \bm{\hat{R}}_{1} (\bm{\Omega}^{(1)} \times \bm{\hat{R}}_{1}) - \notag
\end{displaymath}
\begin{equation}
  \bm{\hat{r}} \cdot (\bm{\Omega}^{(1)} \times \bm{\hat{R}}_{1})  \bm{\hat{R}}_{1} \cdot (\bm{I} -3 \bm{\hat{r}} \bm{\hat{r}})\biggr]
   \label{eq:chain2scatteredfield}
\end{equation}

Calculation of higher-order scattering events proceeds by iteration.  From the linear and angular velocities $\bm{V}^{(n)}$ and $\bm{\Omega}^{(n)}$ of chain $n$ in the sequence, we compute the induced fluid flow field $\bm{u}^{(n)}(\bm{R}_{n})$ at the location of chain $n+1$. From the flow field, we compute the linear and angular velocities $\bm{V}^{(n+1)}$ and $\bm{\Omega}^{(n+1)}$ of chain $n+1$.  We can now repeat the process \emph{ad infinitum}.
The final calculation only needs the part of $\bm{u}^{(3)}$ created by the linear velocity $\bm{V}^{(1)}$ of the first bead, namely
\begin{displaymath}
\bm{u}^{(3)}(\bm{r}) = \frac{27}{64} \frac{R_{h1}R_{h2}R_{h3}R^{2}_{g2} R^{2}_{g3} }{R_{1}^{2}R_{2}^{3}r^{2}} \biggl[(1-3( \bm{\hat{R}}_{1} \cdot \bm{\hat{R}}_{2})^{2}) \times
\end{displaymath}
\begin{equation}
 (1-3  (\bm{\hat{R}}_{2} \cdot \bm{\hat{r}})^{2} - 6( \bm{\hat{R}}_{1} \cdot \bm{\hat{R}}_{2})(\bm{\hat{R}}_{2} \cdot \bm{\hat{r}})
    + \bm{\hat{r}} \cdot[\bm{I}- \bm{\hat{R}}_{2} \bm{\hat{R}}_{2}] \cdot \bm{\hat{R}}_{1})\biggr] ( \bm{\hat{R}}_{1} \cdot \bm{V}_{1}) \bm{\hat{r}}.
\label{eq:u3final}
\end{equation}
This form does not include the contribution to $\bm{u}^{(3)}(\bm{r})$ from $\bm{\Omega}^{(1)}$.

The terms of the mobility tensors $\bm{\mu}_{ii}$ are obtained from the $\bm{u}^{(n)}(\bm{R}_{n})$ or the $\bm{V}^{(n+1)}$ by setting $\bm{R}_{n} = - \bm{R}_{1} - \bm{R}_{2} - \ldots - \bm{R}_{n-1}$ and suppressing the $\bm{V}^{(1)}$. One obtains for the relevant parts of the mobility tensor
\begin{equation}
       \bm{b}_{12} = - \frac{1}{f_{c}} \frac{9}{8} \frac{R_{h1}R_{h2} R_{g2}^{2}}{R_{1}^{4}}  \bm{\hat{R}_{1}}\bm{\hat{R}_{1}}
\label{eq:b2value}
\end{equation}
and
\begin{equation}
      \bm{b}_{123}\cdot \bm{V}_{1} = \bm{u}^{(3)}(\bm{r})|_{\bm{r} \rightarrow - \bm{R}_{1} - \bm{R}_{2}}
\label{eq:b3value}
\end{equation}

Taking appropriate ensemble averages over these tensors leads to a pseudovirial expansion for the self diffusion coefficient, \emph{viz.},
\begin{equation}
        D_{s}(c) = D_{s0} \left(1 - \frac{9}{16} \frac{R_{h1} R_{h2}}{a_{o} R_{g}}\left(\frac{4 \pi}{3}R_{g}^{3} \right) c + 9.3 \cdot 10^{-4} \frac{R_{h1} R_{h2}R_{h3}}{a_{o} R_{g}^{2}}\left(\frac{4 \pi}{3}R_{g}^{3} \right)^{2} c^{2}+\ldots\right) .
\label{eq:Dsseries}
\end{equation}
The numerical coefficient in the $c^{2}$ term was obtained by Monte Carlo integration.

We have now used a generalization of the Kirkwood-Riseman model to treat interchain hydrodynamic interactions.  The motions of each chain set up wakes in the surrounding fluid.  The surrounding fluid drives  the motion of other chains in the fluid, creating fresh wakes which act on still further chains in sequence.  Our generalization has several lacunae.  Intrachain hydrodynamics have not been included in the calculation. The accuracy of the calculation will diminish when chains overlap, due to strong interchain hydrodynamic interactions between pairs of nearly adjacent beads.

\subsection{Short-Range Hydrodynamic Effects\label{ss:shortrange}}

The purpose of this Subsection is to reveal some of the ways in which higher-order hydrodynamic interactions modify polymer dynamics. I follow the results of Phillies and Kirkitelos\cite{phillies1993b}. There are very considerable opportunities for extending the results of ref.\ \cite{phillies1993b}.

Equations \ref{eq:bilspheres}-\ref{eq:Oseen3} introduce short-range hydrodynamic interactions, corrections to the Oseen tensor approximation that become most important when the diffusing bodies are close together.  Consequences of short-range hydrodynamic interactions for the diffusion of colloidal spheres have been studied intensively\cite{phillies2016a}. Because beads of the same polymer are obliged to remain close to each other, effects of short range hydrodynamic interactions are reasonably expected to be at least as important for polymer dynamics as for colloid dynamics. Several authors \cite{freed1974a,freed1976a,freed1978a} have developed multiple scattering approaches for treating polymer-polymer interactions, but none of these developments have included short-range interactions. Freed\cite{freed1983z} has previously identified the use of short-range hydrodynamic interactions as an unexplored possibility in this context.

Some effects of short-range interactions on polymer diffusion have already been examined. The Oseen tensor $\bm{T}_{ij}$ effectively approximates the interacting bodies as points, an approximation conspicuously dubious when treating the diffusion of a linear rod polymer around its major axis. Bernal\cite{bernal1981a} models a rod as a shell of small spheres in order to remove the approximation.  The DeWames-Zwanzig singularity\cite{dewames1967a,zwanzig1968a} in the Kirkwood-Riseman\cite{kirkwood1948a} treatment of translational diffusion by a rigid rod was shown by Yamakawa\cite{yamakawa1970a}  to be eliminated by including the $\mathcal{O}((a/r)^{3})$ corrections to the Oseen tensor.

Phillies and Kirkitelos\cite{phillies1993b} made two applications of the short-range hydrodynamic interaction tensors. First, they calculated the chain-chain hydrodynamic interaction tensors including bead-bead interactions out to the $\mathcal{O}((a/r)^{7})$ level, both for the chain-chain $\bm{T}_{ij}$ and to a higher level for the chain-chain $\bm{b}_{ij}$. They further calculated the effect of the short-range hydrodynamic interactions on the diffusion coefficients of a free monomer and for a monomer bead incorporated into a polymer chain in solution. These effects are entirely distinct from the contribution of short range hydrodynamic interactions to the chain-chain hydrodynamic interaction tensors. Because the beads of a polymer are always close to other beads of the same chain, at no polymer concentration can the diffusion coefficient of a chain monomer be as large as the diffusion coefficient of a free monomer. At concentrations below the overlap concentration, solvent molecules readily penetrate into polymer coils, but polymer chains do not interpenetrate a great deal. As a result, the addition of polymer molecules to a dilute solution is more effective at retarding the motion of free monomers that at retarding the motion of monomer units of a given polymer chain. At polymer concentrations above the chain overlap concentration, the total polymer concentration is the same everywhere in solution, but the correlation hole created by a chain of interest ensures that the concentration of other chains, near the beads of the chain of interest, is never as large as the average concentration of chains in solution. As a result, the effect of interchain interactions on the mobility of a given polymer bead is never as large as the effect of the same interactions on the mobility of a free monomer in solution.

Higher-order hydrodynamic interactions make contributions of the same nature to the drag coefficients of a free monomer and a whole chain. However, the contributions to the free monomer and chain drag coefficients are not equal; nor are they multiplicative, contrary to the core assumption behind the common practice of normalizing polymer transport data with small-molecule diffusion coefficient data as a correction for 'monomer friction effects'. The notion that the concentration dependence for $D_{s}$ for free monomers or solvent molecules reveals the concentration dependence of the mobility of monomer units within a polymer chain is therefore incorrect. However, the effect of interchain interactions on the free monomer mobility and on the mobility of monomer units of polymers can be separately calculated.

\section{Extended Kirkwood-Riseman Model for the Viscosity \label{s:kreta}}

This Section considers the contribution to the solution viscosity $\eta$ from chain-chain hydrodynamic interactions, as obtained from an extended Kirkwood-Riseman model. We obtain the lead terms in a pseudovirial expansion for $\eta(c)$. The underlying hydrodynamic interactions depend on the interchain distance $r$ as $r^{-2}$ or $r^{-3}$, so the convergence of the pseudovirial expansion's cluster integrals is potentially delicate.

The literature includes a considerable number of earlier efforts to compute $\eta(c)$ from some variation on the approach seen here.  Note papers by Brinkman\cite{brinkman1952a}, Riseman and Ullmann\cite{riseman1951a}, Saito\cite{saito1951a,saito1952a}, Yamakawa\cite{yamakawa1961a}, Freed and Edwards\cite{freed1974a,freed1974b,freed1975a}, Freed and Perico\cite{perico1981a}, and Altenberger, et al.\cite{altenberger1984a}. There was appreciable awareness in these reports that the long-range nature of the Oseen tensor can lead to improper integrals during an ensemble averaging process for generating the pseudovirial series. Edwards and Freed proposed\cite{freed1974a,freed1974b,freed1975a} that the integrals were in fact proper due to their hypothesized process of ``hydrodynamic screening''  but later calculations by Freed and Perico\cite{perico1981a} and by Altenberger, et al.\cite{altenberger1984a} conclude that there is no such phenomenon as hydrodynamic screening in polymer solutions.

Our general approach is to apply a velocity field to the solution, and calculate the additional power dissipation caused by the polymer beads as they move with respect to the solvent.

\subsection{Flow Fields from Scattering of a Shear Field \label{ss:ssf}}

We choose to impose a spatially oscillatory flow field
\begin{equation}
     \bm{u}^{(0)}(\bm{r})  = u_{o} \cos(k x) \bm{\hat{j}};
     \label{eq:shearfield}
\end{equation}
$\bm{u}^{(0)}(\bm{r})$ is the bare velocity field and $k$ is the spatial oscillation frequency. The oscillations are not time-dependent, so the shear magnitude is $|\alpha(x)| = u_{0} k  \sin(k x)$. The average shear is $\langle \alpha^{2} \rangle = u_{0}^{2}k^{2}/2 $. The shear is assumed to be sufficiently weak that the average spherical symmetry of the polymer chain is not perturbed.

The effect of the spatial oscillations is to ensure that the total of the external forces, applied to the fluid to create the flow field, vanishes.   At the end of the calculation, we take the limit $k \rightarrow 0$.  As seen below, the scattering of the velocity field by the polymer molecules makes an additional contribution to the flow field, so that the velocity field measured experimentally will not be the field given by eq.\ \ref{eq:shearfield}. The observable shear field will include the contributions due to scattering of the imposed shear field by all the polymers in solution.

The power $P$ dissipated by polymer chains in a solution flow is
\begin{equation}
     P = \left\langle \sum_{i=1}^{M}\sum_{j=1}^{N} f_{ij} (\bm{v}_{ij} - \bm{u}(\bm{r}_{ij}))^{2} \right\rangle.
     \label{eq:powerchaininitial}
\end{equation}
Here the sum proceeds over all $N$ beads of each of the $M$ chains in some volume $V$, $f_{ij}$ is the drag coefficient of bead $j$ of chain $i$, $\bm{v}_{ij}$ is the velocity of that bead, and $\bm{u}(\bm{r}_{ij})$ is the velocity that the solvent would have had, at the location $\bm{r}_{ij}$ of the bead in question, if the bead had been absent.

The viscosity increment is extracted from $P$ via the relationship
\begin{equation}\label{eq:powershear}
  \frac{dP}{dV} = \delta \eta \left(\frac{\partial u_{y}}{\partial x} \right)^{2},
\end{equation}
where the velocity shear has been simplified to correspond to the flow field directions described by eq.\ \ref{eq:shearfield}.

To describe the polymer chains and their motions, we use the same notation as that introduced in the previous section.  Because the fluid motions are not the same as in the self-diffusion problem, the calculational details change.

Each chain's center-of-mass translational velocity is the average of the velocities of its $N$ beads, so
\begin{equation}
      \bm{V}^{(i)} =  \frac{1}{N}\sum_{j=1}^{N} \bm{v}_{j}.
     \label{eq:Vdefeta}
\end{equation}

The translational, rotational, and internal mode components of the chain motion are independent of each other, so the rotational velocity vectors ${\bf \Omega}^{(i)}$ follow from
\begin{equation}
   \frac{1}{N} \sum_{j=1}^{N} \bm{s}_{j}  \times
(\bm{\Omega}^{(i)} \times \bm{s}_{j})
        = \frac{1}{N} \sum_{j=1}^{N} \bm{s}_{j}  \times \bm{v}_{j}.
    \label{eq:omegacalceta}
\end{equation}

As in the previous section, the zero-force and zero-torque equations \ref{eq:zeroforce} and \ref{eq:zerotorque} determine how each chain moves.

The applied solvent flow within chain $n+1$ is obtained from ${\bf u}^{(n)}$ via a Taylor expansion around the center of mass of chain $n+1$, to wit
\begin{equation}
     \bm{u}^{(n)}(\bm{R}_{n}+\bm{s}) =\bm{u}^{(n)}(\bm{R}_{n})
          + (\bm{s} \cdot {\nabla})\bm{u}^{(n)}(\bm{R}_{n})
          + \frac{1}{2} (\bm{s} \cdot \bm{\nabla})^{2}
        {\bf u}^{(n)}(\bm{R}_{n}) + \ldots,
     \label{eq:uexpansion}
\end{equation}
The $\bm{u}^{(n)}$ are in part determined by $\bm{a}_{1}$, the location of the first chain, and
those of the $\bm{R}_{j}$ with $j < n $, these being the displacement vectors taking one from chain $1$ to chain $n$.

For the first chain, after making a Taylor series expansion of the fluid velocity around the chain center of mass $\bm{a}_{1}$ (with $s_{x} = {\bf s} \cdot {\bf i}$ and  $a_{x} = {\bf a}_{1} \cdot {\bf i}$),
the zero-force condition may be written
\begin{multline}
       \int d\bm{s} \, f(\bm{s})g(\bm{s})
      \left(\bm{V}^{(1)} + \bm{\Omega}^{(1)}
      \times \bm{s} + \bm{\dot{w}}(\bm{s}) - u_{0} \cos(k a_{x}) \bm{\hat{j}} \right.  \\  \left. -  \alpha(a_{x}) s_{x} \bm{\hat{j}} - \frac{1}{2} (\bm{s} \cdot \bm{\nabla})^{2} \bm{u}^{(0)}(\bm{a}_{1})-\ldots \right) = 0.
     \label{eq:zeroforce2}
\end{multline}
Because we are discussing weak shear, $f(\bm{s})g(\bm{s})$ is spherically symmetric, so only terms even in $\bm{s}$ survive integration, leading to
\begin{equation}
      \bm{V}^{(1)} = u_{0} \cos(k a_{x}) \bm{\hat{j}} +\mathcal {O}(s^{2})
      \label{eq:V1value}
\end{equation}
Up to terms in $(\bm{s}\cdot \bm{\nabla})^{2}$, the first chain simply moves with the velocity that the solvent would have had, at the chain's center of mass location, if the chain were not present.

Substituting for $\bm{v}^{(1)}$ and $\bm{u}^{(0)}$, the corresponding zero-torque condition is
\begin{multline}
       \int d\bm{s} \, f(\bm{s})g(\bm{s}) \bm{s} \times (\bm{V}^{(1)} +
      \bm{\Omega}^{(1)} \times \bm{s} + \bm{\dot{w}}) =\\
\int d\bm{s} \, f(\bm{s})g(\bm{s}) [\bm{s} \times (u_{0} \cos(k a_{x})
      \bm{\hat{j}} + \alpha(a_{x}) s_{x} \bm{\hat{j}} +  \frac{1}{2} (\bm{s} \cdot \bm{\nabla})^{2}
              \bm{u}^{(0)}(\bm{s})-\ldots) ]
     \label{eq:zerotorque2}
\end{multline}

We denote $\int d\bm{s} \, f(\bm{s})g(\bm{s}) Q(\bm{s}) = F_{o}\langle Q(\bm{s}) \rangle$.  Applying an extended series of identities seen in ref.\ \cite{phillies2002b}, one finally obtains
\begin{equation}
      \bm{\Omega}^{(1)} = \frac{\alpha(a_{x})}{2} \bm{\hat{k}},
      \label{eq:omega1val}
\end{equation}
which is the result of Kirkwood and Riseman\cite{kirkwood1948a} for a single chain in a shear.

Chain 1 cannot at every bead be stationary with respect to the fluid.  For example, it is doing whole-body rotation, so some of its beads are moving in  directions perpendicular to the direction of the fluid flow. The fluid flow, bead velocity, and Oseen tensor then combine to give the  fluid flow $\bm{u}^{(1)}(\bm{r})$ induced by the first polymer chain, namely
\begin{equation}
     \bm{u}^{(1)}(\bm{r}) = \int d\bm{s} \, f(\bm{s})  g(\bm{s})
           \bm{T}(\bm{r} - \bm{s}) \cdot (\bm{v}^{(1)}(\bm{s})
           -\bm{u}^{(0)}(\bm{s})).
      \label{eq:u1general}
\end{equation}
A Taylor-series expansion of the Oseen tensor is
\begin{equation}
  \bm{T}(\bm{r} - \bm{s}) = \bm{T}(\bm{r}) - \bm{s}
            \cdot \bm{\nabla} \bm{T}(\bm{r}) + \mathcal{O}(s^{2})
  \label{eq:TOseenexpand}
\end{equation}
where
\begin{equation}
  \bm{s} \cdot  \bm{\nabla} \bm{T}(\bm{r}) = \frac{1}{8 \pi \eta_{o}}\left(
  \frac{\bm{s}\bm{\hat{r}}}{r^{2}} +\frac{\bm{\hat{r}}\bm{s}}{r^{2}}
   -\frac{\bm{s} \cdot \bm{\hat{r}}}{r^{2}}(\bm{I} + 3 \bm{\hat{r}}
   \bm{\hat{r}})
  \right).
  \label{eq:Tgrad}
\end{equation}

On substituting in eq.\ \ref{eq:u1general} for ${\bf T}$, ${\bf V}^{(1)}$, and
${\bf u}^{(0)}$, and applying identities for integrals over $\bm{s}$, the induced flow is
\begin{equation}
       \bm{u}^{(1)}(\bm{r}) = \frac{F_{o} \alpha S^{2}}{8 \pi \eta_{o}
r^{2}}\frac{xy}{r^{2}}\bm{\hat{r}}.
      \label{eq:u1value}
\end{equation}

The process now advances by iteration. $\bm{u}^{(1)}(\bm{r})$ acts through a vector $\bm{R}_{1}$ on chain 2 inducing in it a translational velocity
\begin{equation}
     \bm{V}^{(2)} =  \frac{F_{o} \alpha S^{2}}{8 \pi \eta_{o}
R_{1}^{2}}\frac{X_{1}Y_{1}}{R_{1}^{2}}\bm{\hat{R}_{1}}
    \label{eq:V2fin}
\end{equation}
and a rotational velocity
\begin{equation}
     \bm{\Omega}^{(2)} = \frac{1}{2} \frac{F_{o} \alpha S^{2}}{8 \pi \eta_{o}
R_{1}^{3}}\left[\left(\frac{X_{1}^{2} - Y_{1}^{2}}{R_{1}^{2}}\right) \bm{\hat{k}}
+\frac{Y_{1}Z_{1}}{R_{1}^{2}} \bm{\hat{j}}
-\frac{X_{1}Z_{1}}{R_{1}^{2}} \bm{\hat{i}}    \right].
     \label{eq:omega2fin}
\end{equation}
Here $\bm{R}_{1} \equiv (X_{1}, Y_{1}, Z_{1})$.

The fluid flow that has been double scattered by chains 1 and 2 is
\begin{equation}
          \bm{u}^{(2)}(\bm{R}_{1},\bm{R}_{2})
= + \alpha \left(\frac{F_{o} S^{2}}{8 \pi
\eta_{o}}\right)^{2}\frac{\bm{\hat{R}}_{2}}{R_{1}^{3} R_{2}^{2}}
 \left[\frac{X_{1}Y_{2} +Y_{1} X_{2}}{R_{1}R_{2} }
(\bm{\hat{R}}_{1} \cdot \bm{\hat{R}}_{2})
         + \frac{X_{1}Y_{1}}{R_{1}^{2}}
     [1-5(\bm{\hat{R}}_{1} \cdot \bm{\hat{R}}_{2})^{2}]
         \right].
         \label{eq:u2finalform}
\end{equation}
Phillies\cite{phillies2002b} supplies the corresponding large expressions for $\bm{V}^{(3)}$, $\bm{\Omega}^{(3)}$, and $\bm{u}^{(3)}$.

\subsection{Power Dissipated by Chains in a Shear Field}

We now advance to calculating the power dissipated by the polymer molecules as they move with respect to the fluid.  The simplest case refers to dilute chains in a shear $\alpha$, for which eq.\ \ref{eq:powerchaininitial} becomes
\begin{equation}\label{eq:powerdilute}
  P = \left \langle M \sum_{i=1}^{N} f_{i} \left(\bm{V}^{(1)} + \frac{\alpha}{2} \bm{\hat{k}} \times \bm{s}_{i} + \bm{\dot{w}}_{i}- \bm{u}^{(0)}(\bm{R}_{i}) - \alpha(x) s_{x} \bm{\hat{j}} \right)^{2} \right\rangle
\end{equation}
$\bm{V}^{(1)}$ and $\bm{u}^{(0)}(\bm{R}_{1})$ cancel. In the model, internal chain modes are neglected, so
the $\bm{\dot{w}}_{i}$ do not modify the viscosity.  Changing variables from $\sum_{i=1}^{N} f_{i}$ to $\int d\bm{s} \, f(\bm{s}) g(\bm{s})$,
applying needed identities for the integrals on $\bm{s}$, and averaging $\langle \cdots
\rangle$ over chain configurations and positions,
\begin{equation}
        P_{1} = N_{c} \frac{F_{o} S^{2}}{6} \alpha^{2}.
     \label{eq:P1simplified}
\end{equation}
The average over chain positions is needed because the shear rate depends on position.  In the above calculation, the limit $k \rightarrow 0$  could have been taken either before or after the positional average.

We calculated above the scattering of the shear field by a specific first chain to a specific second chain, etc.
The flow field acting on a given bead includes the original shear field and also all scattered flows that reach that bead. On the same line, the center-of-mass velocity and rotation rate of a given chain are simply the sums of the center-of-mass velocities and rotation rates
induced by all flows acting on the given chain.

We now introduce a systematical notation that includes all scattering events.  The chain locations are more useful as variables than are the
displacement vectors.  The flow created at ${\bf r}$ by single scattering from a chain at ${\bf a}_{2}$ is
\begin{equation}
    {\bf u}^{(1)}(\bm{R}_{1}) \equiv \bm{u}^{(1)}(\bm{a}_{2}, \bm{r}).
    \label{eq:u1raa}
\end{equation}
Similarly, the double-scattered flow at $\bm{r}$ due to beads 2 and 3 is
$\bm{u}^{(2)}(\bm{a}_{2}, \bm{a}_{3}, \bm{r})$, and so forth.

The total flow field at $\bm{r}$ due to single scattering of the shear field by all chains other than the representative chain 1 is
\begin{equation}
 \bm{u}^{(1T)}(\bm{r}) = \sum_{j=2}^{N_{c}}\bm{u}^{(1)}(\bm{a}_{j}, \bm{r}).
    \label{eq:u1tdef}
\end{equation}
For double-scattered flows, a similar notation arises,
\begin{equation}
    \bm{u}^{(2T)}(\bm{r}) = \sum_{\substack{ j=1\\ k=2\\ j\neq k}}^{N_{c}}
    \bm{u}^{(2)}(\bm{a}_{j},\bm{a}_{k}, \bm{r}).
    \label{eq:u2tdef}
\end{equation}
the restriction on the double sum being that the last chain in the series cannot be chain 1.

What we next do is to calculate all of the flow fields at the representative chain 1.  This includes the original shear field at chain 1, the flow fields created at chain 1 by each of the other chains in the solution, and the flow fields that were created by one chain and scattered by a second chain before reaching chain 1. We then calculate the power dissipation due to chain 1, average over all locations of all chains, calculate the total shear gradient, and finally find the contribution of the representative chain 1 to the viscosity increment.

Chain 1 is a representative chain; it could equally be any chain in the solution. If chain 1 is at $\bm{r}$, so $\bm{r} \equiv \bm{a}_{1}$,
the $\bm{u}^{(1)}(\bm{a}_{j}, \bm{a}_{1})$,  $\bm{u}^{(2)}(\bm{a}_{j}, \bm{a}_{k}, \bm{a}_{1})$,\ldots induce chain motions $\bm{V}^{(2)}(\bm{a}_{j}, \bm{a}_{1})$, $\bm{\Omega}^{(3)}(\bm{a}_{j}, \bm{a}_{k}, \bm{a}_{1})$, etc., as calculated above. The zeroth-scattering-order velocities $\bm{V}^{(1)} \equiv \bm{V}^{(1T)} $ and $\bm{\Omega}^{(1)} \equiv \bm{\Omega}^{(1T)}$ are created by the initial shear field. The higher-order parts of $\bm{V}^{(nT)}$ and $\bm{\Omega}^{(nT)}$, the parts with $n> 1$, are due to scattering by all combinations of other particles, so
\begin{equation}
     \bm{V}^{(2T)}(\bm{a}_{1}) = \sum_{j=2}^{N_{c}} \bm{V}^{(2)}(\bm{a}_{j}, \bm{a}_{1})
     \label{eq:V2totaldef}
\end{equation}
and correspondingly
\begin{equation}
     \bm{\Omega}^{(3T)}(\bm{a}_{1}) = \sum_{\substack{j=1 \\ k=2 \\ j \neq k}}^{N_{c}} \bm{\Omega}^{(3)}(\bm{a}_{j}, \bm{a}_{k}, \bm{a}_{1}).
     \label{eq:Omega2totaldef}
\end{equation}
In these sums, neighboring arguments of a $\bm{u}^{(n)}$, $\bm{V}^{(n)}$ or $\bm{\Omega}^{(n)}$ must be distinct.

The total velocity at chain 1 is
\begin{equation}
     \bm{V} = \sum_{n=1}^{\infty} \bm{V}^{(nT)}
     \label{eq:Vtotaldef}
\end{equation}
and
\begin{equation}
     \bm{\Omega} = \sum_{n=1}^{\infty} \bm{\Omega}^{(nT)}.
     \label{eq:Omegatotaldef}
\end{equation}

The Debye form for the power dissipated by a representative chain is obtained from a sum over the $N$ beads of the chains
\begin{multline}
      P =  \left\langle \sum_{i=1}^{N}  f_{i} \left(\bm{V}^{(1T)} + \bm{\Omega}^{(1T)} \times \bm{s}_{i}+\bm{V}^{(2T)} + \bm{\Omega}^{(2T)} \times \bm{s}_{i}+ \ldots \right. \right. \\  - \left. \left. \bm{u}^{(0)}(\bm{r}_{i}) - \bm{u}^{(1T)}(\bm{r}_{i}) -\ldots \right)^{2} \right\rangle.
       \label{eq:Ptotal1}
\end{multline}
We advance with Taylor series expansions in $\bm{s}_{i}$. As seen above, to lowest order in ${\bf s}$,
${\bf V}^{(n+1T)}$ and ${\bf u}^{(nT)}$ cancel term-by-term  for all $n$, so
\begin{equation}
      P =  \sum_{i=1}^{N}  f_{i} [\bm{\Omega}^{(1T)} \times
\bm{s}_{i} + \bm{\Omega}^{(2T)} \times
\bm{s}_{i}+ \ldots - \bm{s}_{i} \cdot \bm{\nabla} \bm{u}^{(0)}(\bm{a}_{1})
- \bm{s}_{i} \cdot \bm{\nabla} \bm{u}^{(1T)}(\bm{a}_{1})
- \ldots ]^{2}.
       \label{eq:Ptotal3}
\end{equation}

The square generates three sorts of terms. Averaging over chain
configurations,
\begin{multline}
      \langle (\bm{s} \cdot \bm{\nabla}) \bm{u}^{(n)} \cdot
(\bm{s} \cdot \bm{\nabla})
\bm{u}^{(m)} \rangle \equiv
\left\langle \sum_{(i, j, m) = (x, y, z)}
  s_{i}\frac{\partial \bm{u}_{m}^{(a)} }{\partial x_{i}}
  \cdot s_{j}\frac{\partial \bm{u}_{m}^{(b)} }{\partial x_{j}}
\right\rangle
\\ =
\left\langle \frac{S^{2}}{3}\sum_{i,m = (x,y,z)}
 \frac{\partial \bm{u}_{m}^{(a)}}{\partial x_{i}}
 \frac{\partial \bm{u}_{m}^{(b)}}{\partial x_{i}}
\right\rangle.
       \label{eq:Ptotal4}
\end{multline}
Terms in $s_{i}s_{j}$ with $i \neq j$ average to zero.

In addition
\begin{equation}
     \langle (\bm{\Omega}^{(a)} \times \bm{s})\cdot( \bm{\Omega}^{(b)}
\times \bm{s})\rangle = \frac{2}{3} S^{2} \bm{\Omega}^{(a)} \cdot
\bm{\Omega}^{(b)}
     \label{eq:Ptotal5}
\end{equation}
and
\begin{equation}
      \langle \bm{\Omega}^{(a)} \times \bm{s} \cdot (\bm{s} \cdot \bm{\nabla}) \bm{u}^{(b)} \rangle =\langle \bm{\Omega}^{(a)} \cdot \bm{s} \times (\bm{s} \cdot \bm{\nabla} \bm{u}^{(b)}) \rangle,
     \label{eq:Ptotal6}
\end{equation}
while from the zero torque condition
\begin{equation}
  \langle \bm{s} \times (\bm{s} \cdot \bm{\nabla}) \bm{u}^{(b)}\rangle
  = \frac{2 F_{o} S^{2}}{3} \bm{\Omega}^{(b+1)}.
     \label{eq:Ptotal7}
\end{equation}

We obtain the general form for the power dissipation, namely
\begin{equation}
    P =  \sum_{a =0}^{\infty} \sum_{b =0}^{\infty} P_{a,b}
    \label{eq:Ptotalfinal}
\end{equation}
with
\begin{equation}
     P_{a,b} = \left\langle \frac{F_{o} S^{2}}{3} \left[ \sum_{i=1}^{3}\sum_{j=1}^{3}
\left[\bm{u}_{i,j}^{(aT)}  \bm{u}_{i,j}^{(bT)}\right] - 2  \bm{\Omega}^{(a+1T)}
\cdot  \bm{\Omega}^{(b+1T)}\right] \right\rangle.
    \label{eq:Ptotalcomps}
\end{equation}
The Einstein derivative notation
\begin{equation}\label{eq:einsteinderivative}
    \bm{u}_{l,j}^{(aT)}  \equiv
(\partial \bm{u}^{(aT)} \cdot \bm{\hat{l}}/\partial x_{j})
\end{equation}
(where $j,l = 1,2,3$ represent the three Cartesian coordinates) is in use.  The average is over all chain locations. In the first sum,
$a \neq b$ is allowed.  For example, a particle rotating at ${\bf \Omega}^{(2)}$ is moving not only with respect to the driving flow
${\bf u}^{(1)}$ but also with respect to the original imposed shear ${\bf u}^{(0)}$.

\subsection{The Total Shear Field}

In the previous subsection, the bare shear field was $u_{o} \cos(k x) \bm{\hat{j}}$.  The polymer motions and the flow fields that they create can all be traced back to the bare shear field and subsequent scattering events. However, if one does a viscosity measurement, one applies a force, obtains some shear rate, and measures the required force and the corresponding shear field.

We have considered fluid flows and power dissipation created by an imposed shear field $d u^{(0)}_{y}/ dx = u_{0} \sin(kx)$. The imposed field created further flows $\bm{u}^{(1)}$, $\bm{u}^{(2)}$,... via scattering from the polymers in solution. All flows are part of the total
flow ${\bf u}^{(T)}$ and its associated shear, $d u_{y}^{(T)}/ d x$.  Physically, only the total flow can be measured experimentally. The imposed shear is inaccessible to physical observation, so it must be replaced by the total shear. There is here a physical analogy with the replacement made in calculating the dielectric constant, in which the induced dipoles and the total electric field including material contributions must both be calculated, as discussed in this context by Peterson and Fixman\cite{peterson1963a}.

The shear field at $(X,Y,Z)$ due to scattering by a polymer a displacement $-\bm{R}_{1}$ away is
\begin{equation}
   \frac{d u^{(1)}_{y}(\bm{R_{1}})}{d x} = \frac{F_{o} S^{2}}{8 \pi \eta
R_{1}^{3}} \left( \frac{Y_{1}^{2}}{R_{1}^{2}} - \frac{5 X_{1}^{2}
Y_{1}^{2}}{R_{1}^{4}}\right) u_{0} k \sin(k(X-X_{1})).
   \label{eq:u1yshear}
\end{equation}
A similar but more complex form\cite{phillies2002b} gives the shear transmitted from double scattering through $\bm{R}_{1}$ and $\bm{R}_{2}$ to a location $(X,Y,Z)$. An ensemble average over all particle locations, practicable thanks to Mathematica for doing the final integrals, gives the parts of the total shear arising from single and double scattering.  For single scattering one has
\begin{equation}
    \left\langle \frac{d u^{(1)}_{y}}{d x} \right\rangle = \frac{16 \pi}{15} \frac{F_{0}
S^{2}}{8 \pi \eta} c u_{0} k \sin(kx),
    \label{eq:u1yshearaverage}
\end{equation}
$c$ being the number density of polymer molecules. For the double-scattered shear,
\begin{equation}
    \left\langle \frac{d u^{(2)}_{y}}{d x} \right\rangle = - \frac{16
\pi^{2}}{75}\frac{F_{0}^{2} S^{4}}{\eta^{2}} c^{2} u_{o} k \sin(kx)
    \label{eq:u2yshearaverage}
\end{equation}

Integrals of $r^{-3}$ over all space do not converge.  Because we chose a
spatially-oscillatory imposed shear field, in preparation for later taking a small-$\bm{k}$ limit, we obtained
convergent integrals for $\langle \frac{d u^{(1)}_{y}}{d x} \rangle$ and $\langle \frac{d u^{(2)}_{y}}{d x}
\rangle$, at least when $R_{1}$ and $R_{2}$ are integrated over ranges
$[a,b]$, the limits $ b \rightarrow \infty$ and $a \rightarrow 0 $ then
being taken.  From eqs.\ \ref{eq:shearfield}, \ref{eq:u1yshearaverage}, and \ref{eq:u2yshearaverage}, we obtain the total shear through second order concentration contributions, namely
\begin{equation}
     \left\langle \frac{d u^{(T)}_{y} (x)}{d x} \right\rangle
= - u_{o} k \sin(kx) \left[1-
\frac{2}{15} \frac{F_{0} S^{2}}{\eta} c + \frac{16
\pi^{2}}{75}\frac{F_{0}^{2} S^{4}}{\eta^{2}} c^{2} + \mathcal{O}(c^{3})\right]
     \label{eq:utotalyshearaverage}
\end{equation}

\subsection{Linear and Quadratic Terms; Huggins Coefficient}

We now calculate seriatim the contributions $P_{a,b}$ to the dissipated power, eq.\ \ref{eq:Ptotalfinal}. On dividing out the square of the total shear, eq.\ \ref{eq:utotalyshearaverage}, a pseudovirial series for the viscosity is obtained.

The lowest-order term in the series is $P_{0,0}$. Combining results above for ${\bf u}^{(0)}$ and  ${\bf \Omega}^{(1)}$, and taking needed derivatives and integrals
\begin{equation}
       P_{0,0}  = \left\langle N_{c} \frac{F_{o} S^{2}}{3} \left[ (u_{o} k \cos(k
  a_{1x}) \bm{\hat{j}} )^{2} - 2 (\frac{1}{2} u_{o} k \cos(k a_{1x}) \bm{\hat{k}})^{2}   \right]\right\rangle  .
    \label{eq:P00form}
\end{equation}
where  $\langle \cdots \rangle$ is the ensemble average over chain center-of-mass locations. Including contributions by all $N_{c}$ polymer molecules,
\begin{equation}
     P_{0,0} = \frac{N_{c} F_{o} S^{2}}{6} \frac{(u_{o} k)^{2}}{2}.
     \label{eq:P00value}
\end{equation}

The full power series for $P$ is infinite.  To evaluate, we must truncate or resum the series. Here we advance by truncation. There are two obvious choices of truncation variable.  Terms could be ordered by the number of scattering events that they include.  Terms could also be ordered
by how many different particles they include. The lowest order truncation gives the terms with zero scattering events and one polymer chain; these are the terms analyzed by Kirkwood and Riseman.  All higher-order truncations are of mixed order: either they include all terms with a given number of particles but omit some terms involving a given number of scattering events, or alternatively they include all terms involving a given number of scattering events, but omit some terms involving a given number of particles. Higher-order $P_{a,b}$ include terms that only involve a few chains but incorporate many scattering events, because flow fields can be scattered back and forth between two chains an arbitrary number of times.  However, the forms for ${\bf u}^{(1)}$, ${\bf u}^{(2)}$, and ${\bf u}^{(3)}$ show that each scattering event reduces interaction range by an additional factor of $1/r^{3}$.  By analogy with the equilibrium theory of electrolyte solutions, we retain the longest-range interactions, in which a ${\bf u}^{(n)}$ couples $n+1$ distinct chains.  These interactions, the ring diagrams, provide the leading terms of $P_{a,b}$.  They describe scattering by a series of scattering chains at ${\bf a}_{2}, \ldots, {\bf a}_{a}$, finally reaching chain 1 at
${\bf a}_{1}$. Particle 1 is simply a representative particle; we compute all the scattered flows acting on particle 1, and use them to compute the total power dissipated by chain 1.

The model here leads to a power series in $c[\eta]$, thus agreeing with the phenomenological observation that $[\eta]$ is a
good reducing variable for $c$. $P_{0,0}$, evaluated above, is proportional to $(c [\eta])^{1}$.  In $P_{a,b}$, in the factors $u_{i,j}^{(aT)} u_{i,j}^{(bT)}$ and $\bm{\Omega}^{(a+1T)} \cdot \bm{\Omega}^{(b+1T)}$, the chains in the $a$ and $b$ terms may be the
same or may entirely or partly different.  For each independent $\bm{a}_{j}$, the ensemble average yields
a factor $N_{c}$, which is the number of different polymer chains that $j$ could have represented.  Each chain
appearing in one of the $u_{i,j}$ corresponds to a scattering event, each event giving a factor $F_{o} S^{2}/\eta_{o}$.
The leading terms of the $P_{a,b}$  are thus $(N_{c} F_{o} S^{2}/\eta_{o})^{a+b} \sim (c [\eta])^{a+b}$, so the power series for $P$ itself is an expansion in powers of $c [\eta]$.

At long range, the hydrodynamic interaction tensors describing the $\bm{\Omega}^{(n)}$ and $\bm{u}^{(n)}$ depend on interparticle spacings as $r^{-
3}$.  Divergences were avoided because we took a sinusoidal imposed flow $\sim u_{o} \cos(kx)$ and then the long-wavelength $k \rightarrow 0$ limit. The hydrodynamic interaction tensors also diverge at short range. We supply an effective short range cut-off, because the physical $\bm{u}^{(n)}$ and $\bm{\Omega}^{(n)}$ are finite at small $r$.  Peterson and Fixman\cite{peterson1963a} proposed a related cutoff, namely that two overlapped chains were approximated as moving as a rigid dumbbell.

We now compute the $\mathcal{O}(c^{2})$ contributions to $\eta$, these being the $P_{a,b}$ with $a+b=1$ or $a=b=1$. Terms with two chains and more scattering events are allowed by the formalism but will be smaller because the interactions will be shorter-ranged.  For $a+b=1$
\begin{displaymath}
   P_{1,0} = P_{0,1} = \int d\bm{a}_{1} \, d\bm{a}_{2} \, \ldots \, d\bm{a}_{N_{c}} \,
   \exp(-\beta (W_{N_{c}} - A_{N_{c}}) \times
\end{displaymath}
\begin{equation}
\left[\sum_{p \neq q = 1}^{N_{c}}
   \frac{F_{o} S^{2}}{3}(
   - 2  \bm{\Omega}^{(1)}(\bm{a}_{p})
   \cdot  \bm{\Omega}^{(2)}(\bm{a}_{q}, \bm{a}_{p})
 + \sum_{i,j=1}^{3} \left[ u_{i,j}^{(0)}(\bm{a}_{p})
u_{i,j}^{(1)}(\bm{a}_{q},\bm{a}_{p}) \right] )
\right]
   \label{eq:P10start}
\end{equation}
Here $k_{B}$ is Boltzmann's
constant, $\beta = (k_{B}T)^{-1}$, $T$ is the absolute temperature, $W_{N_{c}}$ is the potential energy,
$A_{N_{c}}$ is the normalizing factor, and $p$ and $q$ label chains.
The average over internal chain coordinates gives an $S^{2}$.

All terms of the sum over $p$ and $q$ are identical save  for label.  The
ensemble average is
\begin{displaymath}
   P_{1,0} = \frac{F_{o} S^{2} N_{c}(N_{c}-1)}{3}
\int d{\bf a}_{1} \, d{\bf a}_{2} \,
   \left[ \left(\sum_{i,j=1}^{3} \left[u_{i,j}^{(0)}(\bm{a}_{1})  u_{i,j}^{(1)}(\bm{a}_{2},\bm{a}_{1})\right]
  \right.\right.
\end{displaymath}
\begin{equation}
   \left.\left.
   - 2  \bm{\Omega}^{(1)}(\bm{a}_{1})
   \cdot  \bm{\Omega}^{(2)}(\bm{a}_{2},\bm{a}_{1})\right) \int d\bm{a}_{3} \,
   \ldots d\bm{a}_{M} \, \exp(-\beta (W_{M} - A_{M})\right]
   \label{eq:P10two}
\end{equation}

The non-zero derivative of $\bm{u}^{(0)}$ is
\begin{equation}
   \bm{u}^{(0)}_{,x} = -u_{o} k \sin (k a_{1x}) \bm{\hat{j}}
   \label{eq:u0yx}
\end{equation}
$a_{1x}$ being the $x$ component of $\bm{a}_{1}$. The matching derivative of $\bm{u}^{(1)}$ is
\begin{equation}
     \bm{u}^{(1)}_{,x}=  u_{o} k \sin(k (a_{1x} -X_{1}))\frac{F_{o} S^{2}}{8 \pi
\eta_{o}}  \left[\left(\frac{Y_{1}}{R_{1}^4} - \frac{5
            X_{1}^{2}Y_{1}}{R_{1}^{6}} \right)\bm{\hat{R}}_{1}
+\frac{X_{1}Y_{1}}{R_{1}^{5}} \bm{\hat{i}}  \right].
    \label{eq:u1yx}
\end{equation}
$a_{1x}$ refers to the final particle in the scattering sequence;
$\bm{R}_{1} \equiv (X_{1}, Y_{1},Z_{1})$ points from the penultimate to the ultimate
particle of the scattering sequence.

The angular velocities appear in eqs.\ \ref{eq:omega1val} and \ref{eq:omega2fin}.  In these equations $\alpha$ is the shear at the first particle of the scattering series, namely $-u_{o} k \sin (k a_{1x}) {\bf \hat{j}}$ and $u_{o} k \sin(k (a_{1x} -X_{1})){\bf \hat{j}}$, respectively.
The identity $\sin (k a_{1x})\sin(k (a_{1x} -X_{1})) = (-\cos (2 k a_{1x} - k X_{1}) + \cos(k X_{1}))/2$ is then applied.  The ensemble average only depends on ${\bf a}_{1}$ through $\cos (2 k a_{1x} - k X_{1})$, which vanishes on averaging over ${\bf a}_{1}$.

Recalling the standard form
\begin{equation}
    \frac{g^{(2)}(\bm{r})}{V^{2}} =
    \frac{\int d\bm{a}_{3} \, \ldots \, d\bm{a}_{M} \, \exp(-\beta W(\bm{r},\bm{a}_{3}, \ldots \bm{a}_{M}))}{\int d\bm{a}_{1}\, \ldots \, d\bm{a}_{M_{c}} \exp(-\beta W(\bm{r},\bm{a}_{3}, \ldots \bm{a}_{M_{c}})) }
   \label{eq:g2def}
\end{equation}
for the radial distribution function, here with $\bm{r} = \bm{a}_{2}-\bm{a}_{1}$,
\begin{equation}
     P_{1,0} = - \left( \frac{u_{o}^{2} k^{2}}{2} \right)
\left(\frac{N_{c}(N_{c}-1)(F_{o} S^{2})^{2}}{24 \pi \eta_{o} V}
 \right)
     \int d\bm{R} \,  g^{(2)}(R)  \frac{\cos(kX) }{R^{3}} \left[\frac{X^{2}
+Y^{2}}{R^{2}} - \frac{10 X^{2} Y^{2}}{R^{4}}    \right].
    \label{eq:p10three}
\end{equation}
In the radial integral, the lower cut-off is not required for $P_{1,0}$. Without the $\cos(kx)$, the $\int d\bm{R}$ diverges at large $R$; the angular integral vanishes; the $\int d\bm{R}$ is improper.  The proper long-wavelength limit results from taking $\int d\bm{R}$ and then taking $k \rightarrow 0$. If the shear were linear and not oscillatory in space, $P_{1,0}$ would be undefined, as observed a half-century ago by Saito\cite{saito1952a}.

Choosing $\bm{k}$ to be parallel to the X axis, a useful identity is\cite{jackson1962a}
\begin{equation}
    \cos({\bf k} \cdot {\bf R}) = 4 \pi \sum_{l=0}^{\infty} \frac{i^{l}+(-
i)^{l}}{2}  j_{l}(kr) (4 \pi (2 l + 1))^{1/2} Y_{l0}(\theta).
\label{eq:cosidentity}
\end{equation}
Here $j_{l}$ is a spherical bessel function, and $\theta$ is the angle between ${\bf k}$ and ${\bf R}$.

On invoking spherical coordinates, recourse to Mathematica gives
\begin{equation}
      P_{1,0} = -\eta_{o}
\frac{N_{c}^{2}-N_{c}}{V} \frac{48 \pi}{5} \left(\frac{F_{o} S^{2}}{6
          \eta_{o}}\right)^{2}
      \left(\frac{u_{o}^{2} k^{2}}{2}\right).
      \label{eq:p01almost}
\end{equation}

How can this this term negative?  Mathematically, in the intrinsically positive form $(a-b)^{2}$ the term $- 2 a b$ can be
negative; in the calculation here $P_{1,0}$ can play the role of a $-2ab$. Physically, eq.\ \ref{eq:p01almost} is negative because ${\bf u}^{(1)}$ causes chain 1 to rotate, thereby reducing the
velocity difference between chain 1's beads' velocities and ${\bf u}^{(0)}$, so dissipation is reduced by this term.

We now turn to $P_{1,1}$.  Writing $\bm{\Omega}^{(2T)}$ and $\bm{u}^{(1T)}$ as sums over all the other particles in the system,
\begin{multline}
   P_{1,1} = \left\langle \frac{N_{c} F_{o} S^{2}}{3}
          \left( - 2 \sum_{p,q=2}^{N_{c}}\bm{\Omega}^{(2)}(\bm{a}_{p}, \bm{a}_{1})
          \cdot \bm{\Omega}^{(2)}(\bm{a}_{q}, \bm{a}_{1})
   + \right.\right. \\ \left. \left.\sum_{p,q=2}^{N_{c}} \sum_{i,j=1}^{3}
    \left[u^{(1)}_{j,i}(\bm{a}_{p}, \bm{a}_{1})
    u^{(1)}_{j,i}(\bm{a}_{q}, \bm{a}_{1})\right] \right)
\right\rangle
   \label{eq:P11A}
\end{multline}

Only the self ($p=q$) terms of eq. \ref{eq:P11A} are significant here; the distinct ($p\neq q$) terms give an effect cubic in
concentration.  To $\mathcal{O}(c^{2})$
\begin{displaymath}
    P_{1,1} = \frac{N_{c}(N_{c}-1) F_{o} S^{2}}{3 V}
     \int d\bm{a}_{1} \, d\bm{a}_{2} \,  g^{(2)}(\bm{a}_{1}, \bm{a}_{2})
     \left( \bm{\Omega}^{(2)}(\bm{a}_{2}, \bm{a}_{1})
          \cdot \bm{\Omega}^{(2)}(\bm{a}_{2}, \bm{a}_{1})\right.
\end{displaymath}
\begin{equation}
  \left. +  \sum_{i,j=1}^{3}
    \left[\bm{u}^{(1)}_{j,i}(\bm{a}_{2}, \bm{a}_{1})
    \bm{u}^{(1)}_{j,i}(\bm{a}_{2}, \bm{a}_{1}) \right] \right) +\ldots
     \label{eq:P11B}
\end{equation}

The convergence here at large $R$ is sufficiently strong that
the integrals and the $k \rightarrow 0$ limit can be exchanged, giving
\begin{multline}
    P_{1,1,s} = \frac{c^{2} V  F_{o} S^{2} }{6 }
\alpha^{2} \left(\frac{F_{o}
    S^{2}}{8 \pi \eta_{o}}\right)^{2}
 \times \\ \int_{V} d\bm{R} \, \frac{1}{R^6} \left[
\frac{6 X^{2}Y^{2} -X^{4}-Y^{4}+Z^{4}}{R^{4}} +\frac{2 X^{2}+ 2 Y^{2} -
Z^{2}}{R^{2}}   \right] g^{(2)}(R)
\label{eq:P11s1} \end{multline}

$P_{1,1,s}$ requires a short-range cutoff $a$ for convergence of $\int_{V} d\bm{R}$.  Such a cutoff is physically appropriate.
\ref{eq:u1value} and \ref{eq:omega2fin} are long-range parts of series
expansions.  Short-range terms that prevent divergence are here represented by the cut-off
distance.  Inserting such a cutoff into $P_{1,0}$ has little effect.

Integrating, one obtains
\begin{equation}
     P_{1,1,s} = \left(\frac{F_{o} S^{2}}{8 \pi \eta}\right)^{2} \frac{4 \pi
F_{o} S^{2}}{15 a^{3}} c^{2} \frac{(u_{o} k)^{2}}{2}
    \label{eq:P11sb}
\end{equation}

Combining eqs \ref{eq:utotalyshearaverage}, \ref{eq:P00value},
\ref{eq:p01almost}, and \ref{eq:P11sb},
 \begin{displaymath}
  \eta \left(\left\langle \frac{d u^{(T)}_{y}}{d x} \right\rangle\right)^{2}
= \eta_{o} \left[1 + \frac{F_{o} S^{2}}{6 \eta_{o}} c +\left(- \frac{4 \pi
F_{o}^{2} S^{4}}{15 \eta_{o}^{2}} + \frac{F_{o}^{3} S^{6}}{ 240 \pi
\eta_{o}^{3} a^{3}}    \right)c^{2}
     \right]\times
     \end{displaymath}
\begin{equation}
\left[1-\frac{2}{15} \frac{F_{0} S^{2}}{\eta} c + \frac{16
\pi^{2}}{75}\frac{F_{0}^{2} S^{4}}{\eta^{2}} c^{2}\right]^{-2} \left( \left\langle \frac{d u^{(T)}_{y} (x)}{d x}
\right\rangle\right)^{2}
    \label{eq:P2final}
\end{equation}

In terms of the series
$k_{H}$ of
\begin{equation}
    \eta/\eta_{o} = 1 + [\eta] c + k_{H} [\eta]^{2} c^{2},
    \label{eq:hugginsdef}
\end{equation}
the Huggins coefficient being $ k_{H}$,
\begin{equation}
   [\eta] = \frac{13 F_{o} S^{2}}{30 \eta_{o}}
     \label{eq:etaintrinsicfinal}
\end{equation}
and
\begin{equation}
    k_{H} = \frac{88 -240 \pi - 384 \pi^{2}}{169}+ \frac{225 [\eta]}{4394 \pi
a^{3}}
   \label{eq:kHfinal}
\end{equation}

The cutoff radius $a$  is a crude approximation.  A sound treatment of
hydrodynamics of interpenetrated random coils is needed.  One reasonably expects
$a$ to be moderately smaller than $S^{2}$.

\section{From Pseudovirial Series to Higher Concentrations\label{s:powertostretched}}

The above discussion shows how power series expansions may be used to determine the concentration dependence of $D_{s}$ and $\eta$.  The series approaches face the challenge that at elevated concentrations more and more terms are needed in order to obtain accurate predictions, while at the same time the scale of the calculations required to obtain additional forms becomes larger and larger.  To overcome this difficulty, alternative approaches to computing $D_{s}(c)$ and $\eta(c)$ at large $c$ have been employed.  We here discuss two, namely \emph{self-similarity} and \emph{the Altenberger-Dahler Positive-Function Renormalization Group}.  Self-similarity advances by physical arguments about chain-chain interactions. The Positive-Function Renormalization Group approach proposes to advance by noting that $D_{s}(c)$ and $\eta(c)$ depend both on concentration $c$ and on a coupling parameter $R$, and their values at large $c$ and some $R$ are equal to their values at a smaller $c$ and some other value of $R$, the values of $D_{s}(c)$ and $\eta(c)$ being easier to compute at the smaller $c$ and some other  $R$. The Positive Function Renormalization Group advances by calculating the needed 'other' $R$.

\subsection{Self-Similarity Approach\label{ss:selfsimilarity}}

This Subsection considers the original\cite{phillies1987a} self-similarity derivation of the universal scaling equation for polymer self diffusion. The derivation has several basic assumptions. First, at all concentrations the dominant polymer-polymer interactions are taken to be hydrodynamic, with excluded-volume interactions providing at best secondary corrections.  The interchain hydrodynamic interactions are approximated as being the same, except for numerical coefficients, as the hydrodynamic interactions between hard spheres. Second, the effects of sequential infinitesimal concentration increments on $D_{s}$ are said to be self-similar, whencefrom the name of the derivation. Third, polymer chains in good solvents are taken to contract as polymer concentration is increased.

The form of the hydrodynamic interactions between polymer chains has been calculated above. A velocity $\bm{V}^{(1)}$ of the first polymer chain in a sequence creates a flow field $\bm{u}^{(1)}$ in the solvent.  The flow field acts on chain 2.  The translation and rotation of chain 2 create a further flow  field $\bm{u}^{(2)}$ and so forth.  At every step after chain 2, the final flow field $\bm{u}^{(f)}$ can act back on chain 1, inducing in chain 1 an additional translational velocity $\bm{\delta V} = \bm{u}^{(f)}$, with $\bm{u}^{(f)}$ as evaluated at chain 1. Take $f_{\mathrm{ch}} = 6 \pi \eta R_{h}$ to be the drag coefficient of the first chain. The force the first chain would apply to the solvent, if it moved relative to a quiescent solution, is $f^{o}_{\mathrm{ch}} \bm{V}^{(1)}$. Multiplying through the entire calculation by $f^{o}_{\mathrm{ch}}$, the force the final flow field would exert back on chain 1 in response to chain 1's motions is $f^{o}_{\mathrm{ch}} \bm{V}^{(f)}$. In Brownian motion, no forces external to the polymer-solvent system act on the polymer chains.  The chains move because hydrodynamic fluctuations create flows in the solvent, the chains being moved by the fluctuations, but the fluctuation-dissipation theorem requires that the correlations in the displacements arising from the hydrodynamic fluctuations must be the same as the correlations in the displacements that would appear if chain 1 were subject to an external force that moved chain 1 in the same way with respect to the solvent.

The self-diffusion coefficient of a polymer is determined by its drag coefficient $f_{\mathrm ch}$ via the Einstein equation $D_{s} = k_{B} T/f_{\mathrm ch}$.  $f_{\mathrm ch}$ differs from $f^{o}_{\mathrm{ch}}$ in that it includes contributions to the hydrodynamic drag on a chain due to the chain's interactions with other chains. To determine the concentration dependence of $D_{s}$ it is sufficient to determine the concentration dependence of $f_{\mathrm ch}$.

The ability of chain 2 to affect the drag coefficient of chain 1 is determined by the strength of chain 2's hydrodynamic interactions with the solvent, here approximated by the drag coefficient $f_{\mathrm{ch}}$ and by the coupling coefficient $\alpha$ describing the strength of interchain interactions.   We advance by considering the effect of successive infinitesimal concentration increments on $f_{\mathrm{ch}}$.  The first increment $\delta c$ gives us
\begin{equation}
       f_{\mathrm{ch}}(\delta c) =   f_{\mathrm{ch}}(0) +  \alpha  f_{\mathrm{ch}}(0) \delta c =f_{\mathrm{ch}}(0) (1+\alpha \delta c) .
       \label{eq:eq:fdeltac}
\end{equation}
Here we have applied the approximation that the change in $f_{\mathrm{ch}}$ due to the first concentration increment is proportional to $f_{\mathrm{ch}}$ of the chains in the increment. We now apply a second infinitesimal concentration increment $\delta c$.  The \emph{self-similarity} step is to assert that the chains of the second concentration increment affect not only the chain of interest but also equally the chains of the first concentration increment, so that
\begin{equation}
       f_{\mathrm{ch}}(2\delta c) =   f_{\mathrm{ch}}(0) +  \alpha  f_{\mathrm{ch}}(0) \delta c +\alpha f_{\mathrm{ch}}(\delta c)  \delta c.
       \label{eq:eq:f2deltac}
\end{equation}
On the right-hand-side of the equation, the first two terms are $f_{\mathrm{ch}}(\delta c)$. The third term is the effect of the second concentration increment $\delta c$, written in terms of the drag coefficient $f_{\mathrm{ch}}(\delta c)$ of the chain at concentration $\delta c$. Moving the first two terms from the rhs to the lhs of the equation and dividing by $f_{\mathrm{ch}} \delta c$, one finds
\begin{equation}
       \frac{f_{\mathrm{ch}}(2\delta c) - f_{\mathrm{ch}}(\delta c)}{f_{\mathrm{ch}}(\delta c) \delta c}  =    \alpha.
       \label{eq:f3deltac}
\end{equation}
In the limit $\delta c \rightarrow 0$, the left side is recognized as the logarithmic derivative of $f_{\mathrm{ch}}(c)$, so integration gives
\begin{equation}
      f_{\mathrm{ch}}(c) = f_{\mathrm{ch}}(0) \exp\left[\int_{0}^{c} dc \, \alpha(c) \right].
      \label{eq:f4deltac}
\end{equation}
and correspondingly
\begin{equation}
      D_{s}(c) = D_{s}(0) \exp\left[- \int_{0}^{c} dc \, \alpha(c)\right].
      \label{eq:f5deltac}
\end{equation}

At the time of the original derivation of the Hydrodynamic Scaling Model\cite{phillies1987a} on the basis of self-similarity, the chain-chain hydrodynamic interaction tensors seen above had not yet been obtained.  It was instead proposed that eqs.\ \ref{eq:23} and  \ref{eq:bilspheres}, which describe the mobility $\mu_{ii}$ for pairs of hard spheres, are dimensionally correct for chains even though they do not supply precise numerical coefficients, and are therefore good as a first approximation to the chain-chain hydrodynamic interaction tensors. The conclusion was that
\begin{equation}
    \alpha(c) =  Q R_{h1} R_{g2}^3.
    \label{eq:alphaestimate}
\end{equation}
Here $Q$ includes numerical coefficients and the average of $\bm{\hat{r}}_{ij}\bm{\hat{r}}_{ij}/ {r}_{ij}^4$ over the chain-chain radial distribution function, while the sum over spheres in eq.\ \ref{eq:23} becomes the $\int dc$ of eq.\ \ref{eq:f5deltac}.

The final approximation was to estimate the concentration dependence of the chain radii from the results of Daoud, et al.\cite{daoud1975a}. In the original calculation\cite{phillies1987a}, the radii were taken to scale as
\begin{equation}
    R^{2} \sim M c^{-x},
    \label{eq:chaincontract}
\end{equation}
with $x = 1/4$. The original prediction referred only to long chains with $c$ greater than some overlap concentration $c^{*}$. For long chains at lower concentrations, the degree of chain contraction was predicted to be less.  For short chains, the Daoud, et al.\ model predicts $x \approx 0$. Combining the above three equations, one finds the prediction
\begin{equation}
     D_{s}(c) = D_{0} \exp( - Q'M c^{1-2x}).
     \label{eq:dscderived}
\end{equation}
$Q'$ includes $Q$ and other numerical coefficients arising from the integration. Comparing with the universal scaling equation eq.\ \ref{eq:stretchedexpDs}, if one identifies $1-2x = \nu$, one predicts:

a) For large polymer chains, $\nu = 0.5$, except perhaps at very low concentrations.

b) For short polymer chains at all concentrations, $\nu = 1.0$.

c) For the probe diffusion coefficient $D_{p}$, the radius $R_{h1}$ of the probe does not depend on concentration, so $\nu = 1 - 3x/2 \approx 5/8$.

Finally, identifying $\alpha$ of eq.\ \ref{eq:stretchedexpDs} with $Q' M$, one predicts $\alpha \sim M^{\delta}$ for $\delta = 1.0$.  As discussed below, all of the above predictions have been confirmed experimentally.

\subsection{Positive Function Renormalization Group\label{ss:pfrg}}

This subsection develops the mathematical structure of the Altenberger-Dahler Positive-Function Renormalization Group (PFRG) approach\cite{altenberger1997a,altenberger1997b,altenberger2001a,altenberger2001b,altenberger2002a}. In Section 7, the approach is applied directly to treat the self-diffusion coefficient and the low-shear viscosity.  In Section 8, a fixed-point structure for the viscosity is inferred and then applied via an \emph{ansatz} to infer the frequency dependences of the loss and storage moduli.

Altenberger and Dahler note that renormalization group methods have been invoked in several branches of physics to deal with superficially different mathematical challenges.  Renormalization group methods were inserted into high-energy theory to cope with difficulties arising from cutoff wavelengths and the presence of infinities in series expansions. Renormalization group methods appear in statistical mechanics in applications of self-similarity methods, such as block renormalization, where the methods are used to eliminate insignificant fine detail from descriptions of critical fluctuations. Of more significance here, renormalization group methods can be used to extend the range of validity of lower-order power series expansions. The effort here pursues the last of these uses.  We are not facing divergences or systems with a multiplicity of unimportant short-range length scales.  We have on hand a low-order power-series expansion that would be inordinately tedious to extend to very high order.

Because the Altenberger-Dahler PFRG method has not been used extensively, we first sketch the physical rationales that lead to the method and then consider the mathematical forms.  The starting point is that many physical properties of a solution can be written as a pseudovirial expansion, e.g.,
\begin{equation}
      A(c) = a_{o} + a_{1} c' + a_{2} c'^{2}.
      \label{eq:pseudovirial}
\end{equation}
Here $A$ is the physical property, $c'$ is the solute concentration in physical units, and the $a_{i}$ are the pseudovirial coefficients. The $a_{i}$ are typically obtained from cluster expansions. It is not claimed -- that would be incorrect -- that all concentration-dependent physical properties have pseudovirial expansions.  $A$ is actually a function of two parameters, namely the concentration $c'$ and a coupling parameter $R$, so one may write $A = A(c',R)$.  The coupling parameter $R$ determines the values of the $a_{i}$. Cases in which there are multiple coupling parameters are included by treating $R$ as a vector. The $c \rightarrow 0$ limit of $A$ is simply $a_{o}$.  The limit of noninteracting solute molecules can also be obtained as $R \rightarrow 0$, in which case once again $A = a_{o}$. Introduction of a reference concentration $c_{r}$ and dimensionless concentration units $c = c'/c_{r}$ leads to
\begin{equation}
      A(c) = a_{o} + \left[a_{1} c_{r}\right] c + \left[a_{2}/c_{r}^{2}\right]  c^{2}.
      \label{eq:pseudovirial2}
\end{equation}
At elevated concentrations, the above pseudovirial series become inaccurate. The familiar virial approach is to improve the accuracy of the series by adding additional terms $a_{3}$, $a_{4}$, etc. In the PFRG approach, the series of eq.\ \ref{eq:pseudovirial2} is taken to be exact, but the bare coupling parameter $R$ is replaced with a dressed, concentration-dependent coupling parameter $\bar{R}(R,c)$. The values of  $\bar{R}$ are chosen so that the $a_{i}$ calculated using $\bar{R}$, when inserted into eq.\ \ref{eq:pseudovirial2}, give the correct values for $A$ even at large concentrations.

The Altenberger-Dahler calculation has two major parts. First, constraints on the behavior of $A$ are used to determine functional requirements for the dressed coupling parameter $\bar{R}(R,c)$. Second, at low concentrations $\bar{R} = R$ to high precision. A group of Lie differential equations and infinitesimal generators for the dependences of $A$ and $\bar{R}$ on $c$ are then determined by the group properties of $\bar{R}$. The polymer calculation has three further parts. First, the multichain Kirkwood-Riseman model described above is used to obtain the actual $a_{i}$, including the dependences of the $a_{i}$ on $R$.  These dependences determine Lie group generators and equations needed to compute $\bar{R}$ and $A$  for $D_{s}$ or $\eta$. For an object of fixed $R$, numerical integration determines $A(c)$ (here, either $D_{s}(c)$ or $\eta(c)$) at the level of precision of the input calculations.  Finally, applying the results of Daoud, et al.\cite{daoud1975a} and King, et al.\cite{king1985a}, on chain contraction at elevated polymer concentration, one obtains an approximant to the universal scaling equation for polymer self-diffusion. The approximant is valid to a specific order in the dressed coupling parameter $\bar{R}$.

To open the renormalization group calculation, the constraint on $A$ is that it is positive definite, never zero or negative, so $A$ may always be written in the form $A = \exp(B)$.  It is convenient to transform $A(c',R)$ to dimensionless units by normalizing with respect to $A$ at some non-zero concentration $c_{o}$, namely
\begin{equation}
     \bar{A}(c',R) = A(c',R)/A(c_{o},R).
     \label{eq:AD1}
\end{equation}
Here $\bar{A}(c',R)$ is the normalized and hence dimensionless transformation of $A(c',R)$.  $c_{o}$ and $c_{r}$ are independent, but for simplicity we will choose $c_{o} = c_{r}$ in the following.  Because $\bar{A}$ is also positive definite, it may be written as
\begin{equation}
       \bar{A}(c',R) = \exp(\int_{c_{o}}^{c'} ds \, \mathcal{L}(s,R)
       \label{eq:AD2}
\end{equation}
where
\begin{equation}
       \mathcal{L}(s,R) = \frac{\partial \ln(\bar{A}(s,R)}{\partial s}.
       \label{eq:AD3}
\end{equation}
Equations \ref{eq:AD2} and \ref{eq:AD3} are an identity.  They enforce, and valid because of, the requirement that $ \bar{A}(c',R)$ is positive definite.

In the integral of eq.\ \ref{eq:AD2} we introduce an intermediate concentration $z'$, and divide the one integral into two, giving
\begin{equation}
 \bar{A}(c',R) =  \left( \exp(\int_{c_{o}}^{z'} ds \, \mathcal{L}(s,R)\right)\left(\exp(\int_{z'}^{c'} ds\, \mathcal{L}(s,R)\right)
       \label{eq:AD4}
\end{equation}
We now go to dimensionless concentration units, choosing $c_{o}$ as a reference concentration with  $c = c'/c_{o}$, and make a change of variables $s \rightarrow yz$, finding
\begin{equation}
     \bar{A}(c,R) =   \bar{A}(z,R) \left[ \exp\left( \int_{1}^{c/z} dy \, \mathcal{L}(yz, R)\right)\right]^{z}
       \label{eq:AD5}
\end{equation}
Because $z'$ (in physical units) is intermediate between $c_{o}$ and $c'$, $z$ (dimensionless units) must be $\geq 1$. Here $\exp(az) = (\exp(a))^{z}$ has been applied. The above equation supports the introduction of a dressed coupling parameter $\bar{R}$. The dressed coupling parameter is chosen so that at each $z$ and $R$,
\begin{equation}
    \mathcal{L}(yz, R)  =   \mathcal{L}(y, \bar{R}(z,R)),
    \label{eq:AD6}
\end{equation}
so that $\mathcal{L}$ at an elevated concentration $yz$ can be replaced by $\mathcal{L}$ at a lower concentration $y$ by replacing $R$ with the appropriate $\bar{R}$.  There is an implicit assumption that such an $\bar{R}$ exists. A representative contrary outcome would be that $\mathcal{L}(y,R)$ saturates with changes in $R$, so that there is no value of $\bar{R}$ that satisfies eq.\ \ref{eq:AD6}.  This issue does not arise for the calculation here, but should be kept in mind as a general possibility.  Replacing $R$ with $\bar{R}$ has an analogy in the direct self-similarity calculation of $D_{s}$, namely in those calculations each chain's bare drag coefficient $f_{o}$ was replaced with a dressed drag coefficient $f$ of the chain at the concentration of interest.

On applying eq.\ \ref{eq:AD6} to eq.\ \ref{eq:AD5}, and applying $\exp(\ln(A)) = A$, one has
\begin{equation}
      \frac{\bar{A}(c,R)}{\bar{A}(z,R)}  = \left[ \bar{A}\left(\frac{c}{z},\bar{R}(z,R)\right) \right]^{z}
\label{eq:AD7}
\end{equation}
In order for this equation to be correct, we must be working in dimensionless units, so that the lower bound of the integral in eq.\ \ref{eq:AD5} is unity. Equation \ref{eq:AD7} represents a numerical renormalization of $\bar{A}(c,R)$, in that $\bar{A}(c,R)/\bar{A}(z,R) = 1$ if $c=z$. Eq.\ \ref{eq:AD7} also represents a group property, namely it shows how the effect on $\bar{A}$ of a change in the concentration $c$ can be replaced with a different change in the concentration $c$ together with a corresponding dressed coupling parameter $\bar{R}$.

Multiplying eq.\ \ref{eq:AD7} by ${\bar{A}(z,R)}$, and adopting a new concentration variable via $c \rightarrow cz$, one has
\begin{equation}
      \bar{A}(cz,R) = \bar{A}(z,R)\left[\bar{A}(c,\bar{R}(z,R))   \right]
\label{eq:AD8}
\end{equation}
The form of the left-hand-side of the equation forces the right-hand-side of the equation to be symmetric under the interchange of variables $c$ and $z$. In consequence, severe constraints are placed on the possible functional forms for $\bar{R}$. In particular, as shown by Altenberger and Dahler\cite{altenberger1997b}, their Appendix 1,  eq.\ \ref{eq:AD8} forces the requirement
\begin{equation}
      \bar{R}(c,R) = \bar{R}(c/z, \bar{R}(z,R))
\label{eq:AD9}
\end{equation}
We have now finished the first part of the derivation.  We made two assumptions, the first being that $\bar{A}$ as a variable is never $\leq 0$, and the second being that there is an effective coupling parameter $\bar{R}$ that is consistent with eq.\ \ref{eq:AD7}.

In the second part of the derivation, we show that equations \ref{eq:AD7} and \ref{eq:AD9}  lead to differential equations for $\bar{R}$.  Note that at the reference concentration $c_{o}$ one has $\bar{R}(c_{o},R) = \bar{R}(1,R) = R$, which gives the boundary condition for integrating the differential equations we are about to obtain.  The differential equations are obtained from equations \ref{eq:AD7} and \ref{eq:AD9}, beginning by taking derivatives with respect to $c$. From the derivative of eq.\ \ref{eq:AD7} one sets $c=z$ and notes $\bar{A}(1,\bar{R}(z,R))=1$ (follows directly from eq.\ \ref{eq:AD7}), leading for $u = c/z$ to
\begin{equation}
     \frac{\partial \ln(\bar{A}(z,R))}{\partial z} = \left. \frac{\partial \bar{A}(u, \bar{R}(z,R))}{\partial u} \right|_{u=1} = \gamma(\bar{R}(z,R)).
     \label{eq:AD10}
\end{equation}
$\gamma(\bar{R}(z,R))$ is a differential generator. At $z=1$ the generator becomes
\begin{equation}
     \left. \frac{\partial \bar{A}(z, R)}{\partial z} \right|_{z=1} = \gamma(R)
     \label{eq:AD11}
\end{equation}

      From the derivative of eq.\ \ref{eq:AD9} with respect to $c$, on setting $c=z$ one obtains
\begin{equation}
      \frac{\partial \bar{R}(z,R)}{\partial \ln(z)}  = \left. \frac{\partial \bar{R}(u, \bar{R}(z,R))}{\partial u} \right|_{u=1} = \beta(
\bar{R}(z,R))
   \label{eq:AD12}
\end{equation}
as the definition of $\beta$. Alternatively, the definition in eq.\ \ref{eq:AD12} can be obtained from eq.\ \ref{eq:AD10} by taking a derivative of $\gamma$ with respect to $z$, leading to
\begin{equation}
   \label{eq:AD13}
    \frac{\partial \bar{R}(z,R)}{\partial z} = \frac{(\bar{A}''(z,R)) /\bar{A}(z,R) - (\bar{A}'(z,R)/\bar{A}(z,R))^{2}}{\partial \gamma(\bar{R}(z,R)/\partial \bar{R}(z,R))}.
\end{equation}
Here $\bar{A}'(z,R)= \partial \bar{A}(z,R)/\partial z$  and  $\bar{A}''(z,R) = \partial^{2} \bar{A}(z,R)/\partial z^{2}$. On setting $z=1$, a further result for $\beta$ is obtained from the above two equations, namely
\begin{equation}
     \beta(R) = \frac{\bar{A}''(1,R) /\bar{A}(1,R) - (\bar{A}'(1,R)/\bar{A}(1,R))^{2}}{(\partial \gamma(R)/\partial \bar{R})}.
   \label{eq:AD14}
\end{equation}
This final equation gives $\beta$ as a function of $R$ rather than $\bar{R}$, at least at the initial concentration. $\beta(\bar{R})$ and $\gamma(\bar{R})$ provide the infinitesimal generators for Lie equations for the concentration dependences of $\bar{A}$ and $\bar{R}$.

A variety of methods for integrating these equations are available.  The calculation requires as inputs $\overline{A}$ and its derivatives evaluated at $z=1$.  Altenberger and Dahler\cite{altenberger1997a,altenberger1997b} proceed by approximating $\overline{A}$ with its low-order series expansion.  For reasonable choices of the initial concentration (in physical units) $c_{o}$, this approximation is not very demanding.  Indeed, Altenberger and Dahler use a cubic approximation for $P$ of a hard sphere gas, choose an initial volume fraction $c_{o} =0.16$, and obtain the $P$ predicted by an eight-term virial fraction for $c$ up to 0.62.

\section{From Renormalization Group to Universal Scaling\label{s:rgus}}

In this Section we advance from the hydrodynamic calculations of sections \ref{s:kreds} and \ref{s:kreta} and the Positive Function Renormalization Group approach developed in Subsection \ref{ss:pfrg} to extrapolate the concentration dependence of $D_{s}$ and $\eta$. We invoke the Altenberger-Dahler Positive Function Renormalization Group and equation \ref{eq:Dsseries} for the concentration and chain radius dependences of $D_{s}$ to extrapolate $D_{s}(c)$ to larger concentrations. Equation \ref{eq:Dsseries} includes both $R_{h}$ and $R_{g}$; these are approximated as being a single radius $R'$. We identify the concentration variable of the renormalization group calculation as the physical concentration $c$, and choose $R = R'/R_{o}$ as the dimensionless coupling parameter.  $R_{o}$ is identified as $R'$ at $c=1$.

Eq.\ \ref{eq:Dsseries} is now
\begin{equation}
       D_{s}(c) = D_{o} ( 1 + \overline{a} R^{4} c + \overline{b} R^{7} c^{2})
\label{eq:Dscnormalized}
\end{equation}
    All dependence on $R$ is now explicit. The renormalized pseudovirial coefficients are
\begin{equation}
      \overline{a} = - \frac{9}{16} \frac{4 \pi}{3 a_{o}} R_{o}^{4} c_{r}
\label{eq:abarref}
\end{equation}
and
\begin{equation}
      \overline{b} = - \frac{9.3 \cdot 10^{-4}}{a_{o}} (\frac{4 \pi}{3})^{2} R_{o}^{7} c_{r}^{2}.
\label{eq:bbarref}
\end{equation}
At concentration $c_{r}$, $c=R=1$, $c$ and $R$ both being dimensionless.  These equations differ from expressions employed by Altenberger and Dahler\cite{altenberger1997a,altenberger1997b} in one significant way.  In the earlier calculations, $c$ and $R$ always appeared as the product $cR$, so that the $n^{\rm th}$ term of their virial expansion depended on $R$ as $R^{n}$. Here the $c$ and $R$ dependences are distinct.

$D_{s}$ is transformed to $\overline{D_{s}}$ by dividing by $D_{s}(1) = D_{o} (1 + \overline{a}R^{4}+ \overline{b}R^{7})$. $\overline{A}$ of the prior section is identified as $\overline{D_{s}}$.  All dependence of $\overline{D_{s}}$ on $R$ can be moved to the numerator via the expansion $(1-x)^{-1} \rightarrow 1 + x + x^{2}+\ldots$. So long as one truncates at $R^{7}$, which is the highest-order limit of the original hydrodynamic series, one finds
\begin{equation}
        \overline{D_{s}}(c) = 1 + \overline{a} R^{4} (c-1) + \overline{b} R^{7} (c^{2}-1).
\label{eq:dbarform}
\end{equation}

Identifying the concentration variable $z$ of the prior section with $c$ here, $\gamma(R)$ arises from the logarithmic derivative of $\overline{D_{s}}(c)$ as
\begin{equation}
  \gamma(R) = \overline{a} R^{4} + 2 \overline{b} R^{7}.
   \label{eq:gammaDs}
\end{equation}
The other generator, $\beta(R)$, is determined by $\overline{D_{s}}(c)$, its first and second derivatives evaluated at $c=1$, and $\partial \gamma R/\partial R$ to be
\begin{equation}
      \beta(R) = \frac{2 \overline{b} R^{7} - (\overline{a} R^{4} + 2 \overline{b} R^{7})^{2}}{4\overline{a} R^{3} + 14 \overline{b} R^{6}} \approx \left(\frac{\overline{b} R^{7}}{4\overline{a} R^{3}}\right).
\label{eq:betaDs}
\end{equation}
The final approximation follows from $\overline{a} \gg \overline{b}$  after expanding the denominator in powers of $\overline{b}/\overline{a}$, applying a geometric series expansion, and only retaining terms of order $\mathcal{O}(R^{6})$ and lower. Altenberger and Dahler now offer the approximation that the dependence of $\partial \overline{R}(c, R)/\partial c$ on $\overline{R}$ for $c \neq 1$ is given by the dependence of $\beta(R)$ on $R$ at $c=1$ by replacing $R$ in the latter with $\overline{R}$. With this approximation
\begin{equation}
      \frac{\partial \overline{R}}{\partial \ln c} = \frac{\overline{b} \, \overline{R}^{6}}{2 \overline{a}}
\label{eq:betaprecise}
\end{equation}
Noting $\overline{R}(c)\mid_{c=1} = 1$, an integral with respect to $\ln(c)$ yields
\begin{equation}
      \overline{R}(c) =\left[1 - \frac{3}{2} \frac{\overline{b}}{\overline{a}} \ln(c)    \right]^{-1/3}
    \label{eq:Rbarprecise}
\end{equation}
$\overline{a}$ and $\overline{b}$ have opposite signs, with  $\overline{b}/\overline{a} \ll 1$, and $c \geq 1$, so $\overline{R}(c)$ is well-behaved.  The prediction for $D_{s}(c)$ is
\begin{equation}
        D_{s}(c) = D_{s}(1) \exp\left(\int_{1}^{c} dx \,  (\overline{a} \overline{R}^{4}(x) + 2 \overline{b} \overline{R}^{7}(x))     \right)
\label{eq:Dspredict}
\end{equation}
with the functional behavior of $\overline{R}(x)$ appearing in eq.\ \ref{eq:Rbarprecise}. Ref.\ \cite{phillies1998a} performed a numerical integration of these equations, showing that $D_{s}(c)$ is very nearly a simple exponential in $c$, and that the calculated $D_{s}(c)$ is very nearly independent of $c_{r}$ so long as $c_{r}$ is small enough that $D_{s}(c_{r}) \approx 1$. The reference further noted evidence from viscosity measurements that the renormalization group development could have an interesting fixed point structure, but that here only the fixed point at the origin would be taken into account.

As the final step in the analysis, the issue of the concentration dependence of $R_{g}$ was considered, at the level of approximation of the Daoud formula, eq.\ \ref{eq:chaincontract}. The proposed approach to calculating $D_{s}(c)$ was to imagine using the Positive Function Renormalization Group separately for each final concentration, in each case performing the process with chains whose size was independent of concentration but which were the correct size for the target final concentration. The needed integration of eq.\ \ref{eq:Dspredict} was performed analytically by limiting terms to the $\mathcal{O}(R^{4})$ level, leading to
\begin{equation}
       D_{s} = D_{o} \exp( - \overline{a} R_{g}^{4} c^{1})
       \label{eq:Dsrgsemifinal}
\end{equation}
or finally
\begin{equation}
       D_{s} = D_{o} \exp( - \overline{a} R_{o}^{4} c^{1-2x}).
       \label{eq:Dsrgfinal}
\end{equation}
which is the universal scaling equation.  The above analysis finds that this result is the $\mathcal{O}(R^{4})$ approximant to a  more accurate result.

Reference \cite{phillies1998a} also demonstrates that exponentials and stretched exponentials in $c$ and $R$ are invariants of the Positive Function Group transformation: If you start with a stretched exponential in $c$ and $R$, you end up with a stretched exponential in $c$ and $R$ as the outcome of the renormalization transformation.

\section{Polymer Solution Viscoelasticity from Two-Parameter Temporal Scaling\label{sectiontwoparameter}}

We now make a change of pace.  The use of renormalization group procedures to extrapolate $D_{s}(c)$ and $\eta(c)$ to elevated concentrations suggested using renormalization group approaches to infer the frequency dependences of those properties.  In the above, calculations of hydrodynamic interactions were primary, with self-similarity or the Positive Function Renormalization Group being used to extend those calculations to elevated polymer concentrations.  In this Section,  we focus almost entirely on the renormalization group properties of the calculation, deducing aspects of the fixed-point structure of  the renormalization group for the viscosity from empirical evidence. We then extend this analysis to a two-parameter form, thereby inferring the functional form for the frequency dependence of the loss and storage moduli.

The approach was put into effect in ref.\ \cite{phillies1999a}, which introduced two-parameter temporal scaling to calculate  how the loss modulus $G''(\omega)$ depends on frequency. The approach was entirely successful so far as it went, but has limitations that still need to be overcome.  First, temporal scaling predicts the functional dependence of $G''(\omega)$ and therefore the storage modulus $G'(\omega)$ on $\omega$, but in its current form temporal scaling gives no information on any numerical parameters found in the predicted functions. Temporal scaling does not yet predict how those parameters depend on polymer concentration or molecular weight, let alone what values the parameters have. Second, temporal scaling does not invoke a molecular model of a polymer solution.  As a result, its predictions are substantially noncommunicating with treatments of polymer viscoelasticity that begin with detailed models for molecular motions and intermolecular forces, such as those by Graessley\cite{graessley1965a,graessley1967a}, Bird, et al.\cite{bird1982a,bird1982b}, and Raspaud, et al.\cite{raspaud1995a}.

\subsection{Two-Parameter Temporal Scaling: Fundamental Approaches}

The two-parameter temporal scaling approach has five theoretical parts and an experimental confirmation.

The five theoretical parts lead us to the frequency dependence of $g''(\omega)$. First, the renormalization group derivation of the universal scaling equation for $D_{s}$ is used to treat the low-shear solution viscosity $\eta$.  Second, the phenomenological\cite{phillies2011a} behavior of the solution viscosity is examined.  Third, the experimental phenomenology for $\eta$ is used to infer the fixed point structure of the full renormalization group treatment of $\eta(c)$.  Fourth, we advance from one- to two-parameter scaling by recognizing that $\eta(c)$ is the low-frequency limit of $\eta(c,\omega)$.  Fifth, from the inferred fixed-point structure of the associated renormalization group we infer how $\eta(c,\omega)$ depends on $\omega$ at fixed $c$. Finally, comparison is made with the experimental literature, finding that the two-parameter temporal scaling approach correctly predicts the observed frequency dependences. In more detail:

First, as discussed above, the Hydrodynamic Scaling Model for self-diffusion leads to power series for $D_{s}$, which the positive function renormalization group approach transforms into an exponential concentration dependence for $D_{s}$. The corresponding hydrodynamic calculation for the viscosity, and the same renormalization group approach, leads to an exponential concentration dependence for $\eta$.  In each case, the effect of chain contraction with increasing polymer concentration is to replace the simple-exponential concentration dependence with a stretched-exponential concentration dependence.

The remainder of the analysis invokes only the renormalization group aspect of the calculation, and depends not at all on the assumption of the Hydrodynamic Scaling Model that interchain interactions in solution are dominated by hydrodynamics.  If interchain interactions were instead dominated by chain crossing constraints or by \emph{cryptocrystallites}\cite{rouse1998a}, but the low-concentration behavior was still a power series in concentration, the renormalization group part of the analysis would suffer only quantitative changes.

Second, there is an extensive experimental phenomenology for polymer solution viscosity.  Reviews\cite{phillies2011a,phillies2011b} of nearly the entirety of the phenomenological literature on $\eta(c)$ find that $\eta(c)$ does indeed have the predicted stretched-exponential concentration dependence. In many but not all systems, there is an elevated concentration $c^{+}$ above which $\eta$ instead depends on $c$ as a power law
\begin{equation}
      \eta = \bar{\eta} c^{x}
      \label{eq:etapowerlaw}
\end{equation}
in $c$, and not as a stretched exponential in $c$.  Here $\bar{\eta}$ and $x$ are phenomenological constants. We describe the transition at $c^{+}$ as the \emph{solutionlike-meltlike transition}. When the transition occurs, the transition concentration is typically $c^{+} [ \eta ] \approx 24-40$, with $ [ \eta ]$ being the intrinsic viscosity. In other systems, $c^{+}[ \eta ]$ is found to be as large as 150 or as small as 4. In yet other systems no transition is observed.  In all systems, the stretched-exponential curve admits eq.\ \ref{eq:etapowerlaw} as a local tangent.  This local tangential behavior is not the solutionlike-meltlike transition.

Milas, et al.\cite{milas1990a} and Graessley, et al.\cite{graessley1976a} report viscoelastic parameters, including in various studies the low-shear viscosity, steady-state compliance $J_{e}^{o}$ and characteristic shear rate $\dot{\gamma}_{r}$ for non-Newtonian behavior for solutions of linear\cite{milas1990a,graessley1976a} and star polymers\cite{graessley1976a}. These results were systematically reanalyzed\cite{phillies1995b}. For each system examined, all viscoelastic parameters measured consistently showed either the same solutionlike behavior or the same meltlike behavior.  When a solutionlike-meltlike transition occurred, it occurred at the same concentration for all parameters measured.

Third, the transition at $c^{+}$  might be envisioned to have either a physical or a mathematical basis.  As a physical transition, there could at $c^{+}$ be a crossover in the nature of the dominant force controlling the solution dynamics.  For example, the crossover could be from domination by hydrodynamic interactions at lower concentrations to domination by chain crossing/entanglement interactions at elevated concentrations. On the other hand, as a mathematical transition, there could be a change in the nature of the mathematical solutions, for example because the identity of the fixed point controlling the renormalization process changes with increasing $c$.

A physical transition might plausibly: occur at the same $c [\eta]$ in different systems (because chain crossing constraints are not sensitive to chemical details of the chain structure), cover a wide range of concentration (because near $c^{+}$ the dominant forces would be competitive), and show near $c^{+}$ a discontinuity in $d \eta/dc$ (because there is no reason for the different forces that dominate below and above $c^{+}$ to give the same slope as $c \rightarrow c^{+}$).  On the other hand, a mathematical transition might plausibly occur over a narrow range of concentrations, occur at very different $c [\eta]$ in different systems, and be analytic (first derivative $d \eta/dc$ continuous) at $c^{+}$.

As it happens, the transition in the concentration dependence of $\eta$ shows precisely the traits expected for a mathematical transition. Furthermore, a transition that is rather similar  to the solutionlike-meltlike transition is seen for $\eta(c)$ of spherical microgel melts and hard-sphere colloids, so the transition cannot be due to any hypothetical crossover to chain reptation at elevated polymer concentrations. After all, spheres cannot reptate. \emph{For the purpose of motivating the investigations of the remainder of this Section, we take as a postulate that $c^{+}$ marks a mathematical fixed point transition.} The low-concentration stretched-exponential behavior corresponds to the fixed point at $c=0$, while the elevated-concentration power-law behavior corresponds to a fixed point at some large concentration.

Fourth, the discussion thus far has taken $\eta$ to be a function of the single variable $c$.  However, polymer solutions are viscoelastic. Their viscoelastic responses are characterized by a frequency-dependent loss modulus $G''(\omega)$ and a frequency-dependent storage modulus $G'(\omega)$. The moduli are also concentration-dependent but in the usual convention one writes $G''(\omega)$ and not $G''(\omega, c)$.  The viscosity $\eta(c)$ is the low-frequency limit of a frequency-dependent viscosity
\begin{equation}
     \eta(c,\omega) = \frac{G''(\omega)}{\omega}.
     \label{eq:etaomega}
\end{equation}
The discussion so far considers on $\eta(c)$, so when we extend to frequency dependence we consider $G''(\omega)/\omega$ and $G'(\omega)/\omega^{2}$ and not $G''(\omega)$ or $G'(\omega)$. $G''(\omega)/\omega$ and $G'(\omega)/\omega^{2}$ have the appropriate property that they go to constants when $\omega \rightarrow 0$.

Fifth, it is assumed that $\eta(c, \omega)$ is dominated by the same fixed points that determine the behavior of $\eta(c,\omega)|_{\omega \rightarrow 0}$.  Consider a $(c, \omega)$ plane, the $c$ axis being horizontal and the $\omega$ axis being vertical. If one proceeds away from $(c,\omega) = (0,0)$ by moving along the line $\omega=0$, one observes the dependence of $\eta(c,0)$ on $c$.  With increasing $c$, one eventually encounters the  solutionlike-meltlike transition. At smaller $c$, $\eta(c,0)$ depends on $c$ as a stretched exponential, corresponding to a renormalization group fixed point at the origin. Above the transition, $\eta(c,0)$ depends on $c$ as a power law in $c$, corresponding to the dominance of a second fixed point that is not at the origin.  If, instead of staying at $\omega = 0$, one instead advanced away from the $\omega = 0$ axis by moving perpendicular to the $\omega=0$ axis, thereby staying at fixed $c$, the same fixed points would control the behavior of $\eta(c,\omega)$. The $\omega$ dependence would then be a stretched exponential at smaller $\omega$, due to the fixed point at $(0,0)$, and a power law at larger $\omega$, due to the second fixed point at some larger $(c,\omega)$.
\begin{equation}
    \frac{G''(\omega)}{\omega} =
    \begin{cases}
      G_{20} \exp(- \alpha \omega^{\delta}) , & \text{if $\omega \leq \omega_{t}$}\\
       \overline{G}_{20} \omega^{-x}, &   \text{if $\omega \geq \omega_{t}$}.
    \end{cases}
    \label{eq:frequencies}
\end{equation}
Similar two-case formulae are expected to describe $G'(\omega)/\omega^{2}$ and $\eta(\kappa)$, $\kappa$ being the shear rate. The six parameters $G_{20}$, $\alpha$, $\delta$, $\omega_{t}$, $\overline{G_{20}}$, and $\omega_{t}$ are numerical constants appropriate to the particular polymer, its molecular weight, and its solution concentration.  Because the transition is predicted and found to be continuous and analytic (functions and first derivatives the same at the transition frequency $\omega_{t}$), these six parameters are not all independent from each other. Between them, there are are only four independent parameters.

The \emph{ansatz} given here is not a complete derivation. However, as is shown in the original paper\cite{phillies1999a} and in ref.\ \cite{phillies2011a}, chapter 13, equation \ref{eq:frequencies} and the corresponding equations for $G'(\omega)/\omega^{2}$, and separately for $\eta(\kappa)$, are in excellent agreement with experimental studies of the storage and loss moduli and of shear thinning. Furthermore, as would be expected for physically significant variables, the six parameters found in these equations all show smooth dependences, often power laws, on $c$.on

\section{Brief Description of Individual Historical Papers}

In this Section we present short summaries of the papers that developed and tested the hydrodynamic scaling model.  The original paper in the series was ref.\ \cite{phillies1985y}, \emph{Phenomenological Scaling Laws for `Semidilute' Macromolecule Solutions from Light Scattering by Optical Probe Particles}, which was the first systematic review of optical probe particles diffusing through matrix polymer solutions. The probes were polystyrene latex spheres and bovine serum albumin, diffusing through solutions of several water-soluble polymers and bovine serum serum albumin. Probe motion was determined using quasielastic light scattering. The paper set themes for later work, identified areas that were later explored, and made clear that experiment did not match some contemporary models. The probe diffusion coefficient $D_{p}$ was found to follow the stretched exponential
\begin{equation}
     D_{p} = D_{p0} \exp(- \alpha c^{\nu} M^{\gamma} R^{\delta}),
     \label{eq:scaling}
\end{equation}
with $c$, $M$, and $R$ being the polymer concentration and molecular weight, and the probe radius, respectively, $D_{p0}$ and $\alpha$ being constants. The exponents were found to be $\nu = 0.6-1.0$, $\gamma = 0.8 \pm 0.1$, and $\delta$ in the range 0 to -0.1, contrary to some theoretical expectations\cite{langevin1978a} that one should find  $\gamma = 0$ and $\delta = 1$.  For small polymers $D_{s}$ tracked the solution fluidity $\eta^{-1}$. For large ($M \geq 100$ kDa) polymers, probes diffuse faster than expected from $\eta^{-1}$, even when the probes were extremely large ($R \approx 0.62 \mu$m).

The successor paper\cite{phillies1987a} \emph{Universal Scaling Equation for Self-Diffusion by Macromolecules in Solution} extended the work in the previous paper to the polymer self-diffusion coefficient $D_{s}$.  It was shown that the then-available measurements of $D_{s}$ at elevated concentration uniformly fit stretched exponentials in $c$, but did not fit the power laws predicted by some scaling models. Also, the stretched exponential described $D_{s}(c)$ accurately over a full range of concentrations, with no indication of a discontinuity at some elevated 'entanglement' concentration $c^{*}$.  These results were not widely expected on the basis of other polymer models, leading to the criticism that the finding was purely phenomenological, and the emphatic suggestion\cite{ware} that a derivation of the universal scaling equation was needed.

The skeleton of a derivation for the universal scaling equation for polymer self-diffusion was soon supplied\cite{phillies1987a}.  \emph{Dynamics of Polymers in Concentrated Solution: The Universal Scaling Equation Derived} obtained the stretched exponential for $D_{s}$. The model was non-reptational; collective (hydrodynamic) modes were taken to dominate local (entanglement) modes. The derivation reached the stretched exponential via a self-similarity argument, an assumed form for chain-chain hydrodynamic interactions, and the known tendency of random-coil polymer coils to contract in concentrated polymer solutions. The model predicted that $\alpha \sim M^{1}$, and that $\nu$ changes from $1$ to $1/2$ with increasing polymer molecular weight, these results being confirmed by review of the literature.

The mathematical structure of the derivation did not rely on transport properties unique to self-diffusion.  The same approach, with numerical modifications reflecting the quantity being calculated, was therefore expected to be applicable to other transport properties.  Indeed, this author and Peczak\cite{phillies1988a} showed in \emph{The Ubiquity of Stretched-Exponential Forms in Polymer Dynamics} that polymer solution transport properties generally follow stretched-exponential concentration dependences.

A simplest form of Kirkwood-Riseman model\cite{phillies1988c} for polymer dynamics appears in \emph{Quantitative Prediction of $\alpha$ in the Scaling Law for Self-Diffusion}, where it was used to compute $\alpha$.  For 1 MDa polystyrene, the calculation gave $\alpha = 2$, while $\alpha \approx 0.7$ is the experimental number.  As $M$ is reduced, the calculated $\alpha$  decreases more rapidly than the measured $\alpha$, so the error in the calculated $\alpha$ at smaller $M$  is closer to 50\% than it is to a factor of three. However, the calculation did not incorporate intrachain hydrodynamics, and took the distance of closest approach of two polymer chains to be twice the monomer radius, both of these approximations tending to increase $\alpha$, so it is not surprising that our approximation for $\alpha$ gave a value larger than the experimental one.

The calculation of ref.\ \cite{phillies1988c} was extended to treat self-diffusion by star polymers.  In \emph{Chain Architecture in the Hydrodynamic Scaling Model}, it was shown\cite{phillies1990a} that if one compares self diffusion of linear polymers and of many-armed star polymers, the polymers being of equal \emph{total} molecular weight, a solution of matrix polymers is modestly more effective at retarding the linear polymer.  However, if one compares self-diffusion of linear and star polymers at equal arm molecular weight, a linear polymer being a two-armed star, the matrix polymer is far more effective at retarding the star polymer than at retarding the linear polymer.

The short paper \emph{The Hydrodynamic Scaling Model for Polymer Dynamics}\cite{phillies1991a} notes a series of experimental tests distinguishing between the hydrodynamic scaling and reptation-scaling models, including (i) presence or absence of multiple dynamic regimes, (ii) difference or lack of difference between sphere and random-coil polymer diffusion, (iii) power-law or stretched-exponential concentration and molecular weight dependences of $D_{s}$ and $\eta$, strong or weak effect of probe radius on $D_{p}/D_{p0}$, and effect of chain architecture on $D_{s}$.  For every test, hydrodynamic scaling is preferred to reptation-scaling, showing that the ongoing theoretical project to refine the Hydrodynamic Scaling Model was on the right track. The paper echoed analysis in the extended article \emph{The Hydrodynamic Scaling Model for Polymer Self-Diffusion}\cite{phillies1989a}, in particular that there was a need for substantial additional measurements on solutions of small-$M$ (say, $\leq 100$ kDa) polymers and a requirements that measurements should carried systematically down systematically to a zero matrix concentration.  Furthermore, based in particular on studies of how $D_{p}$ depends on $R$, it was clear that polymer solutions are qualitatively not like chemically crosslinked gels, even on short time scales.

The paper \emph{Range of Validity of the Hydrodynamic Scaling Model}\cite{phillies1992a} observes that solvent-mediated interactions are absent in the melt (except as one views the melt as its own solvent), and therefore with increasing concentration there should be a transition in dynamic behavior. In this paper, the transition was identified with the change from stretched-exponential to power-law concentration dependences.  This interpretation was not sustained by more modern work, but the solutionlike-meltlike transition still had an important role in understanding polymer dynamics.

In their paper \emph{Higher-Order Hydrodynamic Interactions in the Calculation of Polymer Transport Properties}, Phillies and Kirkitelos\cite{phillies1993b} examined consequences of higher-order bead-bead interactions.  Bead-bead interactions are usually modelled using the Oseen tensor.  However, it is entirely clear from calculations of the self and mutual diffusion coefficients of colloidal spheres\cite{carter1985a} that the Oseen tensor is totally inadequate as an approximation for the true hydrodynamic interaction tensor, and that higher-order (in $a/R$, $a$ being the bead radius and $R$ being the distance between beads) terms must be included. Phillies and Kirkitelos included higher-order terms in calculating $D_{s}$.  Furthermore, they calculated the effect of interchain interactions on polymer bead and free monomer mobilities, showing that the effects are not the same. Inferring a concentration dependence for the friction coefficient of individual polymer beads from the concentration dependence of the friction coefficient of free small molecules in solution (the \emph{monomer friction coefficient correction}) is therefore fundamentally invalid.

Phillies and Quinlan\cite{phillies1995a}, in \emph{Analytic Structure of the Solutionlike-Meltlike Transition in Polymer Solution Dynamics}, report a high-precision detailed study of the analytic structure of the solutionlike-meltlike transition in the viscosity.  $\eta(c)$ of several hydroxypropylcellulose samples shows a stretched-exponential concentration dependence at at smaller $c$ and a power-law concentration dependence at larger $c$.  Phillies and Quinlan showed that the viscosity transition is analytic -- the functions and their first derivatives are both continuous -- at the transition concentration. A later analysis of the literature in \emph{Viscosity of Hard Sphere Suspensions}\cite{phillies2002c} shows that hard and soft-sphere suspensions show the same solutionlike-meltlike transition in $\eta(c)$, at a concentration well below the concentration of the known phase transition, thus demonstrating that the solutionlike-meltlike transition does not arise from topological effects unique to long linear polymers.

Writing in \emph{Hydrodynamic Scaling of Viscosity and Viscoelasticity of Polymer Solutions, Including Chain Architecture and Solvent Quality Effects}, Phillies\cite{phillies1995b} applied the universal scaling equation and power law forms to the concentration and molecular weight dependences of various viscoelastic parameters, including results on linear and star polymers and systems having various solvent qualities.  This paper was primarily a phenomenological study; model calculations corresponding to the viscoelastic parameters have not yet been made.

The paper \emph{Quantitative Experimental Confirmation of the Chain Contraction Assumption of the Hydrodynamic Scaling Model}\cite{phillies1997a} takes advantage of a unique feature of dielectric relaxation spectroscopy, namely with some polymers the technique can measure both the rotation of the chain end-to-end vector and also the length of that vector. The Hydrodynamic Scaling Model asserts that the deviation of the concentration dependence of transport properties from a simple exponential in concentration is caused by chain contraction at elevated polymer concentrations. The dielectric relaxation measurements confirm that the model's assertion is quantitatively exact.

Phillies, et al.'s\cite{phillies1997b} paper \emph{Probe Diffusion in Sodium Polystyrene Sulfonate - Water: Experimental Determination of Sphere-Chain Binary Hydrodynamic Interactions} made a quantitative test of the hydrodynamic model used here. $D_{c}$ of three sizes of polystyrene sphere was determined in solutions of seven different monodisperse polystyrene sulphonates ($1.5 \leq M \leq 1188$ kDa), each at ten or more concentrations up to 20 g/L.  The initial slopes $dD_{p}/dc|_{c \to 0}$ were compared with theory. The one uncertainty is the diameter to be assigned to the polymer's monomeric subunits. Fortunately, for large probes the slopes are very nearly independent of the assumed subunit diameter.  For $M > 10$ kDa quantitative agreement between measurement and theory was obtained, confirming the validity of the hydrodynamic calculation.

The paper \emph{Derivation of the Universal Scaling Equation of the Hydrodynamic Scaling Model via Renormalization Group Analysis}\cite{phillies1998a} replaced the self-similarity approach of Ref. \cite{phillies1987a} with a calculation based on the positive-function renormalization group\cite{altenberger1997a,altenberger1997b}.  The Hydrodynamic Scaling Model's universal scaling equation was again obtained, via  a very different approach.

Use of renormalization group techniques in the prior paper led to a much more radical paper\cite{phillies1999a}  \emph{Polymer Solution Viscoelasticity from Two-Parameter Temporal Scaling}, which proposed to find the frequency dependence of the loss modulus from a consideration of the fixed points of a hypothetical renormalization group derivation of the viscosity $\eta(c)$, made on the lines of ref.\ \cite{phillies1998a}, after identifying $\eta(c)$ as  a one-dimensional slice of $\eta(c,\omega)$.  The \emph{ansatz} in the paper leads to the conclusion that $G''(\omega)/\omega$ is a stretched exponential in $\omega$ at lower frequencies and a power law in $\omega$ at elevated frequencies, the transition between the two regimes being continuous and analytic. Preliminary tests against literature data were highly satisfactory.  Further tests reported in \emph{Temporal Scaling Analysis: Viscoelastic Properties of Star Polymers}\cite{phillies1999b}, \emph{Temporal Scaling Analysis: Linear and Crosslinked Polymers}\cite{phillies2002a}, and  \emph{Viscosity of Hard Sphere Suspensions}\cite{phillies2002c} were equally satisfactory.  In particular, it was confirmed that the predicted forms, with fitted parameters, agreed with Kronig-Kramers relations.  Furthermore, the forms that describe well $G''(\omega)$ and $G'(\omega)$ of linear polymers also describe well  $G''(\omega)$ and $G'(\omega)$ of spherical microgel melts, showing that chain topology does not make a qualitative contribution to the functional forms of the loss and storage moduli.

In \emph{Low-Shear Viscosity of Non-Dilute Polymer Solutions from a Generalized Kirkwood-Riseman Model}\cite{phillies2002b}, the model of ref.\ \cite{phillies1988c} was employed to calculate the concentration dependence of the viscosity, including interacting pairs and trios of polymer chains. An extended computation leads to values for the initial slope $d \eta / dc$ and for the Huggins coefficient. The results of this calculation were used in \emph{Self-Consistency of Hydrodynamic Models for the Low-Shear Viscosity and the Self-Diffusion Coefficient}\cite{phillies2002d} to calculate $\alpha$ of equation \ref{eq:stretchedexpDs}. Taking the Huggins coefficient from the viscosity calculation as an experimental input, $\alpha$ for self-diffusion was determined, as a function of polymer molecular weight, with no free parameters.  Comparison with experimental determinations $\alpha$ found almost exact agreement over four orders of magnitude in $M$ and $\alpha$.

Finally, the renormalization group treatment of $D_{s}(c)$ requires as input a power-series expansion for $D_{s}(c)$.  As the first step toward advancing on that expansion, in \emph{Fourth-Order Hydrodynamic Contribution to the Polymer Self-Diffusion Coefficient}\cite{merriam2004a} Merriam and Phillies used a hydrodynamic multiple-scattering approach to compute the chain-chain-chain-chain-chain hydrodynamic interaction tensor, which could be used to calculate the $c^{3}$ correction to $D_{s}$.

\section{Phenomenological Evidence}

This section presents phenomenological evidence on polymer dynamics.  As will be seen, the evidence supports the Hydrodynamic Scaling Model.
In understanding the evidence and what it means, it is worthwhile to begin with the philosophical observations of Thomas Kuhn\cite{kuhn1962a} on how theories compete. Kuhn's fundamental thesis is that one's model of the world influences which experiments need to be made, which quantities need to be calculated or elsewise predicted, and which sorts of data are important. An experiment that is viewed as a critical test of one model may for a different model appear to be only of marginal relevance. In the end, a successful model predicts all experimental observations within its scope, but at the earlier stages of adoption some sorts of measurements taken to be central and others are taken to be marginal, to be considered later.

As a an example of the above matter, Kuhn treated the early-19th century competition between phlogiston models for chemical structure and Dalton's Law of Multiple Proportions. The phlogiston model was widely accepted because it was extremely successful.  It ordered and explained vast amounts of descriptive chemical information, for example, the model explained why pure metals are more similar to each other than their oxides are similar to each other. While it was entirely possible to weigh the amount of each element needed to form a particular chemical compound, within the phlogiston picture such measurements did not appear to matter. Dalton's Law of Multiple Proportions put that interpretation on its head, proposing that the weight of each element in a pure compound was the central chemical fact. The descriptive material explained well by the phlogiston models, such as the colors of the metallic oxides, were set aside, to be explained as it turned out a century and a half later with quantum mechanics. Materials that were in fact solid solutions of several compounds, leading to material substances in which the Law of Multiple Proportions was not followed, were viewed as anomalous special cases not as disproofs of the model. In moving from the phlogiston model to Dalton's Law, not only did science change how one described matter, but it also changed which experimental findings were to be treated as marginal results, and which experimental findings were to be treated as central tests of the theory.

The difference in world view between the hydrodynamic scaling and reptation-scaling models arises already in data presentation and experimental plans. The models predict that transport coefficients depend on concentration and molecular weight as stretched exponentials and as power laws, respectively. Furthermore, the Hydrodynamic Scaling Model, if correct, is valid from dilute solution up to some large concentration, while reptation-scaling models, if correct, are only valid at concentrations above some overlap concentration $c^{*}$ and extending to the melt. In consequence, an experiment whose plan arises from hydrodynamic scaling concepts includes measurements on dilute as well as concentrated solution behavior.  Experiments whose plans arise from reptation-scaling concepts often only report measurements on solutions having $c > c^{*}$, and are therefore not always helpful for testing hydrodynamic scaling.  Correspondingly, in making graphical presentations of transport coefficients against $c$ or $M$, data testing reptation-scaling models are usefully set out on log-log plots, while data arising from Hydrodynamic Scaling Models are necessarily set out on linear or semilog plots.

The following sections treat experimental tests of various aspects of the Hydrodynamic Scaling Model. Section  \ref{ss:tensors}presents an experimental test of the accuracy of the generalized Kirkwood-Riseman model for intermacromolecular hydrodynamic interactions. Section \ref{ss:dependences} considers tests of the predicted functional dependences of $D_{p}$ on $c$ and $M$.  Predictions for $D_{s}$ and $\eta$ require a notional bead radius $a$.  Section \ref{ss:beada} demonstrates that values of $a$ from $D_{s}$ and from $\eta$ are mutually consistent.  In the model, non-exponentiality arises from chain contraction; section \ref{ss:drs} uses the literature on dielectric relaxation of polymers in solution to demonstrate that chain contraction accounts quantitatively for the non-exponential concentration dependence of the dielectric relaxation time. Chain contraction and expansion are also affected by solvent quality. Section \ref{ss:etaQ} examines results of Dreval, et al.\cite{dreval1973a} that compare concentrated-solution viscosity and solvent quality, and results of Phillies and Clomenil\cite{phillies1993a} on probe diffusion through polymers in good and theta solvents. Finally, at some concentration the simple model presented here is obliged to become inapplicable, because the polymer molecules are too close together to treat the solvent as a continuum.  When this concentration is exceeded, a transition must take place. Section \ref{ss:tomelt} presents evidence that this predicted transition has been observed.

\subsection{Measurements of the Hydrodynamic Interaction Tensor\label{ss:tensors}}

The hydrodynamic scaling treatment is based on an extended Kirkwood-Riseman model.  Is the extended Kirkwood-Riseman model accurate? Phillies, Lacroix, and Yambert\cite{phillies1997b} made a quantitative experimental test of the model of Section \ref{s:extendedKRg}.  The test was successful.  They used quasi-elastic light scattering spectroscopy to measure the diffusion coefficient $D_{p}$ of polystyrene latex spheres of three known sizes through solutions of seven polystyrene sulphonates, $1.5 \leq M \leq 1190$ kDa, at a series of polymer concentrations. Anomalous polyelectrolyte effects were suppressed by working in 0.2M NaCl.  The initial slopes $\alpha = \lim_{c \rightarrow 0} d D_{p}/dc$ were determined.

Comparison was then made with the hydrodynamic calculations of Phillies and Kirkitel\-os\cite{phillies1993b}, in which bead-bead hydrodynamic interactions were truncated at the $r^{-3}$ (Rotne-Prager) level. Probe spheres were treated as a single polymer bead having a known large radius. What size $a$  should be assigned to the monomer beads in the polymer? The algebraic answer is a function of the sum of the monomer and probe radii. The polymer beads were much large than the polymeric monomer units, so the exact size assigned to the monomers has little effect on the calculation. Polymer radii of gyration and hydrodynamic radii were calculated from their molecular weights using the data of Pietzsch, et al.\cite{pietzsch1992a}. The Phillies-Kirkitelos calculation thus  \emph{has no free parameters}. It predicts numerical values of $\alpha$. For $M < 10$ kDa, it appears inaccurate to model the somewhat rigid polystyrene sulphonate as a gaussian-random cloud of monomer beads. For polymer $M > 10$ kDa, nearly quantitative agreement between calculated and measured values of $\alpha$ was found, as seen in ref.\ \cite{phillies1997b}, Figure 4.  $\alpha$ for a 1 MDa polymer was calculated to be $\approx 0.15$, and was found experimentally to be $\approx 0.13$. Over nearly a hundred-fold range of polymer molecular weights, $\alpha \sim M^{\gamma}$ for $\gamma=1$ was found, in agreement with the Phillies and Kirkitelos calculation.

These experimental results directly confirm the validity of the hydrodynamic approach for calculating interchain hydrodynamic interactions, at least for the self terms $b_{ii}$ of the calculated mobility. These experiments did not test self-similarity, the Positive-Function Renormalization group process, or the size of the two-chain tensor $T_{ij}$.

\subsection{Concentration and Molecular Weight Dependences\label{ss:dependences}}

The Hydrodynamic Scaling Model, via either self-similarity or the Positive Function Renormalization Group, predicts that transport coefficients have stretched-exponential dependences on concentration and molecular weight. The scaling prefactor $\alpha$ and scaling exponent $\nu$ of the stretched exponential are predicted by the model to depend on $M$ but to be independent of polymer concentration. The PFRG mathematical structure has a route permitting a transition to a power-law concentration dependence at large $c$.

Experimentally\cite{phillies2011a-89}, $D_{s}$, $D_{p}$, and $\eta$ follow stretched exponentials in $c$, beginning at extreme dilution and extending out to elevated polymer concentrations, often $c [\eta] \gg 1$. For $D_{s}$ and $D_{p}$, there are no indications of a discontinuity or change in slope of the concentration dependence for some $c[\eta]$ near unity.  The lack of a discontinuity agrees with the Hydrodynamic Scaling Model presumption that the same dynamics apply in dilute and non-dilute solutions.  The same lack of a discontinuity is inconsistent with proposals that polymer dynamics change qualitatively at some concentration $c^{*} \approx [\eta]^{-1}$ at which polymer chains overlap and entangle.

The Hydrodynamic Scaling Model for $D_{p}$ predicts that $\alpha \sim M^{\gamma}$ for $\gamma \approx 1$; also, with increasing $M$, $\nu$ should decrease from $1$ to $5/8$ or $1/2$. Phillies and Quinlan measured $\eta$ and $D_{p}$ of 20.4 and 230 nm polystyrene spheres, for dextran solutions having $M$ in the range 10-500 kDa and a range of concentrations. Values for $\alpha$ and $\nu$ were extracted from each set of measurements.  Over nearly two orders of magnitude in $M$, $\alpha \sim M^{0.84}$, while as predicted by the model $\nu$ decreased from $1$ to $5/8$.

If the matrix polymer is replaced with a globular matrix species such as a protein, $D_{p}(c)$ for probe spheres diffusing through the protein solution continues to have a stretched-exponential form.  This result is consistent with the hydrodynamic scaling expectation that replacing a random-coil matrix with a hard-sphere matrix changes numerical coefficients in the hydrodynamic interaction tensor, but has no qualitative effect on $D_{p}(c)$.  This result is inconsistent with reptation-tube model expectations that the dynamics of entangling and non-entangling matrix species should not be similar at high concentrations.\cite{ullmann1985a}

A comparison\cite{phillies1990a} of $D_{s}$ for linear and star polymers diffusing through a matrix solution of dissolved linear polymers as studied by Wheeler and Lodge\cite{lodge1989a} finds that a large concentration of linear polymers is approximately equally effective at retarding the motion of linear and star polymers having the same total molecular weight.  However, comparing linear and 12-armed star polymers having the same arm molecular weight, a linear polymer being a two-armed star, the same matrix solutions are far more effective at retarding the motion of the 12-armed star than at retarding the motion of a linear polymer. These measurements of Wheeler and Lodge\cite{lodge1989a} were shown by this author\cite{phillies1990a} to be consistent with the Hydrodynamic Scaling Model.

At elevated polymer concentrations, $D_{p}$ often shows non-Stokes-Einsteinian behavior, i.e., $\kappa \equiv D_{p}(c)\eta(c)/D_{p}(0)\eta(0) \gg 1$, up to $\kappa \approx 10^{3}$, even for large (e.g. 1 $\mu$m diameter) probes. Non-Stokes-Einsteinian behavior is equally found for probes in non-entangling bovine serum albumin solutions\cite{phillies2011a-89,ullmann1985a}, showing that non-Stokes-Einsteinian behavior is not an indicator for the presence of reptational motion by the matrix.

For $\eta(c)$, $\dot{\gamma}_{r}$, and $J_{e}^{o}$, in some systems but not others, the concentration dependences show at elevated $c$ a transition from a stretched-exponential to a power-law concentration dependence.\cite{phillies1995b} On one hand, reptation-tube models do predict scaling (power law) behavior at large $c$.  On the other hand, reptation-tube models indicate that the transition should be universal, appreciably independent of the details of polymer chemical structure, and therefore should consistently appear at about the same $c[\eta]$.  Experimentally, the transition is not uniform and occurs at greatly different concentrations $c[\eta]$ in different systems, contrary to expectations from tube-type models.

Tube-type models for concentrated solutions ascribe to polymer chains a mode of motion -- reptation -- that is inaccessible to large spheres, implying that chains will diffuse through concentrated solutions of large polymer molecules considerably more rapidly than will spheres that have the same hydrodynamic radius.  Brown and Zhou\cite{brown1989a,phillies1992z}  compared $D_{p}$ of spheres and $D_{s}$ of random coil probes through solutions of the same polymer.  For large probes and chains in solutions of a smaller matrix polymer,  $D_{p}/D_{s}$ was approximately independent of concentration as $D_{p}(c)/D_{p}(0)$ declined more than two orders of magnitude. For smaller probes in solutions of a large matrix polymer, with increasing matrix $c$ the matrix polymer was much more effective at retarding motions of the random coil polymer than at retarding the motion of spherical probes. Their findings are consistent with Hydrodynamic Scaling Models that view hydrodynamic radii  as the central variable, but are inconsistent with tube-type models.

\subsection{The Bead Diameter $a$\label{ss:beada}}

Calculations of the concentration dependences of the self-diffusion coefficient and the viscosity lead to outcomes determined in part by a notional bead diameter $a$. It could be proposed that in each of these calculations there is a free parameter, so quantitative comparisons between data and the theoretical model are impossible. (For probe diffusion, this difficulty does not arise, because $a$ has little effect on $d D_{p}/dc$, as discussed in Section \ref{ss:tensors}.)

$a$ can be estimated from the viscosity calculation.  This $a$ will be denoted $s_{\eta}$.  Pearson\cite{pearson1987a} and Yamakawa\cite{yamakawa1971a} report that $k_{H}$ is in the range 0.3-0.6.  Noting for nondraining spheres $[\eta] = 2.5 \bar{v}$ and in appropriate units $\bar{v} = 4 \pi R^{3} /3$, one finds from eq.\  \ref{eq:kHfinal} $a_{\eta} = 0.18 R$.  The estimated $a_{\eta}$ does not depend strongly on the assumed $k_{H}$.

Second, in ref.\ \cite{phillies1998a} the same hydrodynamic approach was applied to the self-diffusion coefficient $D_{s}$, obtaining\cite{phillies1998a}
\begin{equation}
       \alpha = - \frac{9}{16} \frac{R_{h1}^{2}}{R_{g} a_{D}} \frac{4 \pi
R_{g}^{3}}{3}  \frac{N_{A}}{M}.
      \label{eq:alphaDsvalue}
\end{equation}
Here $R_{g}$  is the radius of gyration, $R_{h}$ is the hydrodynamic radius, $a_{D}$ is a notional bead size inferred from self-diffusion, $N_{A}$ is Avogadro's
number, and $M$ is the polymer molecular weight.  For\cite{adam1977a} $1.27 \times10^{6}$ Da polystyrene in benzene\cite{adam1977a},  $R_{g} \approx 620$\AA, $R_{h} \approx 380$\AA, and from a systematic review\cite{phillies2011a} of the published literature $\alpha \approx - 0.6$ with $c$ in g/L at this molecular weight.   Combining these findings, $a_{D} \approx 0.17 R_{g} $.

The two paths to estimating $a$ indicate that $a$ is $0.18 R$ or $0.17 R_{g} $.  Given the approximations needed to reach this point, the two estimates of $a$ agree to within experimental error and calculational imprecision. Determinations of $a$ from two separate types of data after separate calculations\cite{phillies1987a,phillies1988c,phillies1993b,phillies1998a} based on the Hydrodynamic Scaling Model lead to about the same notional bead diameter. This outcome would be expected if the Hydrodynamic Scaling Model supplied the legitimate physical treatment, but is unlikely if $a_{\eta}$ and $a_{D}$ were simply fitting parameters that had no physical meaning.

\subsection{Dielectric Relaxation Spectroscopy\label{ss:drs}}

Dielectric spectroscopy is sensitive to the size and temporal behavior of polymer dipole moments.  Understanding of polymer dipole moments may be traced back to Stockmayer\cite{stockmayer1967a}, who identified three classes of polymer dipole moment and their relaxations, namely: (1) dipoles whose orientation is determined by the orientation of pendant side groups, and which therefore change direction on very short time scales, (2) dipoles whose direction is determined by the chain contour, and which are aligned perpendicular to the chain contour, so that they are relaxed via local segmental motion, and (3) dipoles associated with the chain contour, that point along the chain contour, so that the magnitude of the dipole moment is determined by the end-to-end-vector of the polymer chain, and which change direction on the slow time scales on which the polymer and its end-to-end vector rotate in space. Stockmayer classed polymers with type 3 dipoles as \emph{type-A polymers}. Dielectric relaxation spectroscopy associated with solutions of type-A polymers has been reviewed by Adachi and Kotaka\cite{adachi1993a}.

A type-A polymer can be viewed as a series of short segments, each segment $i$ having a dipole moment $\bm{d}_{i}$ aligned parallel to the segment. The time-dependent dipole moment $\bm{M}(t)$ of the polymer is the sum of the moments of the $N$ segments
\begin{equation}
      \bm{M}(t) = \sum_{i=1}^{N} \bm{d}_{i}(t).
      \label{eq:dipoleA}
\end{equation}
For identical segments, $\bm{d}_{i}(t) \sim \bm{r}_{i}(t)$, $\bm{r}_{i}(t) $ being the segment end-to-end vector. The mean-square of the polymer end-to-end vector $\langle \bm{R}_{e} \rangle$  is therefore $\langle R_{e}^{2} \rangle \sim \langle \bm{M}(t) \cdot \bm{M}(t) \rangle$. The dipole relaxation function
\begin{equation}
     \Phi(t) = \frac{\langle \bm{M}(t) \cdot \bm{M}(0)\rangle}{\langle \bm{M}(0) \cdot \bm{M}(0)\rangle}
     \label{eq:dipoletime}
\end{equation}
describes the relaxation of the polymer end-to-end vector.  $\Phi(t)$ is usually said to describe rotational diffusion. However, for a random-coil polymer as opposed to a rigid body, the length of the end-to-end vector and hence $| \bm{M}(t)|$ fluctuates in time, so $\Phi(t)$ must also capture the relaxation of fluctuations in the magnitude of $ \bm{M}(t)$.

With dielectric relaxation one can measure both $\Phi(t)$ (and, hence the polymer rotational diffusion coefficient) and $\langle R_{e}^{2} \rangle$. As demonstrated by by Adachi and Kotaka\cite{adachi1993a} and by ref.\ \cite{phillies1997a}, the end-to-end vectors of pairs of polymer molecules are almost certainly very nearly uncorrelated in direction, so dielectric relaxation spectroscopy measures single-chain properties, even in concentrated solutions. Adachi, et al.\cite{adachi1985aDR,adachi1988aDR,adachi1989aDR} exploited their demonstration by measuring the dielectric relaxation strength $\Delta \epsilon$ and relaxation time $\tau_{n}$ of \emph{cis}-polyisoprenes of multiple molecular weights in good and theta solvents at concentrations 0-500 g/l. $\tau_{n}$ increases by as much as several hundredfold over this range. In the same systems $\Delta \epsilon$ decreased with increasing $c$, in some cases by as much as 50\%.

Phillies\cite{phillies1997a} reconsidered the experimental findings of Adachi, et al.\cite{adachi1985aDR,adachi1988aDR,adachi1989aDR}.  Simple phenomenological forms that describe quantitatively the concentration dependence of $\langle R_{e}^{2} \rangle$ were identified. The chain-chain hydrodynamic interaction tensor for rotation-rotation coupling was proposed, based on the comparable sphere-sphere coupling, to scale as $R^{6}/r^{6}$, $R$ being a chain radius and $r$ being a distance between chains. The 1997 analysis proposed that the ensemble average over chain positions had an effective lower limit $a \sim R$, so that the self term of the rotation-rotation coupling had a strength $\sim R^{3}$. Invoking the self-similarity rationale, the renormalization group treatment not having been developed at the time of the paper, ref.\ \cite{phillies1997a} proposed
\begin{equation}
       \tau_{n}(c) = \tau_{n0} \exp(\alpha c \langle (R_{e}(c))^{2}\rangle ^{3/2}).
\label{eq:dielectric}
\end{equation}
The effect of the concentration-dependent chain contraction is then to determine the curvature of $ \tau_{n}(c)$.

For dielectric relaxation $\tau_{n}$ and $\langle (R_{e}(c))^{2}\rangle$ have been measured directly\cite{adachi1985aDR,adachi1988aDR,adachi1989aDR}. A fit of eq.\ \ref{eq:dielectric} to the experimental measurements has two free parameters, namely $\tau_{n0}$ and $\alpha$.  On  a semilog plot, these parameters give an intercept and an initial slope. However, any deviation of $\tau_{n}(c)$ from pure exponential behavior is determined by the known quantity $\langle (R_{e}(c))^{2}\rangle$.

There is quantitative agreement between eq.\ \ref{eq:dielectric}, in its determination of the dependence of $\tau_{n}(c)$ on $R_{e}$,  and experiment.  In theta solvents, polymer sizes are independent from polymer concentration, so $\tau_{n}(c)$ is predicted to be a simple exponential, precisely as found.  In good solvents, $\langle (R_{e}(c))^{2}\rangle$ decreases with increasing $c$, leading to a  $\tau_{n}(c)$ that increases less rapidly than a pure exponential, also as observed. The degree of non-exponential behavior is determined by the amount of chain contraction, as confirmed quantitatively in ref.\ \cite{phillies1997a}, precisely as predicted by the Hydrodynamic Scaling Model and eq.\ \ref{eq:dielectric}. The results in ref.\ \cite{phillies1997a} may be seen as a significant advance over the self-similarity and renormalization group treatments in one key respect, namely an assumed theoretical dependence of $R_{g}$ on $c$ has been replaced with the experimental dependence, thereby creating quantitative agreement of calculation and experiment.

\subsection{Viscosity and Solvent Quality\label{ss:etaQ}}

Dreval, et al.\cite{dreval1973a} review an extremely extensive set of viscoelastic studies not readily available in the Western literature. They consider a reduced viscosity $\tilde{\eta} = (\eta-\eta_{0})/(\eta_{0} c [\eta])$.  For a series of homologous polymers in the same solvent, $c [\eta]$ was found to be a good reducing variable, the intrinsic viscosity $[\eta]$ collapsing $\eta(c)$ for different $M$ onto a single curve.  However, when the same polymer samples were dissolved in several different solvents, plots of $\tilde{\eta}$ against $c[\eta]$ were found to lie on different curves.  Dreval, et al., showed that the various curves could all be reduced onto each other by introducing a new variable $K_{M}$ and plotting $\tilde{\eta}$ against $K_{M} c [\eta]$.  With a correct choice of $K_{M}$ for each solvent:polymer pair, all measurements of $\tilde{c}$ of the same polymer in different solvents could be reduced to a single master curve.

Dreval, et al., then introduced a chain expansion parameter $\alpha_{\eta}$, defined via
\begin{equation}
     \alpha_{\eta}^{3} = \frac{[\eta]_{s}}{[\eta]_{\mathrm[theta]}}.
     \label{eq:alphaetadef}
\end{equation}
Here $[\eta]_{s}$ is the intrinsic viscosity of the polymer in the solvent of interest, and $[\eta]_{\mathrm[theta]}$ is the intrinsic viscosity of the same polymer in a theta solvent, where it is unexpanded.  Dreval, et al, then demonstrated that $K_{M} \alpha_{\eta}^{3}$ is approximately a constant, i.e., $K_{M} \sim \alpha_{\eta}^{-3}$.  Solvent quality thus enters $\eta(c)$ exactly as predicted by the Hydrodynamic Scaling Model and seen in eq.\ \ref{eq:dielectric}, namely $\eta(c)$ is a function of $c \langle (R_{e}(c))^{2}\rangle ^{3/2}$, and $R_{e}$ is determined in part by the solvent quality.  The reducing variable $K_{M}$ divides out the effect of solvent quality on $R_{e}$, so that plots for a given polymer of $\eta(c)$ against $c$ in different solvents are all reduced to the same master curve.

Phillies and Clomenil\cite{phillies1993a} measured the diffusion of 67 nm polystyrene spheres through aqueous solutions followed of 139 kDa hydroxypropylcellulose at temperatures of 10 and 41 C, these being good and near-theta solvent conditions. At  both temperatures, the diffusion coefficient of the spheres followed a stretched exponential $\exp(- \alpha c^{\nu})$, with $\nu=3/4$ under good solvent conditions and $\nu = 1$ under theta conditions.  Recalling the hydrodynamic scaling prediction $\nu = 1-2x$, $x$ being the concentration exponent for chain contraction, with $x > 0$ under good solvent conditions and $x \approx 0$ under theta conditions, one sees that the observed values of $\nu$ were consistent with the hydrodynamic scaling predictions for the effect of solvent quality on $D_{p}(c)$.

\subsection{Transition to the Melt\label{ss:tomelt}}

 In its present form, the Hydrodynamic Scaling Model refers to dilute and concentrated solutions, not to melts or plasticized melts.  As polymer concentration is increased from dilute solution, less and less solvent is present.  At some concentration, one can no longer invoke the image of a polymer solution as lines of polymer beads separated from each other by a continuum fluid. With increasing concentration, the average gap between polymer chains eventually becomes smaller than the size of a polymer molecule, rendering continuum hydrodynamic descriptions inapplicable. When continuum descriptions become inapplicable, the system switches over  from solution behavior (molecules floating in a solvent) to plasticized melt behavior (solvent molecules in pockets intercalated within a mesh of polymer coils).  The hydrodynamic scaling model thus predicts that there should be a qualitative change in polymer dynamic properties at some very high concentration.

 Such a transition has actually been observed.  Studies showing this effect include work by von Meerwall, et al.\cite{meerwall1985y}, Pickup and Blum\cite{pickup1985z}, and Kosfeld and Zumkley\cite{kosfeld1979z}.  The diffusion coefficient of solvent molecules in polymer solutions typically follows a simple exponential dependence on concentration, for polymer concentrations up to $\approx 400$ g/l.  At larger polymer concentrations, the solvent diffusion coefficient decreases considerably more rapidly, namely as a stretched exponential in concentration with exponents in the range 2.4-3.8.  For the actual fits showing this transition, see \emph{Phenomenology of Polymer Solution Dynamics}, Section 5.3.

 A similar transition has been seen in segmental reorientation times\cite{viovy1986y,tardiveau1980y}  $\tau$ in some but not other\cite{johnson1992y} systems. For concentrations out to 0.3 g/g polystyrene in toluene, Viovy and Monnerie\cite{viovy1986y} and Tardiveau\cite{tardiveau1980y} found that $\tau$ increased gradually, perhaps by two-fold, with increasing $c$; between 0.3 and 0.5 g/g $\tau$ increased far more rapidly nearly 30-fold. On the other hand, Johnson, et al.\cite{johnson1992y}, studied rotational diffusion by center-labelled polyisoprene and a small-molecule probe in polyisoprene: tetrahydrofuran for polymer matrix volume fractions ranging from 0.0 to 1.0.  The matrix polymer was much more effective at retarding rotation of the small-molecule probe than at retarding motion of the labelled polymer.  However $\tau$ was consistently a stretched exponential in matrix concentration, with no indication of a transition at some elevated concentration.

A transition is also found near volume fraction 0.4 in the viscosity of hard sphere systems. An analysis of the literature supporting this observation is found in Phillies\cite{phillies2002c}. The viscosity shows a stretched-exponential increase at lower concentrations and a much sharper power-law increase at larger concentrations. The dynamic crossover is found at $0.4 \leq \phi \leq 0.45$ and $5 \leq \eta_{r} \leq 15$, so the viscosity crossover is very clearly not the same as the hard sphere melting transition found at $\phi \cong 0.5$ and $\eta_{r} \cong 50 \pm 5$.

\section{Summary and Directions for Future Development\label{s:future}}

This review has presented a comprehensive treatment of the Hydrodynamic Scaling Model of polymer solution dynamics. The Hydrodynamic scaling model differs from some other treatments of non-dilute polymer solutions in that it takes polymer dynamics up to high concentrations to be dominated by solvent-mediated hydrodynamic interactions. Many other models take the contrary stand, namely that chain crossing constraints dominate the dynamics of nondilute polymer solutions.  We began by examining single-chain behavior, emphasizing the Kirkwood Riseman model that forms the basis of our calculations. We then developed an extended Kirkwood-Riseman model that gives interchain hydrodynamic interactions. The model was used to generate pseudovirial series for the concentration dependences of the self-diffusion coefficient and the low-shear viscosity.  Extrapolations to large polymer concentration were made, based either on self-similarity or on the Altenberger-Dahler Positive-Function Renormalization Group. The observed stretched-exponential concentration dependences were predicted.  An inferred fixed-point structure for the renormalization group led to a two-parameter temporal scaling \emph{ansatz} from which the frequency dependences of the storage and loss moduli were inferred..

This Section notes possible directions for future research. Topics needing further consideration include: (i) inclusion of intrachain hydrodynamic interactions, (ii) incorporation of segmental dynamics, (iii) explanation of the solutionlike-meltlike viscosity transition and (perhaps a different reflection of the same phenomenon) the neutral polymer slow mode, (iv) extension, refinement, or replacement of the Positive-Function Renormalization Group approach to extrapolating to elevated concentrations, and (v) direct treatment of frequency or time dependence, leading to results for the plateau modulus, steady state compliance, characteristic shear rate, and Cox-Merz rule.  On a longer time scale, note (vi) treatment of the polymer as its own (viscoelastic) solvent, leading to modeling of melt systems, and (vii) extension to polyelectrolyte systems.

The hydrodynamic calculations considered above only treat interactions between polymer chains.  Intrachain hydrodynamic interactions have not yet been incorporated into the model, thought Kirkwood and Riseman\cite{kirkwood1948a} show how this incorporation might advance.  These interactions are important because the motion of each bead partially entrains the surrounding solvent, so that beads near the center of a polymer coil are less effective than might have been expected at creating fluid motion in the surrounding solvent.

The model as treated above remains true to the Kirkwood-Riseman spirit, namely it focuses on whole-body translations and rotations, but neglects internal chain motions. Experimental techniques including dielectric relaxation and NMR have been used to examine segmental dynamics, the relative motions of parts of a longer polymer chain, but theoretical interpretation of these measurements in terms of the Hydrodynamic Scaling Model is lacking.

The experimental development that led to the Hydrodynamic Scaling Model began with probe diffusion, measurements of the motion of rigid spheres or other particles through polymer solutions. The current theoretical treatment is appropriate when the probe and polymer coils are similar in size. If the probe is much smaller than the matrix polymers, probe diffusion may be enhanced by the relative motion of parts of nearby chains, i.~e., by segmental dynamics.  Extension of the Hydrodynamic Scaling Model to treat segmental dynamics would advance understanding of both of these issues.

Some concentration dependence issues are not yet resolved. Phillies and Quinlan demonstrated for hydroxypropylcellulose the existence of a solutionlike-meltlike transition, in which the concentration dependence of the viscosity switches over from a stretched-expo\-nential to a power law in concentration.  Similar transitions can be observed for $\eta(c)$ of many but not all other polymers.  However, the transition is not universal, in the sense that it occurs in different polymers at very different concentrations, no matter whether the transition concentration is expressed as a polymer weight concentration (approximately, a volume fraction) or as a concentration in natural units $c [\eta]$, $[\eta]$ being the polymer's intrinsic viscosity.  Furthermore, the transition is observed for solutions of hard and soft spheres in small-molecule solvents. Experiment thus rules out interpretations of the transition as arising from long-chain topological effects. Quasielastic light scattering from polymer and probe solutions finds at elevated concentrations the appearance of a slow mode in the light scattering spectrum.  The slow mode appears at approximately the concentration at which the solutionlike-meltlike transition is observed in the viscosity, suggesting that the two effects have a common origin, perhaps arising from the fixed-point structure of the Positive-Function Renormalization Group for this problem.

Extension of the hydrodynamic calculations to elevated concentrations was based on the Positive-Function Renormalization Group.  The Renormalization Group calculation does bring one to answers that agree with experiment,  but the path is obscured by the indirect nature of renormalization group methods. A clearer calculation might also permit one to calculate higher-order corrections to the basic renormalization group forms.

The model as presented gives an average diffusion coefficient or a low-frequency viscosity.  A two-variable renormalization group \emph{ansatz} was introduced to predict the functional form of the frequency dependences of the storage and loss moduli and the dependence of shear thinning on shear rate.  The prediction is an \emph{ansatz}, not a theoretical derivation. Replacement of the \emph{ansatz} with a full calculation should replace qualitative statements with quantitative predictions. In particular, the predicted functional forms for the frequency dependences embody a series of parameters, each with concentration and molecular weight dependences.  A direct calculation should give numerical values for these parameters. Also, polymer solution dynamics includes linear and non-linear viscoelastic effects. The analytic calculation would predict the plateau modulus, steady state compliance, and characteristic shear rate for shear thinning. One might also reasonably expect that such calculations would both explain the approximate validity of the Cox-Merz rule and predict quantitative corrections to that rule.

Calculations on melts and polyelectrolyte systems remain well into the future.


\begin{thebibliography}{100.}
\bibitem{philliesdaoud1978}  Indeed, the author once specifically asked one of the leading developers of the reptation/scaling model to justify this implicit assumption.  The question was understood, and the developer in question promised to get back to me with an answer.
\end{thebibliography}
\end{document}